\documentclass[twocolumn,superscriptaddress,footnotebib]{revtex4-2}%
\usepackage{amsmath}%
\usepackage{amsfonts}%
\usepackage{amssymb}%

\usepackage{graphicx}

\usepackage{dsfont}
\usepackage{hyperref}
\usepackage{bm}
\usepackage{multirow}
\usepackage{color}
\usepackage{float}
\usepackage{hyperref}

\begin{document}

\title{Strain engineering of transient exciton diffusion in WSe$_2$ monolayers\\ at cryogenic temperatures}

\author{Roberto Rosati}
\email{rosatir@staff.uni-marburg.de}
\affiliation{Department of Physics, Philipps-Universit\"at Marburg, Renthof 7, D-35032 Marburg, Germany}

\author{Mohammed Adel Aly}
\affiliation{Department of Physics, University of M\"unster, Wilhelm-Klemm-Strasse 10, D-48149 M\"unster, Germany}
\affiliation{Department of Physics, Faculty of Science, Ain Shams University, 11566 Cairo, Egypt}

\author{Robert Schmidt}
\affiliation{Department of Physics, University of M\"unster, Wilhelm-Klemm-Strasse 10, D-48149 M\"unster, Germany}

\author{Rudolf Bratschitsch}
\affiliation{Department of Physics, University of M\"unster, Wilhelm-Klemm-Strasse 10, D-48149 M\"unster, Germany}

\author{Ermin Malic}
\affiliation{Department of Physics, Philipps-Universit\"at Marburg, Renthof 7, D-35032 Marburg, Germany}

\begin{abstract}
Tungsten-based transition metal dichalcogenides exhibit dark excitons as the energetically lowest states. These  are crucial for exciton thermalization and propagation and they dominate low-temperature photoluminescence via the emergence of pronounced phonon sidebands. After a resonant excitation, highly mobile hot dark excitons are formed, which quickly thermalize into an equilibrium distribution. The application of strain modifies the exciton energy landscape and, in particular, the relative energy separation between bright and dark exciton states. The impact of strain on the transient photoluminescence and diffusion of non-equilibrium excitons has remained largely unexplored so far. 
In this work, we  investigate the spatiotemporal exciton dynamics in strained hBN-encapsulated WSe$_2$ monolayers at cryogenic temperatures. We demonstrate that tensile strain abruptly increases the excess energy of hot excitons, thereby accelerating their transient diffusion. We trace this back to suppressed phonon-mediated scattering from bright to dark excitons. Furthermore, we predict a periodic modulation of the transient exciton diffusion in the presence of a compressive strain resulting from strain-driven emission of M phonons. The gained microscopic insights illustrate how strain can be used to engineer transient photoluminescence and exciton diffusion in technologically promising 2D semiconductors.

\end{abstract}

\maketitle

\section{Introduction}

Two-dimensional semiconductors, such as transition metal dichalcogenides (TMD), are characterized by tightly bound excitons. The charge neutrality of these Coulomb-bound electron-hole pairs inhibits their control by external electric fields \cite{Malic23}. Only in vertical \cite{Unuchek18,Ciarrocchi19,Leisgang20} or lateral heterostructures \cite{Lau18,Rosati23,Vandoolaeghe26,Sonowal26,Kitaura26}, the appearance of spatially separated dipolar excitons allows electrical control.  The application of 
external mechanical strain presents a strategy to overcome this limitation and obtain directional control of exciton propagation even in TMD monolayers. Strain modifies the exciton energy landscape  in a valley-dependent way \cite{Castellanos13,Conley13,Zhu13,Ghorbani13,Niehues18,Khatibi18,Schmidt16,Deilmann19,Blundo20,Li23,Faria23}. In particular,  the energy of bright excitons shifts to the red with tensile strain, allowing funneling of excitons toward spatial, high-strain regions \cite{Cordovilla18,Moon20,Harats20,Su22,Gelly22,Lee22}, where intense photoluminescence (PL) is observed \cite{Branny17,Palacios17,Kern16}. 
Recently, it was demonstrated that interestingly strain can even result in exciton \textit{anti}-funneling toward spatial, low-strain regions in  tungsten-based TMDs \cite{Rosati21e}. This is driven by dark K$\Lambda$ excitons, which shift blue with strain - opposite to bright KK excitons. Similarly, strain has been shown to  accelerate exciton diffusion at equilibrium \cite{Rosati21a,Uddin22} as well as to enhance PL intensity \cite{Kumar24} and polarization \cite{Kumar25} or to modify exciton \textit{g}-factors \cite{Blundo22} by bringing dark and bright excitons in resonance. 

\begin{figure}[t]
    \centering
	\includegraphics[width=0.97\columnwidth]{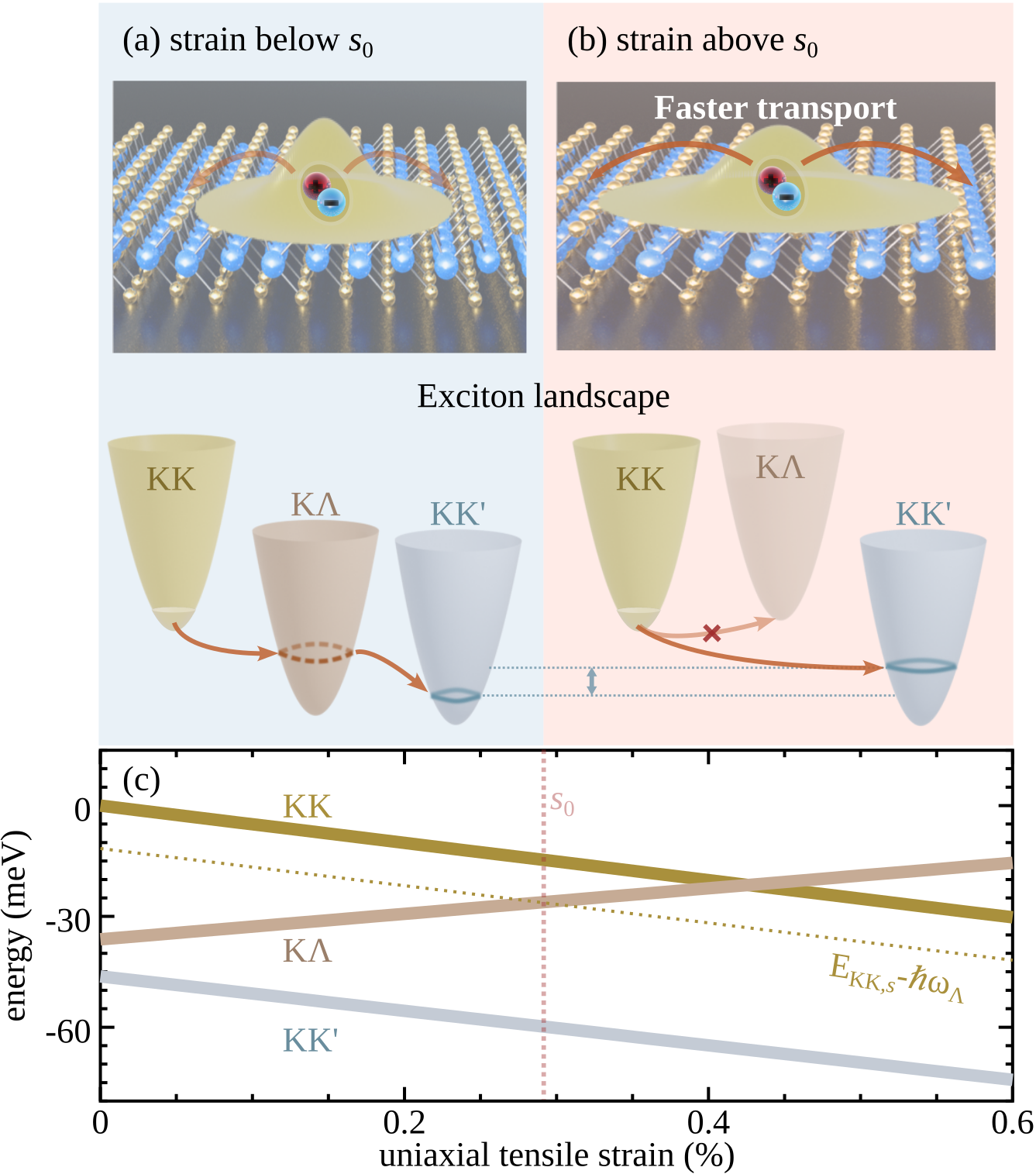}
    \vspace{-0.4cm}
	\caption{\textbf{Strain-engineering of exciton landscape in WSe$_2$ monolayers:} (a,b) Tensile strain leads to a speed-up of exciton transport due to a spectral reordering of KK, K$\Lambda$ and KK$^\prime$ excitons. Strain particularly modifies possible scattering channels and exciton thermalization. Strain larger than the critical value $s_0\approx 0.3\,\%$ deactivates the phonon-mediated KK-to-K$\Lambda$ scattering resulting in a direct formation of hot dark KK$^\prime$ states with a large excess energy. 
   (c) The state reordering is a result of opposite strain gauge factors for KK (KK$^\prime$)  and K$\Lambda$ excitons.
       }
    \vspace{-0.7cm}
	\label{fig_1}
\end{figure}

Dark excitons also have a crucial impact on the transient spatiotemporal exciton dynamics through the formation of \textit{non-equilibrium} dark excitons \cite{Rosati20,Rosati20b,Rosati21c}. In tungsten-based TMD monolayers the optically excited bright KK excitons scatter into the energetically lower dark states by emitting intervalley phonons, cf. Figs. \ref{fig_1}(a,b).
Due to the spectral mismatch between the bright-dark exciton separation and intervalley phonon energies \cite{Jin14},  \textit{hot} dark excitons are formed with a finite excess energy. 
Subsequently, they gradually lose their energy by scattering with quasi-elastic intravalley phonons, whose efficiency decreases linearly with temperature. As a result, the thermalization time of hot dark excitons increases from tens of femtoseconds to tens of picoseconds when  moving toward cryogenic temperatures \cite{Rosati20}. In hBN-encapsulated WSe$_2$, this gradual thermalization has been directly tracked via time-resolved PL \cite{Rosati21c}. Since hot excitons have a higher group velocity, their formation leads to a transient diffusion which is orders of magnitude faster than the equilibrium diffusion \cite{Rosati21e} - similar to the behaviour demonstrated  in two- and three-dimensional perovskites \cite{Ziegler20,Rosati26,Saris26}.

Strain can be used to tailor the dark-bright exciton energy landscape and to control the excess energy of hot dark excitons and thereby the transient exciton PL and exciton diffusion in TMDs.  This strain engineering of the ultrafast spatiotemporal exciton dynamics at cryogenic temperatures has  remained largely unexplored so far. In this work, we bridge this knowledge gap by performing a material-specific and predictive theoretical study on the impact of tensile and compressive strain on the transient PL as well as on the temporal exciton thermalization and the spatial exciton propagation in the exemplary case of strained, hBN-encapsulated WSe$_2$ monolayers. We show how tensile strain leads to an abrupt increase in the excess energy of hot dark KK$^\prime$ excitons resulting in  a speed-up of exciton diffusion for strain values larger than the critical strain $s_0\approx 0.3\,\%$. This can be traced back to the deactivation of phonon-mediated scattering channels into  K$\Lambda$ excitons. 
While intervalley scattering with optical phonons is already closed in only weakly strained TMDs due to their large energies in the range of 27-33\,meV \cite{Jin14}, intervalley scattering with acoustic phonons with energies of about 12-14\,meV is energetically not allowed  for strain values with E$_{\text{KK}, s}-\hbar\omega_\Lambda<\text{E}_{\text{K}\Lambda, s}$, where 
$\text{E}_{v, s}$ is the minimum energy of excitons in valley $v$ at strain $s$ and
$\hbar\omega_\Lambda\equiv \hbar\omega^{\text{TA},\Lambda}_0$ is the energy of the intervalley TA acoustic phonons. As shown in Fig. \ref{fig_1}(c), this takes place at $s_0\approx 0.3\,\%$ for hBN-encapsulated WSe$_2$ monolayers, while larger $s_0$ values are expected for lower dielectric screening due to a generally increased energy separation between KK and K$\Lambda$ excitons \cite{Kumar24,Su22}.
Furthermore, we predict a qualitatively different behaviour in the presence of compressive strain that is characterized by a strain-induced periodic modulation of exciton diffusion due to the activation of M-phonon emission. Overall, our microscopic study sheds light on the impact of tensile and compressive strain on exciton thermalization and transient exciton diffusion in TMD monolayers.

\section{Theoretical approach}

First, we evaluate the strain-dependent exciton energy landscape. 
Starting from the unstrained single-particle dispersion relation \cite{Kormanyos15} and strain-induced variations of effective masses and band extrema \cite{Khatibi18}, we apply a continuum approach to solve the Wannier
equation  \cite{Haug09,Selig16,Selig18,Brem18} 
 with a non-local Coulomb screening \cite{Brem19b}. In this way, we obtain a set of excitonic states 
$\vert\mathbf{Q},v;s\rangle=\hat{X}^{\dagger}_{v,\textbf{Q}, s}\vert 0 \rangle$ with the strain-dependent energy $E^{v}_{\textbf{Q},s}=\text{E}_{v, s}+\hbar^2|\mathbf{Q}|^2/(2M_{v,s})$, where $\mathbf{Q}$ denotes the center-of-mass momentum and $M_{v,s}$ the total exciton mass in valley $v$ at strain $s$. Here, we have introduced $\hat{X}^{\dagger}_{v,\textbf{Q}, s}$ as the exciton creation operator. 
Due to a considerable energetic separation to higher excitonic states and the limited range of considered strain values,  we restrict our investigations to the lowest $1s$ state of the bright KK as well as of momentum-dark KK$^\prime$ and K$\Lambda$ excitons \cite{Malic18}. Note that other valleys, such as $\Gamma$K and $\Gamma$K$^\prime$, become only relevant at larger strain values $s\gtrsim 1\,\%$ \cite{Rosati21a}. 

Now, we introduce the excitonic intravalley Wigner function 
$N^{v}_{\mathbf{Q}, s}(\mathbf{r},t)=\sum_{\textbf{Q}^\prime} \langle \hat{X}^{\dagger}_{v,\textbf{Q}+\textbf{Q}^\prime/2, s} \hat{X}_{v,\textbf{Q}-\textbf{Q}^\prime/2, s} \rangle_t e^{\imath \textbf{Q}^\prime\cdot \textbf{r}}$ with $\textbf{r}$ indicating the center-of-mass position while $\langle \hat{X}^{\dagger}_{v,\textbf{Q}, s} \hat{X}_{v,\textbf{Q}^\prime, s} \rangle_t$ is the time-dependent single-exciton density matrix after tracing out the other degrees of freedom \cite{Rosati14}. We then numerically determine the evolution of $N^{v}_{\mathbf{Q}, s}(\mathbf{r},t)$ within the continuum approach.
Exploiting the Heisenberg equation, in the low-excitation \cite{Kulig18,Perea19} and undoped regime \cite{Gao16,Kato16,Cadiz17,Titze18} the associated spatiotemporal exciton dynamics reads \cite{Hess96,Rosati20}
\begin{align}
\begin{split}\label{SBE}
\dot{N}^{v}_{\mathbf{Q},s}(\mathbf{r},t)=& \left(-\frac{\hbar \mathbf{Q}}{M_{v,s}}\cdot \nabla - \gamma \delta_{\mathbf{Q},0}\delta_{v,\text{KK}} \right)N^{v}_{\mathbf{Q},s}(\mathbf{r},t)\\ 
&+\Gamma^{\text{KK}\,v}_{0\,\mathbf{Q}, s} |p_{0,s}(\mathbf{r},t)|^2\!\!+\!\!\left.\dot{N}^{v}_{\mathbf{Q},s}(\mathbf{r},t)\right|_{\text{sc}}\,.\\
\end{split}
\end{align}
The first term indicates the free spatial evolution of excitons, while the second term takes into account the losses due 
to the radiative recombination of bright excitons $\vert\text{KK},0;s\rangle$ within the light cone via a strain-independent rate $\gamma_s\approx \gamma$ corresponding to a radiative linewidth of $\hbar\gamma$=2.5\,meV \cite{Khatibi18}. 
The second line of Eq. (\ref{SBE}) provides the effects induced by exciton-phonon scattering. 
The first contribution describes the formation of incoherent excitons due to phonon-driven transfer from 
excitonic polarization $\vert p_{\mathbf{Q}\approx 0, s}(\mathbf{r}, t)\vert^2$ (referred to in literature as coherent excitons \cite{Selig18}). This polarization follows the spatiotemporal laser profile, for which we take a Gaussian with a full-width half maximum of 0.2\,ps in time and 850\,nm in space, as realized in a previous experiment \cite{Rosati21c}. The last term in Eq. (\ref{SBE}) provides the scattering-induced dynamics, which we describe with the conventional Markov approach as a local Boltzmann collision rate
$\left.\dot{N}^{v}_{\mathbf{Q},s}(\mathbf{r},t)\right|_{\text{sc}}=\Gamma^{\text{in},v}_{\mathbf{Q},s}(\mathbf{r},t)\!-\!\Gamma^{\text{out},v}_{\mathbf{Q},s}N^{v}_{\mathbf{Q},s}(\mathbf{r},t)$  in the Boltzmann transport equation with $\Gamma^{\text{in},v}_{\mathbf{Q},s}(\mathbf{r},t)=\sum_{v^\prime,\textbf{Q}^\prime}\Gamma^{v^\prime v}_{\textbf{Q}^\prime\textbf{Q},s}N^{v^\prime}_{\textbf{Q}^\prime,s}(\mathbf{r},t)$ and $\Gamma^{\text{out},v}_{\mathbf{Q},s}=\sum_{v^\prime,\textbf{Q}^\prime}\Gamma^{v v^\prime}_{\textbf{Q}\textbf{Q}^\prime,s}$.
Both terms depend on the scattering coefficients $\Gamma^{vv^\prime}_{\textbf{Q}\textbf{Q}^\prime, s}=\frac{2 \pi}{\hbar} \sum_{\alpha,\pm} 
\left|G_{\alpha \boldsymbol{Q}^{\prime}-\boldsymbol{Q}, s}^{v v^{\prime}}\right|^2 \eta^{\pm}_{\alpha \boldsymbol{Q}^{\prime}-\boldsymbol{Q}}\delta\left(E^{v^\prime}_{\boldsymbol{Q}^{\prime}, s}-E^{v}_{\boldsymbol{Q},s}\pm\hbar \omega^{\alpha}_{\boldsymbol{Q}^{\prime}-\boldsymbol{Q}, s}\right)$,
which describe the scattering from exciton $\vert v,\textbf{Q};s\rangle$ to $\vert v^\prime,\textbf{Q}^\prime;s\rangle$ via interaction with phonon modes $\alpha$  with the energy $\hbar \omega^{\alpha}_{\boldsymbol{Q}^{\prime}-\boldsymbol{Q}}$ (neglecting the weak variation of phonon energy with strain \cite{Khatibi18,Dadgar18}). Here, $G_{\alpha \boldsymbol{Q}, s}^{v v^{\prime}}$ are the coefficients of the exciton-phonon Hamiltonian, derived from the analogous electron-phonon coefficients \cite{Jin14} through the excitonic form factors \cite{Brem18}, while $\eta^{\pm}_{\alpha \boldsymbol{Q}}=\left(\frac{1}{2} \pm \frac{1}{2}+n_{\alpha \boldsymbol{Q}}\right)$
are the factors for the emission (+) and absorption (-) of phonons with the Bose-Einstein occupation $n_{\alpha \boldsymbol{Q}}$.
Exciton-phonon scattering crucially depends on strain, mostly via the modification of energies of the involved initial and final exciton states \cite{Niehues18,Khatibi18,Dadgar18,Aslan18}. 
In particular, strain-induced energy shifts can open or close specific intervalley  scattering channels, which strongly affects the formation of hot excitons and their excess energy (see  Figs. \ref{fig_1}(a,b)). 

The resulting spatiotemporal dynamics can be experimentally accessed via space-, time- and energy-resolved photoluminescence $I_s(E,\mathbf{r},t)$. 
Extending the generalized Elliott formula introduced in Ref. \cite{Brem20} and adapted for time-resolved PL in Ref. \cite{Rosati20b}, this can be described as 
\begin{equation}\label{PL}
I_s(E,\textbf{r},t)= \frac{2 |M|^2\left[
I_{\text{d},s}(\textbf{r},t)+I_{\text{ind},s}(E,\textbf{r},t)\right]}{(\text{E}_{\text{KK},s}-E)^2+\hbar^2(\gamma+\Gamma^{\text{out,KK}}_{0,s})^2}
\end{equation}
including the direct $I_{\text{d},s}(\textbf{r},t)=\gamma N^{\text{KK}}_{0, s}(\textbf{r},t)$ and the phonon-assisted indirect contribution to the emission 
$I_{\text{ind}, s}(E,\textbf{r},t)=\sum_{\mathbf{Q},v,\alpha,\pm} |G^{v\,\text{KK}}_{\alpha;\mathbf{Q}, s}|^2\eta^{\pm}_{\alpha,\textbf{Q}} N^{v}_{\mathbf{Q}, s}(\mathbf{r},t)\frac{2\Gamma^{\text{out},v}_{\mathbf{Q},s}}{4\left(E^{v}_{\mathbf{Q}, s}\mp \hbar \omega^\alpha_{\textbf{Q}} - E\right)^2 + \left(\hbar\Gamma^{\text{out},v}_{\mathbf{Q}, s}\right)^2}$.
The coefficient $|M|^2$ is the squared modulus of the exciton-photon matrix element in the lightcone \cite{Brem20} and it directly relates to the radiative rate $\gamma$ \cite{Rosati26}.
  The spatiotemporal dynamics of Eq. (\ref{SBE}) allows us to track excitonic diffusion, which induces a spatial broadening of the squared width of the PL profile 
 $\sigma_s^2(t)=\int \int \mathbf{r}^2I_s(E,\mathbf{r},t) d\mathbf{r} dE/2\int \int I_s(E,\mathbf{r},t) d\mathbf{r}dE$.
  At equilibrium, this evolves linearly in time, while in the transient non-equilibrium regime,     the effective diffusion coefficient is given by $D_s(t)=\frac{1}{2}\partial_t \sigma_s^2(t)$, see Sec. 1 in the Supplemental Material (SM) \footnote{\label{SI}See Supplemental Material at [URL will be inserted by publisher] for additional details on the microscopic evaluation of photoluminescence and diffusion as well as for additional time-resolved spectra and diffusion at temperature $T=50\,$K or with artificially stronger bright-to-KK$^\prime$ scattering.} for more details.
While Eq. (\ref{PL}) describes the local photoluminescence as a well-established techniques for measuring spatiotemporal exciton dynamics \cite{Kulig18,Cordovilla19,Moon20,Rosati21e,Ginsberg20}, the more general  Eq. (\ref{SBE}) can be adapted to other experimental approaches, such as local pump-probe \cite{Yuan20,Su22,LoGerfoMorganti25}, raster scanning \cite{Beret22,Lamsaadi2023,Rosati25}, transient gratings \cite{Rieland24,Bi25} and space- and time-resolved ARPES \cite{Schmitt25,Madeo20,Schmitt22,Aeschlimann20,Dong21,Bange24,Boschini24,Meneghini26}.
 
\section{Results}
\subsection{Strain-engineering of transient exciton optics}

The direct radiative recombination of bright excitons gives rise to a peak in the PL spectrum at  the energy E$_{\text{KK},s}$. The  phonon-assisted recombination of  dark excitons dominates the PL at low temperatures, although the efficiency of this process is orders of magnitude smaller. This is compensated for by the much higher occupation of the energetically lowest dark states at cryogenic temperatures. The resulting phonon sidebands appear energetically below E$_{\text{KK},s}$ and allow to track dark excitons. In unstrained hBN-encapsulated WSe$_2$ monolayers, phonon sidebands have been demonstrated to provide direct optical access to the thermalization of non-equilibrium hot dark excitons \cite{Rosati21c}. In a recent study, it was shown that strain can be used to fingerprint the dark exciton landscape via the emergence of their phonon sidebands \cite{Lopez22,Kumar24,Kumar25}. However, the interplay of strain and non-equilibrium exciton dynamics has remained unexplored so far.

 \begin{figure}[t]
    \centering
	\includegraphics[width=\columnwidth]{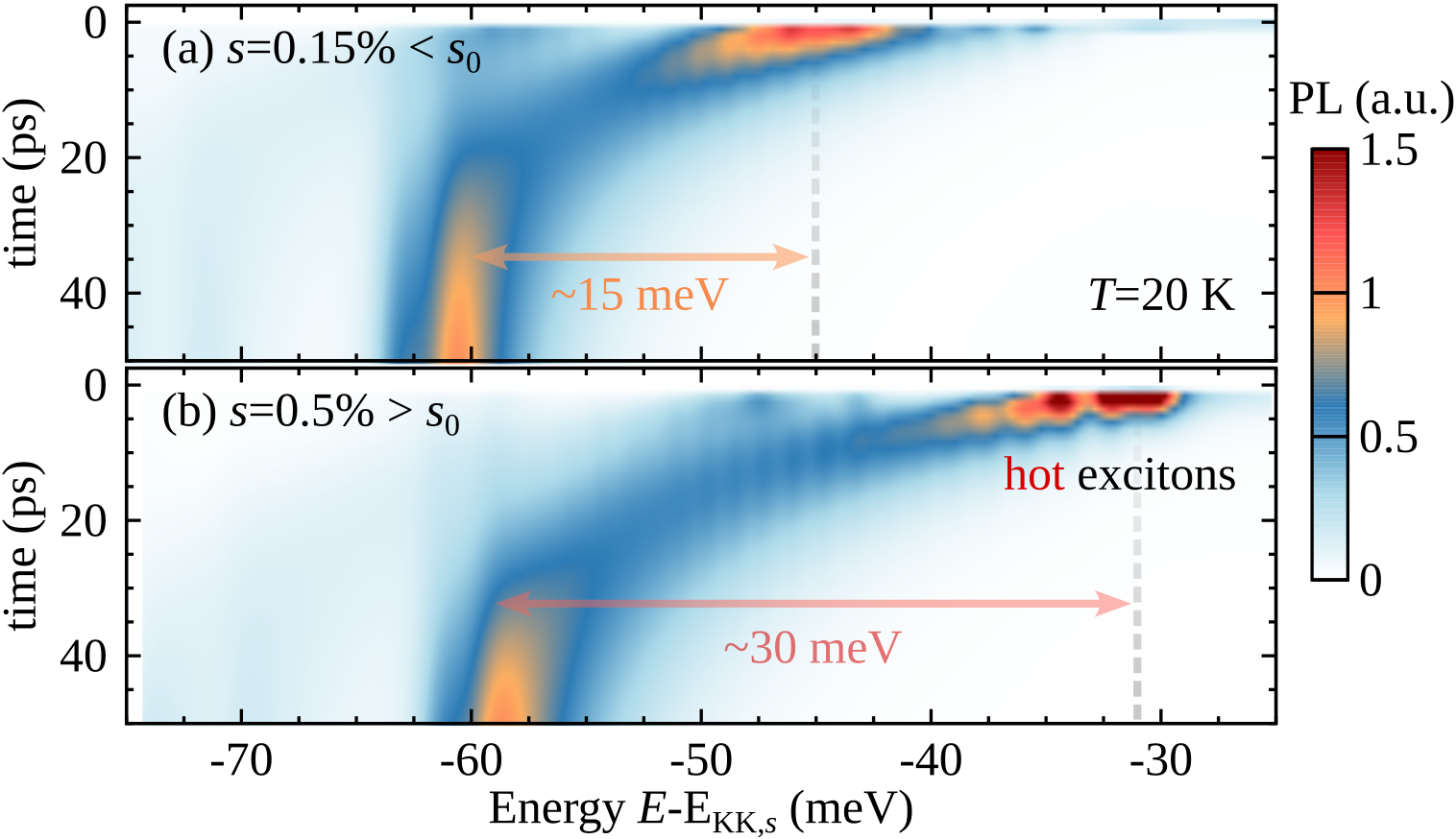}
	\caption{\textbf{Strain-tuning of exciton thermalization:} Time- and energy-resolved photoluminescence of hBN-encapsulated WSe$_2$ at $T=$20\,K under an external uniaxial tensile strain of (a) $s=0.15\%<s_0$ and (b) $s=0.5\%>s_0$ with $s_0\approx0.3\%$ being the critical strain value, see Fig. \ref{fig_1}(c). An initial high-energy signal redshifts toward the equilibrium emission within few tens of picoseconds. This is a hallmark of the thermalization of hot dark excitons (Fig. \ref{fig_1}(a,b)). 
    The energy separation between the initial and the equilibrium resonance increases abruptly from 15 meV to almost 30\,meV at the critical strain value of $s_0$  reflecting  the suppression of the scattering channel from KK to K$\Lambda$ excitons (Fig. \ref{fig_1}(b)). 
	}
    \vspace{-0.5cm}
	\label{fig_2}
\end{figure}

In Fig.~\ref{fig_2}, we show the energy- and time-resolved emission of a hBN-encapsulated WSe$_2$ monolayer at $T=$\,20\,K under an external uniaxial tensile strain of $s=0.15\,\%$ and $s=0.5\,\%$, respectively. In both cases, an initially higher emission energy  gradually redshifts on a timescale of  few tens of picoseconds. The final equilibrium resonance lies approximately 60\,meV below   E$_{\text{KK},s}$ \cite{Brem20,Rosati20b}.
Similarly to the case of unstrained WSe$_2$ \cite{Rosati20b}, such a redshift indicates the initial formation of hot dark excitons and their subsequent thermalization to their equilibrium distribution. 
Following an optical excitation of bright excitons, hot dark excitons in the KK$^\prime$ valley are generated on a sub-picosecond timescale via emission of intervalley phonons, see Fig. \ref{fig_1}(a,b). The energy separation between bright and dark excitons does not match the energy of the involved phonons resulting in the formation of \textit{hot} dark excitons with an excess energy with respect to band minimum at E$_{\text{KK}^\prime,s}$. 
The excess energy is smaller than the energy of intravalley optical phonons, whose emission is hence energetically forbidden. As a consequence,  hot dark excitons lose their  energy by scattering with intravalley acoustic phonons. Since the latter carry only a small amount of energy, this relaxation requires multiple scattering events resulting in a gradual loss of excess energy within few tens of picoseconds. This is directly visible in the calculated time-resolved PL spectra via a temporal redshift of the excitonic resonance, cf. Fig. \ref{fig_2}, in agreement with a recent joint theory-experiment study on unstrained WSe$_2$ monolayers \cite{Rosati20b}.

In unstrained TMDs and for strain values up to $s_0\approx 0.3\%$, the optically excited bright excitons first form hot K$\Lambda$ excitons before they scatter into the energetically lowest KK$^\prime$ states, see Fig. \ref{fig_1}(a). Above the critical strain value $s_0$, the relative separation between bright and dark excitons changes such that the KK-to-K$\Lambda$ scattering channel  is turned off, see Fig. \ref{fig_1}(b). A direct consequence is that optically excited KK excitons can only scatter directly into KK$^\prime$ excitons forming hot dark excitons with a much larger excess energy. Therefore, we find an abrupt increase in the energy of the initial PL signature in Fig. \ref{fig_2}(b). In Fig. S1 of the SM, we show more strain values in the range below and above $s_0$.
When energetically allowed, the scattering process from KK into KK$^\prime$ through K$\Lambda$ is faster than the direct scattering from KK to KK$^\prime$. This occurs mostly due to the three-fold degeneracy of K$\Lambda$ states together with the very efficient scattering between K$\Lambda$ and KK$^\prime$ excitons via M phonons \cite{Jin14,Rosati21a}. However, strain closes this fast channel by inducing a blueshift of K$\Lambda$ excitons (Fig. \ref{fig_1}(c)).

\subsection{Strain engineering of ultrafast exciton diffusion}

\begin{figure}[t]
    \centering
	\includegraphics[width=\columnwidth]{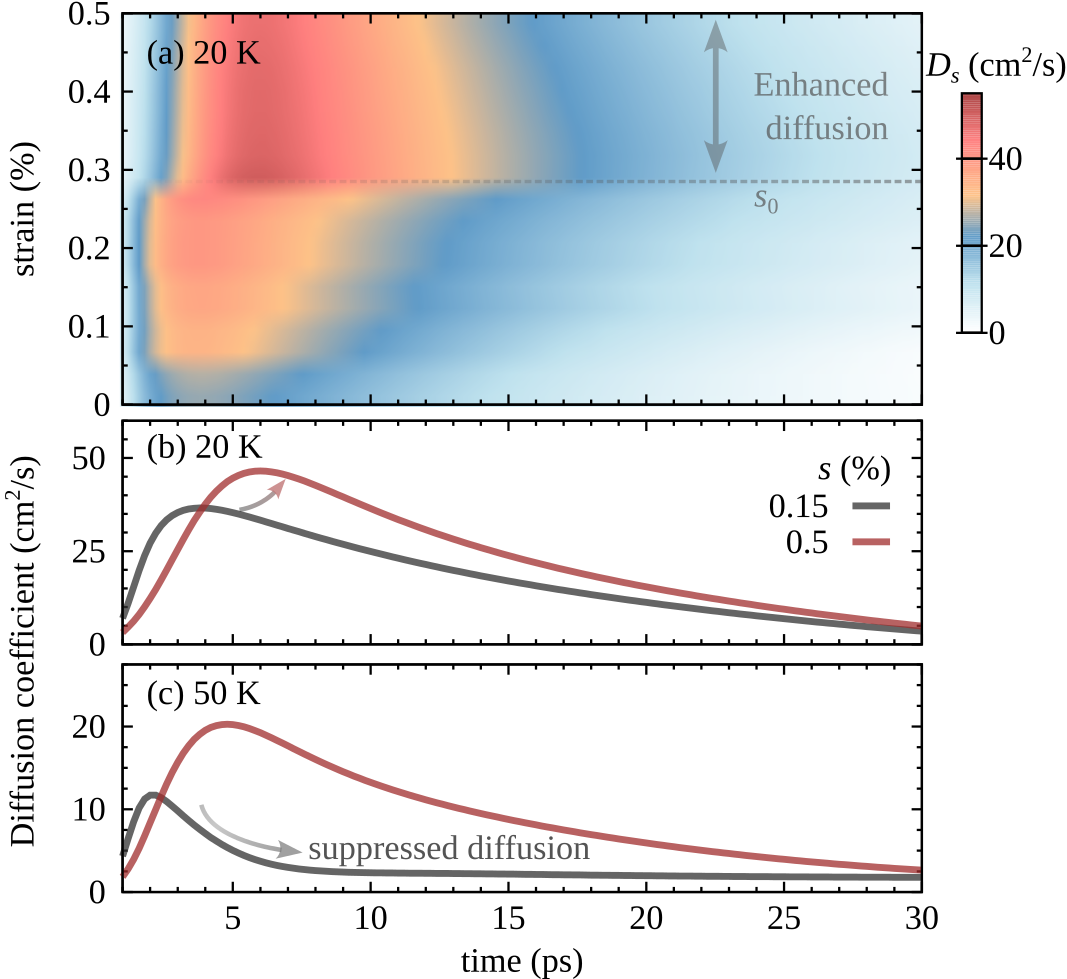}
	\caption{\textbf{Strain engineering of exciton propagation:} (a) Time-resolved transient exciton diffusion coefficient of hBN-encapsulated WSe$_2$ as a function of tensile strain at 20 K. We find an abrupt enhancement above the critical strain value of $s_0\approx 0.3\,\%$. (b,c) Cuts at two specific strain values below and above $s_0$ at 20\,K and 50K, respectively. Interestingly, at the larger temperature, the duration of the enhanced transient diffusion is considerably shortened to only few picoseconds for strain values smaller than $s_0$. This is due to the thermal activation of phonon absorption inducing  the KK$^\prime$-to-K$\Lambda$  back-scattering.
	}
	\label{fig_3}
\end{figure}

Since hot excitons exhibit larger group velocities, they typically give rise to a faster transient diffusion than at equilibrium \cite{Rosati21c,Rosati26,Saris26}. 
As strain can be used to control the excess energy of excitons (Fig. \ref{fig_2}), strain engineering of exciton diffusion is possible. Figure \ref{fig_3} shows the time-resolved effective exciton diffusion coefficient $D_s(t)$ at $T=$\,20\,K for different values of strain. We find that the maximum effective diffusion coefficient abruptly increases from approx. 35\,cm$^2$/s to about 45\,cm$^2$/s above the critical strain value $s_0\approx0.3\,\%$.
This maximum diffusion appears shortly after the optical excitation, before gradually decreasing as excitons thermalize via quasi-elastic scattering with acoustic phonons. The diffusion coefficient decreases on a timescale of a few tens of picoseconds (Fig. \ref{fig_3}(b)), reflecting the gradual reduction of the excess energy shown in Fig. \ref{fig_2}.

The connection between the transient diffusion and the exciton excess energy can be understood from a generalized Fick's law \cite{Rosati26}, which neglecting the occupation of the high-energy valleys results in $D_s(t)\approx\frac{\tau_s}{M}\langle E_s\rangle_{t}$ with the scattering time $\tau_s=1/\Gamma^{\text{out,KK}^\prime;s}_{0}$, exciton mass $M\equiv M_{\text{KK}^\prime,s}$, and the averaged exciton energy $\langle E_s\rangle_{t}$. 
This relation follows under the assumptions of a weak spatial inhomogeneity of $N^{\text{KK}^\prime}_{\textbf{Q},s}(\textbf{r},t)$. This implies timescales larger than $\tau_s$, where first scattering even takes place and randomizes the momentum distribution \cite{Rosati21d,Fitzgerald25}.  Furthermore, in the simplified equation we neglect effects stemming from an intervalley bottleneck or back-scattering \cite{Rosati21d} and assume a negligible momentum dependence of the scattering rate $\Gamma^{\text{out},\text{KK}^\prime}_{\textbf{Q},s}\approx\Gamma^{\text{out},\text{KK}^\prime}_{0,s}$ \cite{Rosati26}.
The averaged exciton energy depends on the spatiotemporal exciton occupation via $\langle E_s\rangle_{t}=\int d\textbf{r}\sum_\textbf{Q}E^{\text{KK}^\prime}_{\textbf{Q}, s}N^{\text{KK}^\prime}_{\textbf{Q}, s}(\textbf{r},t)/\int d\textbf{r}\sum_\textbf{Q}N^{\text{KK}^\prime}_{\textbf{Q}, s}(\textbf{r},t)$. 
For times $t$ larger than the thermalization time $t_{\text{th}}$, excitons equilibrate in a local Boltzmann distribution with $\langle E_s\rangle_{t_{\text{th}}}=k_BT$, leading to the well-known conventional diffusion coefficient $D_s=\tau_s k_BT/M$ \cite{Rosati26}. 
In contrast, before thermalization, the averaged energy $\langle E_s \rangle_{t}$ is determined by the excitonic excess energy.  This is approximately one order of magnitude larger than the thermal energy at 20\,K, leading to considerably faster transient exciton diffusion.  The abrupt increase of the excess energy for tensile strain values above $s_0$ (Fig.~\ref{fig_2}) is reflected by an abrupt enhancement of the maximum diffusion above this critical strain value, see Fig.~\ref{fig_3}(a). 
Interestingly, the opposite trend is found during the first 2--3\,ps, when diffusion becomes slower at larger strain values, see Fig.~\ref{fig_3}(b). This can be traced back to the suppressed scattering into K$\Lambda$ excitons, thereby slowing down the formation of KK$^\prime$ excitons, i.e. the direct KK-to-KK$^\prime$ scattering process is slightly slower, hence requiring more time until highly mobile hot KK$^\prime$ excitons become crucial.  Note that there is also an enhancement in diffusion for strain values above $s\approx0.05\,\%$ due to the suppression of a phonon bottleneck between KK$^\prime$ and K$\Lambda$ excitons \cite{Rosati20b}, see Fig. S1 in the SM for more details. 

Higher temperatures slow down the transient exciton diffusion due to the increased efficiency of exciton–phonon scattering. Surprisingly, we find that this occurs in a highly asymmetric way for strain values larger and smaller than $s_0$. 
For $s=0.5\,\%>s_0$, the evolution of $D_s(t)$ is 
qualitatively similar at 20 and 50\,K, see the red lines in Figs. \ref{fig_3}(b,c). The two results differ only by the expected quantitative reduction of the maximum diffusion by a factor of approx. 2.5. This originates from the linear increase of the exciton-phonon scattering rates with temperature, which at large strain are dominated by the interaction with acoustic $\Gamma$ phonons.
For the smaller strain $s=0.15\,\%<s_0$, at 50\,K the transient diffusion is strongly suppressed already after 5\,ps - contrary to the case at 20\,K, see the gray lines in Figs. \ref{fig_3}(b,c) as well as the surface plot in Fig. S2 of the SM. This results from the thermal activation of the intervalley scattering from hot KK$^\prime$ into K$\Lambda$ states  via absorption of acoustic M phonons. 
These modes have low energies (15–16 meV \cite{Jin14}) and strongly couple the KK$^\prime$ and K$\Lambda$ excitons. Their thermal activation, already significant at 50 K, rapidly transfers hot KK$^\prime$ excitons back the K$\Lambda$ states, suppressing the transient diffusion. Note that the same mechanism also governs the strain dependence of the equilibrium diffusion at room temperature \cite{Rosati21a}.
For larger strain values with $s>0.4-0.5\,\%$,  no thermal activation of  M phonons is possible and the enhanced transient diffusion is preserved for a longer time, since here K$\Lambda$ excitons are shifted too far away from KK$^\prime$ states (Fig. \ref{fig_1}(c)). 
In the intermediate range of strain values $s_0\lesssim s\lesssim 0.4-0.5\,\%$, the scattering from KK to K$\Lambda$ is energetically suppressed, while hot KK$^\prime$ excitons can still scatter into K$\Lambda$ states. Together with the larger thermal broadening of transient distributions, this leads to
a variation of the transient diffusion with strain that is less abrupt at $T=50\,$K than at 20\,K. In particular, the maximum diffusion smoothly increases from 10 to 20\,cm$^2$/s for strain values going from $s_0$ to 0.4-0.5\,\%, see Sec. 1 of the SM for more details.

\subsection{Periodic strain modulation of exciton diffusion}

In conventional semiconductors, the excess energy of photo-excited electrons can be tuned simply by changing the excitation energy. When the excess energy exceeds the optical-phonon energy $\hbar \omega_{\text{op}}$, successive phonon-induced scattering reduces the energy by multiples of  $\hbar \omega_{\text{op}}$  \cite{Kuhn92,Rosati15b}.
In contrast, in TMD monolayers the exciton oscillator strength is strongly centered at the bright-exciton energy, which cannot be tuned by laser energy. As a result, exciton excess energy is largely insensitive to excitation energies close to resonance. 
Here, we show how compressive strain allows a continuous tuning of exciton excess energy - in analogy to the effect of laser excitation energy on electrons. 

\begin{figure}[t]
    \centering
	\includegraphics[width=\columnwidth]{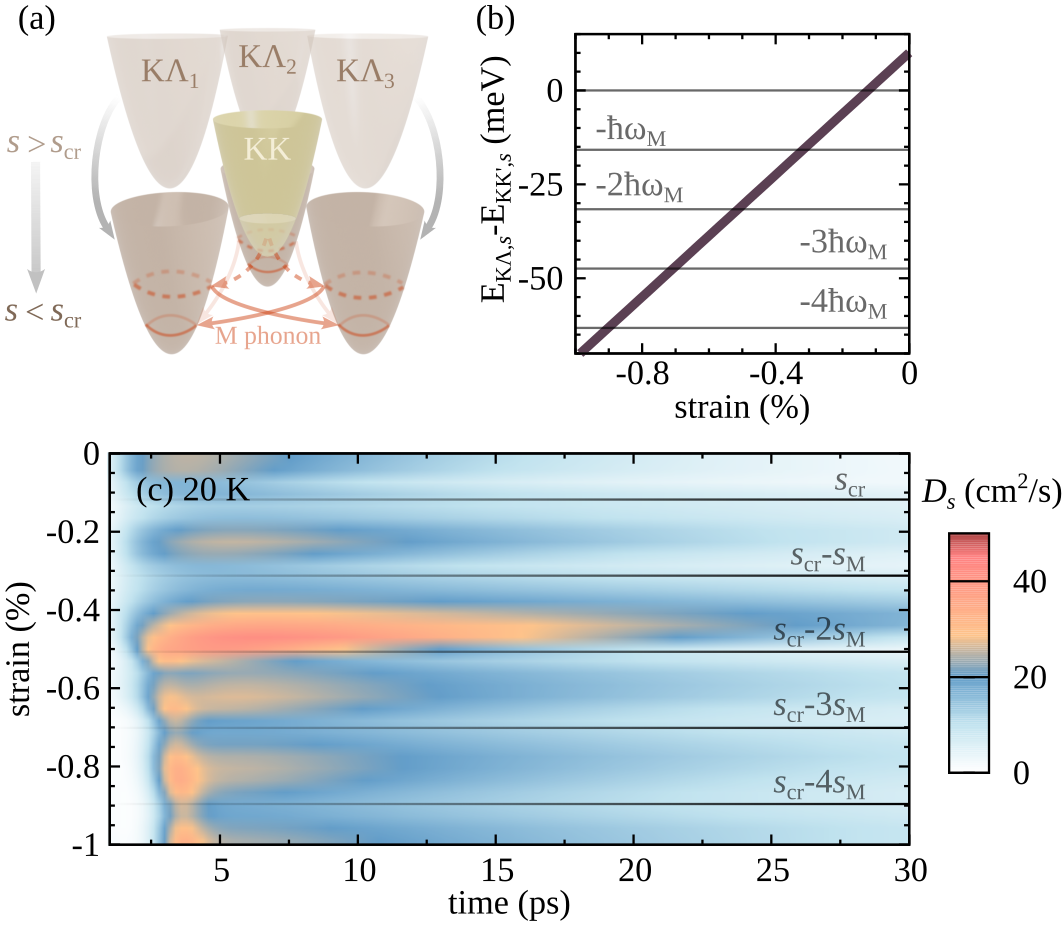}
	\caption{\textbf{Strain-induced periodic modulation of exciton diffusion:}
    (a) Activation of K$\Lambda$ excitons in presence of compressive strain. Here, hot dark excitons scatter between the three degenerate K$\Lambda$ valleys by emitting acoustic M phonons with the energy $\hbar \omega_\text{M}$. (b) For strain values smaller than $s_{\text{cr}}\approx-0.1\,\%$,  K$\Lambda$ excitons become energetically lower  than KK$^\prime$ excitons. The large strain gauge factor of E$_{{\text{K}\Lambda}, s}-$E$_{\text{KK}^\prime, s}$ allows emission of acoustic M phonons with a  multiple of $\hbar \omega_\text{M}$ (horizontal lines), which corresponds to a strain periodicity of $s_\text{M}\approx0.2\,$meV/\%. (c) The resulting periodic strain modulation of the transient exciton diffusion.}
	\label{fig_4}
\end{figure}

Recently, compressive strain has been experimentally realized using piezoelectric support \cite{Iff19,An23} or thermal expansion of the substrate \cite{Gant19,Henriquez23}. The key qualitative difference between tensile and compressive strain is that the lowest state changes from KK$^\prime$ to K$\Lambda$ for uniaxial strain smaller than $s_{\text{cr}}\approx -0.1\,\%$. This change has two important consequences:
First, K$\Lambda$ excitons are three-fold degenerate, contrary to KK$^\prime$ states. This opens up new scattering channels for hot excitons. While hot  KK$^\prime$ excitons can only interact with intravalley $\Gamma$ phonons,  hot K$\Lambda$ excitons can also interact with M phonons and scatter between degenerate K$\Lambda$ valleys, see Fig. \ref{fig_4}(a). 
Second, in contrast to only a small change in the energy separation 
between bright KK and  dark KK$^\prime$  excitons  
 with tensile strain \cite{Dirnberger21} (see 
quasi-parallel strain-dependent lines for KK and KK$^\prime$ excitons in Fig. \ref{fig_1}(c)),
compressive strain induces a large variation of bright-dark energy separation. This is comparable to the variation of the energy separation $\text{E}_{\text{K}\Lambda,s}-\text{E}_{\text{KK}^\prime,s}$ between the two energetically lowest valleys, which decreases with compressing strain with a gauge factor $d_{\Lambda-\text{K}^\prime}\approx$\,80\,meV/\%, see Fig. \ref{fig_4}(b). This is due to the opposite strain-induced energy shift of KK and K$\Lambda$ excitons, see Fig. \ref{fig_1}(c). These two qualitative changes lead to drastically different transient exciton diffusion with compressive and tensile strain.

At the critical strain value $s_{\text{cr}}\approx-0.1\,\%$,  K$\Lambda$ and KK$^\prime$ excitons are quasi-degenerate (Fig. \ref{fig_4}(b)), leading to a small  transient diffusion due to the enhanced intervalley scattering, similar to the case of equilibrium diffusion at high temperatures\cite{Rosati21a}. Here, hot KK$^\prime$ excitons are formed with an excess energy of approximately 15\,meV  (Fig. \ref{fig_2}(a)), which is in the range of the energy of M acoustic phonons \cite{Jin14}, and hence hot KK$^\prime$ excitons can  relax into K$\Lambda$ states by emitting an M phonon.   This leads to a suppressed diffusion for $s\approx s_{\text{cr}}$.
A tuning of compressive strain beyond this critical value induces a periodic change of fast and slow exciton diffusion, see Fig. \ref{fig_4}(c). This indicates a cyclic variation of exciton excess energy. The initial excess energy corresponds to the energy separation between the two lowest dark excitons, hence it increases with compressive strain with the gauge factor $d_{\Lambda-\text{K}^\prime}$. Once a certain excess energy is reached, emission of intervalley M phonons drives the system back to the initial excess energy. At first approximation, this process is quantized in multiples of the average acoustic phonon energy $\hbar \omega_\text{M}\equiv 0.5\hbar (\omega^{\text{M,TA}}_0+\omega^{\text{M,LA}}_0)=15.8$\,meV, because exciton scattering  scales with one over the phonon energy and acoustic M phonons possess the smallest energy. 
The horizontal lines in Fig. \ref{fig_4}(b) indicate when the energy separation between the two dark valleys is large enough to activate the emission of M  phonons. This takes place
with a strain periodicity $s_\text{M}=\hbar \omega_\text{M}/d_{\Lambda-\text{K}^\prime}\approx 0.2$\,\%. 
Interestingly, the same periodicity appears in the strain- and time-dependent exciton diffusion in Fig. \ref{fig_4}(c). First, increased compressive strain induces a higher excess energy of hot dark excitons resulting in a larger exciton diffusion. Whenever the condition is reached for the emission of M phonons inducing scattering between the three K$\Lambda$ valleys, the diffusion is significantly reduced. This results in a phonon-induced periodic modulation of exciton diffusion. Note that there is a deviation from this expected periodicity at higher compressive strain values, which can be ascribed to the small contribution of other phonon modes and exciton valleys, in particular other M phonons and KK$^\prime$ excitons. For large compressive strain $s\lesssim s_{\text{cr}}-$2$s_\text{M}$ there is a even a competition between two- and one-phonon processes.

\section{Conclusions}

Based on a microscopic, material-specific and predictive approach, we study the possibility of strain engineering of exciton emission and diffusion in  hBN-encapsulated WSe$_2$ monolayers at cryogenic temperatures.  We predict an abrupt increase of the excess energy of hot dark KK$^\prime$ excitons above the critical strain of $s_0\approx 0.3\,\%$.  This is explained by the deactivation of the phonon-mediated scattering to K$\Lambda$ excitons, which considerably slows down exciton thermalization and increases the energy of transient phonon sidebands that can be monitored in time-resolved photoluminescence spectra.
Furthermore, we show that the maximum exciton diffusion coefficient increases abruptly from about  35 to about 45\,cm$^2$/s above the critical strain $s_0$, which is
one order of magnitude faster than equilibrium diffusion.  The enhanced transient diffusion reflects the suppressed phonon-mediated scattering from bright to dark K$\Lambda$ excitons. Finally, we predict a qualitatively different diffusion behaviour in presence of compressive strain that is characterized by a periodic strain modulation. This can be traced back to an interplay of a strain-induced increase in the excess energy of hot dark excitons and its abrupt release via emission of M phonons.  Overall, our work provides new microscopic insights into strain engineering of exciton emission and propagation in strained 2D semiconductors.\\

\begin{acknowledgements}
We acknowledge financial support by the Deutsche Forschungsgemeinschaft (DFG) via the regular project 512604469. We thank A. Chernikov for discussions.
\end{acknowledgements}


\begin{thebibliography}{88}%
\makeatletter
\providecommand \@ifxundefined [1]{%
 \@ifx{#1\undefined}
}%
\providecommand \@ifnum [1]{%
 \ifnum #1\expandafter \@firstoftwo
 \else \expandafter \@secondoftwo
 \fi
}%
\providecommand \@ifx [1]{%
 \ifx #1\expandafter \@firstoftwo
 \else \expandafter \@secondoftwo
 \fi
}%
\providecommand \natexlab [1]{#1}%
\providecommand \enquote  [1]{``#1''}%
\providecommand \bibnamefont  [1]{#1}%
\providecommand \bibfnamefont [1]{#1}%
\providecommand \citenamefont [1]{#1}%
\providecommand \href@noop [0]{\@secondoftwo}%
\providecommand \href [0]{\begingroup \@sanitize@url \@href}%
\providecommand \@href[1]{\@@startlink{#1}\@@href}%
\providecommand \@@href[1]{\endgroup#1\@@endlink}%
\providecommand \@sanitize@url [0]{\catcode `\\12\catcode `\$12\catcode
  `\&12\catcode `\#12\catcode `\^12\catcode `\_12\catcode `\%12\relax}%
\providecommand \@@startlink[1]{}%
\providecommand \@@endlink[0]{}%
\providecommand \url  [0]{\begingroup\@sanitize@url \@url }%
\providecommand \@url [1]{\endgroup\@href {#1}{\urlprefix }}%
\providecommand \urlprefix  [0]{URL }%
\providecommand \Eprint [0]{\href }%
\providecommand \doibase [0]{https://doi.org/}%
\providecommand \selectlanguage [0]{\@gobble}%
\providecommand \bibinfo  [0]{\@secondoftwo}%
\providecommand \bibfield  [0]{\@secondoftwo}%
\providecommand \translation [1]{[#1]}%
\providecommand \BibitemOpen [0]{}%
\providecommand \bibitemStop [0]{}%
\providecommand \bibitemNoStop [0]{.\EOS\space}%
\providecommand \EOS [0]{\spacefactor3000\relax}%
\providecommand \BibitemShut  [1]{\csname bibitem#1\endcsname}%
\let\auto@bib@innerbib\@empty
\bibitem [{\citenamefont {Malic}\ \emph {et~al.}(2023)\citenamefont {Malic},
  \citenamefont {Perea-Causin}, \citenamefont {Rosati}, \citenamefont
  {Erkensten},\ and\ \citenamefont {Brem}}]{Malic23}%
  \BibitemOpen
  \bibfield  {author} {\bibinfo {author} {\bibfnamefont {E.}~\bibnamefont
  {Malic}}, \bibinfo {author} {\bibfnamefont {R.}~\bibnamefont {Perea-Causin}},
  \bibinfo {author} {\bibfnamefont {R.}~\bibnamefont {Rosati}}, \bibinfo
  {author} {\bibfnamefont {D.}~\bibnamefont {Erkensten}},\ and\ \bibinfo
  {author} {\bibfnamefont {S.}~\bibnamefont {Brem}},\ }\bibfield  {title}
  {\bibinfo {title} {Exciton transport in atomically thin semiconductors},\
  }\href {https://doi.org/10.1038/s41467-023-38556-9} {\bibfield  {journal}
  {\bibinfo  {journal} {Nat. Commun.}\ }\textbf {\bibinfo {volume} {14}},\
  \bibinfo {pages} {3430} (\bibinfo {year} {2023})}\BibitemShut {NoStop}%
\bibitem [{\citenamefont {Unuchek}\ \emph {et~al.}(2018)\citenamefont
  {Unuchek}, \citenamefont {Ciarrocchi}, \citenamefont {Avsar}, \citenamefont
  {Watanabe}, \citenamefont {Taniguchi},\ and\ \citenamefont
  {Kis}}]{Unuchek18}%
  \BibitemOpen
  \bibfield  {author} {\bibinfo {author} {\bibfnamefont {D.}~\bibnamefont
  {Unuchek}}, \bibinfo {author} {\bibfnamefont {A.}~\bibnamefont {Ciarrocchi}},
  \bibinfo {author} {\bibfnamefont {A.}~\bibnamefont {Avsar}}, \bibinfo
  {author} {\bibfnamefont {K.}~\bibnamefont {Watanabe}}, \bibinfo {author}
  {\bibfnamefont {T.}~\bibnamefont {Taniguchi}},\ and\ \bibinfo {author}
  {\bibfnamefont {A.}~\bibnamefont {Kis}},\ }\bibfield  {title} {\bibinfo
  {title} {Room-temperature electrical control of exciton flux in a van der
  {W}aals heterostructure},\ }\href {https://doi.org/10.1038/s41586-018-0357-y}
  {\bibfield  {journal} {\bibinfo  {journal} {Nature}\ }\textbf {\bibinfo
  {volume} {560}},\ \bibinfo {pages} {340} (\bibinfo {year}
  {2018})}\BibitemShut {NoStop}%
\bibitem [{\citenamefont {Ciarrocchi}\ \emph {et~al.}(2019)\citenamefont
  {Ciarrocchi}, \citenamefont {Unuchek}, \citenamefont {Avsar}, \citenamefont
  {Watanabe}, \citenamefont {Taniguchi},\ and\ \citenamefont
  {Kis}}]{Ciarrocchi19}%
  \BibitemOpen
  \bibfield  {author} {\bibinfo {author} {\bibfnamefont {A.}~\bibnamefont
  {Ciarrocchi}}, \bibinfo {author} {\bibfnamefont {D.}~\bibnamefont {Unuchek}},
  \bibinfo {author} {\bibfnamefont {A.}~\bibnamefont {Avsar}}, \bibinfo
  {author} {\bibfnamefont {K.}~\bibnamefont {Watanabe}}, \bibinfo {author}
  {\bibfnamefont {T.}~\bibnamefont {Taniguchi}},\ and\ \bibinfo {author}
  {\bibfnamefont {A.}~\bibnamefont {Kis}},\ }\bibfield  {title} {\bibinfo
  {title} {Polarization switching and electrical control of interlayer excitons
  in two-dimensional van der {W}aals heterostructures},\ }\href
  {https://doi.org/10.1038/s41566-018-0325-y} {\bibfield  {journal} {\bibinfo
  {journal} {Nat. Photonics}\ }\textbf {\bibinfo {volume} {13}},\ \bibinfo
  {pages} {131} (\bibinfo {year} {2019})}\BibitemShut {NoStop}%
\bibitem [{\citenamefont {Leisgang}\ \emph {et~al.}(2020)\citenamefont
  {Leisgang}, \citenamefont {Shree}, \citenamefont {Paradisanos}, \citenamefont
  {Sponfeldner}, \citenamefont {Robert}, \citenamefont {Lagarde}, \citenamefont
  {Balocchi}, \citenamefont {Watanabe}, \citenamefont {Taniguchi},
  \citenamefont {Marie}, \citenamefont {Warburton}, \citenamefont {Gerber},\
  and\ \citenamefont {Urbaszek}}]{Leisgang20}%
  \BibitemOpen
  \bibfield  {author} {\bibinfo {author} {\bibfnamefont {N.}~\bibnamefont
  {Leisgang}}, \bibinfo {author} {\bibfnamefont {S.}~\bibnamefont {Shree}},
  \bibinfo {author} {\bibfnamefont {I.}~\bibnamefont {Paradisanos}}, \bibinfo
  {author} {\bibfnamefont {L.}~\bibnamefont {Sponfeldner}}, \bibinfo {author}
  {\bibfnamefont {C.}~\bibnamefont {Robert}}, \bibinfo {author} {\bibfnamefont
  {D.}~\bibnamefont {Lagarde}}, \bibinfo {author} {\bibfnamefont
  {A.}~\bibnamefont {Balocchi}}, \bibinfo {author} {\bibfnamefont
  {K.}~\bibnamefont {Watanabe}}, \bibinfo {author} {\bibfnamefont
  {T.}~\bibnamefont {Taniguchi}}, \bibinfo {author} {\bibfnamefont
  {X.}~\bibnamefont {Marie}}, \bibinfo {author} {\bibfnamefont {R.~J.}\
  \bibnamefont {Warburton}}, \bibinfo {author} {\bibfnamefont {I.~C.}\
  \bibnamefont {Gerber}},\ and\ \bibinfo {author} {\bibfnamefont
  {B.}~\bibnamefont {Urbaszek}},\ }\bibfield  {title} {\bibinfo {title} {Giant
  {S}tark splitting of an exciton in bilayer {MoS}$_2$},\ }\href
  {https://doi.org/10.1038/s41565-020-0750-1} {\bibfield  {journal} {\bibinfo
  {journal} {Nat. Nanotechnol.}\ }\textbf {\bibinfo {volume} {15}},\ \bibinfo
  {pages} {901} (\bibinfo {year} {2020})}\BibitemShut {NoStop}%
\bibitem [{\citenamefont {Lau}\ \emph {et~al.}(2018)\citenamefont {Lau},
  \citenamefont {Calvin}, \citenamefont {Gong}, \citenamefont {Yu},\ and\
  \citenamefont {Yao}}]{Lau18}%
  \BibitemOpen
  \bibfield  {author} {\bibinfo {author} {\bibfnamefont {K.~W.}\ \bibnamefont
  {Lau}}, \bibinfo {author} {\bibnamefont {Calvin}}, \bibinfo {author}
  {\bibfnamefont {Z.}~\bibnamefont {Gong}}, \bibinfo {author} {\bibfnamefont
  {H.}~\bibnamefont {Yu}},\ and\ \bibinfo {author} {\bibfnamefont
  {W.}~\bibnamefont {Yao}},\ }\bibfield  {title} {\bibinfo {title} {Interface
  excitons at lateral heterojunctions in monolayer semiconductors},\ }\href
  {https://doi.org/10.1103/PhysRevB.98.115427} {\bibfield  {journal} {\bibinfo
  {journal} {Phys. Rev. B}\ }\textbf {\bibinfo {volume} {98}},\ \bibinfo
  {pages} {115427} (\bibinfo {year} {2018})}\BibitemShut {NoStop}%
\bibitem [{\citenamefont {Rosati}\ \emph {et~al.}(2023)\citenamefont {Rosati},
  \citenamefont {Paradisanos}, \citenamefont {Huang}, \citenamefont {Gan},
  \citenamefont {George}, \citenamefont {Watanabe}, \citenamefont {Taniguchi},
  \citenamefont {Lombez}, \citenamefont {Renucci}, \citenamefont {Turchanin},
  \citenamefont {Urbaszek},\ and\ \citenamefont {Malic}}]{Rosati23}%
  \BibitemOpen
  \bibfield  {author} {\bibinfo {author} {\bibfnamefont {R.}~\bibnamefont
  {Rosati}}, \bibinfo {author} {\bibfnamefont {I.}~\bibnamefont {Paradisanos}},
  \bibinfo {author} {\bibfnamefont {L.}~\bibnamefont {Huang}}, \bibinfo
  {author} {\bibfnamefont {Z.}~\bibnamefont {Gan}}, \bibinfo {author}
  {\bibfnamefont {A.}~\bibnamefont {George}}, \bibinfo {author} {\bibfnamefont
  {K.}~\bibnamefont {Watanabe}}, \bibinfo {author} {\bibfnamefont
  {T.}~\bibnamefont {Taniguchi}}, \bibinfo {author} {\bibfnamefont
  {L.}~\bibnamefont {Lombez}}, \bibinfo {author} {\bibfnamefont
  {P.}~\bibnamefont {Renucci}}, \bibinfo {author} {\bibfnamefont
  {A.}~\bibnamefont {Turchanin}}, \bibinfo {author} {\bibfnamefont
  {B.}~\bibnamefont {Urbaszek}},\ and\ \bibinfo {author} {\bibfnamefont
  {E.}~\bibnamefont {Malic}},\ }\bibfield  {title} {\bibinfo {title} {Interface
  engineering of charge-transfer excitons in {2D} lateral heterostructures},\
  }\href {https://doi.org/10.1038/s41467-023-37889-9} {\bibfield  {journal}
  {\bibinfo  {journal} {Nat. Commun.}\ }\textbf {\bibinfo {volume} {14}},\
  \bibinfo {pages} {2438} (\bibinfo {year} {2023})}\BibitemShut {NoStop}%
\bibitem [{\citenamefont {Vandoolaeghe}\ \emph {et~al.}(2026)\citenamefont
  {Vandoolaeghe}, \citenamefont {Fortuna}, \citenamefont {Chakraborty},
  \citenamefont {Nayak}, \citenamefont {Taniguchi}, \citenamefont {Watanabe},
  \citenamefont {Sahoo}, \citenamefont {Chervy},\ and\ \citenamefont
  {Murthy}}]{Vandoolaeghe26}%
  \BibitemOpen
  \bibfield  {author} {\bibinfo {author} {\bibfnamefont {E.}~\bibnamefont
  {Vandoolaeghe}}, \bibinfo {author} {\bibfnamefont {F.}~\bibnamefont
  {Fortuna}}, \bibinfo {author} {\bibfnamefont {S.~K.}\ \bibnamefont
  {Chakraborty}}, \bibinfo {author} {\bibfnamefont {B.}~\bibnamefont {Nayak}},
  \bibinfo {author} {\bibfnamefont {T.}~\bibnamefont {Taniguchi}}, \bibinfo
  {author} {\bibfnamefont {K.}~\bibnamefont {Watanabe}}, \bibinfo {author}
  {\bibfnamefont {P.~K.}\ \bibnamefont {Sahoo}}, \bibinfo {author}
  {\bibfnamefont {T.}~\bibnamefont {Chervy}},\ and\ \bibinfo {author}
  {\bibfnamefont {P.~A.}\ \bibnamefont {Murthy}},\ }\bibfield  {title}
  {\bibinfo {title} {Dipolar excitonic quantum wires at atomically sharp
  lateral interfaces},\ }\bibfield  {journal} {\bibinfo  {journal} {Nano
  Lett.}\ }\href {https://doi.org/10.1021/acs.nanolett.6c02624}
  {10.1021/acs.nanolett.6c02624} (\bibinfo {year} {2026})\BibitemShut {NoStop}%
\bibitem [{\citenamefont {Sonowal}\ \emph {et~al.}(2026)\citenamefont
  {Sonowal}, \citenamefont {Erkensten}, \citenamefont {Malic},\ and\
  \citenamefont {Rosati}}]{Sonowal26}%
  \BibitemOpen
  \bibfield  {author} {\bibinfo {author} {\bibfnamefont {K.}~\bibnamefont
  {Sonowal}}, \bibinfo {author} {\bibfnamefont {D.}~\bibnamefont {Erkensten}},
  \bibinfo {author} {\bibfnamefont {E.}~\bibnamefont {Malic}},\ and\ \bibinfo
  {author} {\bibfnamefont {R.}~\bibnamefont {Rosati}},\ }\bibfield  {title}
  {\bibinfo {title} {Coulomb-mediated interactions of charge-transfer excitons
  in {TMD} lateral heterostructures},\ }\href
  {https://doi.org/10.1103/wm33-yhyl} {\bibfield  {journal} {\bibinfo
  {journal} {Phys. Rev. B}\ }\textbf {\bibinfo {volume} {114}},\ \bibinfo
  {pages} {L171407} (\bibinfo {year} {2026})}\BibitemShut {NoStop}%
\bibitem [{\citenamefont {Kitaura}(2026)}]{Kitaura26}%
  \BibitemOpen
  \bibfield  {author} {\bibinfo {author} {\bibfnamefont {R.}~\bibnamefont
  {Kitaura}},\ }\href {https://arxiv.org/abs/2609.32392} {\bibinfo {title}
  {Electrically switchable one-dimensional quadrupolar excitons in lateral
  double heterojunctions}} (\bibinfo {year} {2026}),\ \Eprint
  {https://arxiv.org/abs/2609.32392} {arXiv:2609.32392 [cond-mat.mes-hall]}
  \BibitemShut {NoStop}%
\bibitem [{\citenamefont {Castellanos-Gomez}\ \emph {et~al.}(2013)\citenamefont
  {Castellanos-Gomez}, \citenamefont {Rold{\'{a}}n}, \citenamefont
  {Cappelluti}, \citenamefont {Buscema}, \citenamefont {Guinea}, \citenamefont
  {van~der Zant},\ and\ \citenamefont {Steele}}]{Castellanos13}%
  \BibitemOpen
  \bibfield  {author} {\bibinfo {author} {\bibfnamefont {A.}~\bibnamefont
  {Castellanos-Gomez}}, \bibinfo {author} {\bibfnamefont {R.}~\bibnamefont
  {Rold{\'{a}}n}}, \bibinfo {author} {\bibfnamefont {E.}~\bibnamefont
  {Cappelluti}}, \bibinfo {author} {\bibfnamefont {M.}~\bibnamefont {Buscema}},
  \bibinfo {author} {\bibfnamefont {F.}~\bibnamefont {Guinea}}, \bibinfo
  {author} {\bibfnamefont {H.~S.~J.}\ \bibnamefont {van~der Zant}},\ and\
  \bibinfo {author} {\bibfnamefont {G.~A.}\ \bibnamefont {Steele}},\ }\bibfield
   {title} {\bibinfo {title} {Local strain engineering in atomically thin
  {MoS}$_2$},\ }\href {https://doi.org/10.1021/nl402875m} {\bibfield  {journal}
  {\bibinfo  {journal} {Nano Lett.}\ }\textbf {\bibinfo {volume} {13}},\
  \bibinfo {pages} {5361} (\bibinfo {year} {2013})}\BibitemShut {NoStop}%
\bibitem [{\citenamefont {Conley}\ \emph {et~al.}(2013)\citenamefont {Conley},
  \citenamefont {Wang}, \citenamefont {Ziegler}, \citenamefont {Haglund},
  \citenamefont {Pantelides},\ and\ \citenamefont {Bolotin}}]{Conley13}%
  \BibitemOpen
  \bibfield  {author} {\bibinfo {author} {\bibfnamefont {H.~J.}\ \bibnamefont
  {Conley}}, \bibinfo {author} {\bibfnamefont {B.}~\bibnamefont {Wang}},
  \bibinfo {author} {\bibfnamefont {J.~I.}\ \bibnamefont {Ziegler}}, \bibinfo
  {author} {\bibfnamefont {R.~F.}\ \bibnamefont {Haglund}}, \bibinfo {author}
  {\bibfnamefont {S.~T.}\ \bibnamefont {Pantelides}},\ and\ \bibinfo {author}
  {\bibfnamefont {K.~I.}\ \bibnamefont {Bolotin}},\ }\bibfield  {title}
  {\bibinfo {title} {Bandgap engineering of strained monolayer and bilayer
  {MoS}$_2$},\ }\href {https://doi.org/10.1021/nl4014748} {\bibfield  {journal}
  {\bibinfo  {journal} {Nano Lett.}\ }\textbf {\bibinfo {volume} {13}},\
  \bibinfo {pages} {3626} (\bibinfo {year} {2013})}\BibitemShut {NoStop}%
\bibitem [{\citenamefont {Zhu}\ \emph {et~al.}(2013)\citenamefont {Zhu},
  \citenamefont {Wang}, \citenamefont {Liu}, \citenamefont {Marie},
  \citenamefont {Qiao}, \citenamefont {Zhang}, \citenamefont {Wu},
  \citenamefont {Fan}, \citenamefont {Tan}, \citenamefont {Amand},\ and\
  \citenamefont {Urbaszek}}]{Zhu13}%
  \BibitemOpen
  \bibfield  {author} {\bibinfo {author} {\bibfnamefont {C.~R.}\ \bibnamefont
  {Zhu}}, \bibinfo {author} {\bibfnamefont {G.}~\bibnamefont {Wang}}, \bibinfo
  {author} {\bibfnamefont {B.~L.}\ \bibnamefont {Liu}}, \bibinfo {author}
  {\bibfnamefont {X.}~\bibnamefont {Marie}}, \bibinfo {author} {\bibfnamefont
  {X.~F.}\ \bibnamefont {Qiao}}, \bibinfo {author} {\bibfnamefont
  {X.}~\bibnamefont {Zhang}}, \bibinfo {author} {\bibfnamefont {X.~X.}\
  \bibnamefont {Wu}}, \bibinfo {author} {\bibfnamefont {H.}~\bibnamefont
  {Fan}}, \bibinfo {author} {\bibfnamefont {P.~H.}\ \bibnamefont {Tan}},
  \bibinfo {author} {\bibfnamefont {T.}~\bibnamefont {Amand}},\ and\ \bibinfo
  {author} {\bibfnamefont {B.}~\bibnamefont {Urbaszek}},\ }\bibfield  {title}
  {\bibinfo {title} {Strain tuning of optical emission energy and polarization
  in monolayer and bilayer {MoS}$_{2}$},\ }\href
  {https://doi.org/10.1103/PhysRevB.88.121301} {\bibfield  {journal} {\bibinfo
  {journal} {Phys. Rev. B}\ }\textbf {\bibinfo {volume} {88}},\ \bibinfo
  {pages} {121301} (\bibinfo {year} {2013})}\BibitemShut {NoStop}%
\bibitem [{\citenamefont {Ghorbani-Asl}\ \emph {et~al.}(2013)\citenamefont
  {Ghorbani-Asl}, \citenamefont {Borini}, \citenamefont {Kuc},\ and\
  \citenamefont {Heine}}]{Ghorbani13}%
  \BibitemOpen
  \bibfield  {author} {\bibinfo {author} {\bibfnamefont {M.}~\bibnamefont
  {Ghorbani-Asl}}, \bibinfo {author} {\bibfnamefont {S.}~\bibnamefont
  {Borini}}, \bibinfo {author} {\bibfnamefont {A.}~\bibnamefont {Kuc}},\ and\
  \bibinfo {author} {\bibfnamefont {T.}~\bibnamefont {Heine}},\ }\bibfield
  {title} {\bibinfo {title} {Strain-dependent modulation of conductivity in
  single-layer transition-metal dichalcogenides},\ }\href
  {https://doi.org/10.1103/PhysRevB.87.235434} {\bibfield  {journal} {\bibinfo
  {journal} {Phys. Rev. B}\ }\textbf {\bibinfo {volume} {87}},\ \bibinfo
  {pages} {235434} (\bibinfo {year} {2013})}\BibitemShut {NoStop}%
\bibitem [{\citenamefont {Niehues}\ \emph {et~al.}(2018)\citenamefont
  {Niehues}, \citenamefont {Schmidt}, \citenamefont {Dr{\"u}ppel},
  \citenamefont {Marauhn}, \citenamefont {Christiansen}, \citenamefont {Selig},
  \citenamefont {Bergh{\"a}user}, \citenamefont {Wigger}, \citenamefont
  {Schneider}, \citenamefont {Braasch}, \citenamefont {Koch}, \citenamefont
  {Castellanos-Gomez}, \citenamefont {Kuhn}, \citenamefont {Knorr},
  \citenamefont {Malic}, \citenamefont {Rohlfing}, \citenamefont {Michaelis~de
  Vasconcellos},\ and\ \citenamefont {Bratschitsch}}]{Niehues18}%
  \BibitemOpen
  \bibfield  {author} {\bibinfo {author} {\bibfnamefont {I.}~\bibnamefont
  {Niehues}}, \bibinfo {author} {\bibfnamefont {R.}~\bibnamefont {Schmidt}},
  \bibinfo {author} {\bibfnamefont {M.}~\bibnamefont {Dr{\"u}ppel}}, \bibinfo
  {author} {\bibfnamefont {P.}~\bibnamefont {Marauhn}}, \bibinfo {author}
  {\bibfnamefont {D.}~\bibnamefont {Christiansen}}, \bibinfo {author}
  {\bibfnamefont {M.}~\bibnamefont {Selig}}, \bibinfo {author} {\bibfnamefont
  {G.}~\bibnamefont {Bergh{\"a}user}}, \bibinfo {author} {\bibfnamefont
  {D.}~\bibnamefont {Wigger}}, \bibinfo {author} {\bibfnamefont
  {R.}~\bibnamefont {Schneider}}, \bibinfo {author} {\bibfnamefont
  {L.}~\bibnamefont {Braasch}}, \bibinfo {author} {\bibfnamefont
  {R.}~\bibnamefont {Koch}}, \bibinfo {author} {\bibfnamefont {A.}~\bibnamefont
  {Castellanos-Gomez}}, \bibinfo {author} {\bibfnamefont {T.}~\bibnamefont
  {Kuhn}}, \bibinfo {author} {\bibfnamefont {A.}~\bibnamefont {Knorr}},
  \bibinfo {author} {\bibfnamefont {E.}~\bibnamefont {Malic}}, \bibinfo
  {author} {\bibfnamefont {M.}~\bibnamefont {Rohlfing}}, \bibinfo {author}
  {\bibfnamefont {S.}~\bibnamefont {Michaelis~de Vasconcellos}},\ and\ \bibinfo
  {author} {\bibfnamefont {R.}~\bibnamefont {Bratschitsch}},\ }\bibfield
  {title} {\bibinfo {title} {Strain control of exciton--phonon coupling in
  atomically thin semiconductors},\ }\href
  {https://doi.org/10.1021/acs.nanolett.7b04868} {\bibfield  {journal}
  {\bibinfo  {journal} {Nano Lett.}\ }\textbf {\bibinfo {volume} {18}},\
  \bibinfo {pages} {1751} (\bibinfo {year} {2018})}\BibitemShut {NoStop}%
\bibitem [{\citenamefont {Khatibi}\ \emph {et~al.}(2018)\citenamefont
  {Khatibi}, \citenamefont {Feierabend}, \citenamefont {Selig}, \citenamefont
  {Brem}, \citenamefont {Linder\"alv}, \citenamefont {Erhart},\ and\
  \citenamefont {Malic}}]{Khatibi18}%
  \BibitemOpen
  \bibfield  {author} {\bibinfo {author} {\bibfnamefont {Z.}~\bibnamefont
  {Khatibi}}, \bibinfo {author} {\bibfnamefont {M.}~\bibnamefont {Feierabend}},
  \bibinfo {author} {\bibfnamefont {M.}~\bibnamefont {Selig}}, \bibinfo
  {author} {\bibfnamefont {S.}~\bibnamefont {Brem}}, \bibinfo {author}
  {\bibfnamefont {C.}~\bibnamefont {Linder\"alv}}, \bibinfo {author}
  {\bibfnamefont {P.}~\bibnamefont {Erhart}},\ and\ \bibinfo {author}
  {\bibfnamefont {E.}~\bibnamefont {Malic}},\ }\bibfield  {title} {\bibinfo
  {title} {Impact of strain on the excitonic linewidth in transition metal
  dichalcogenides},\ }\href {https://doi.org/10.1088/2053-1583/aae953}
  {\bibfield  {journal} {\bibinfo  {journal} {2D Mater.}\ }\textbf {\bibinfo
  {volume} {6}},\ \bibinfo {pages} {015015} (\bibinfo {year}
  {2018})}\BibitemShut {NoStop}%
\bibitem [{\citenamefont {Schmidt}\ \emph {et~al.}(2016)\citenamefont
  {Schmidt}, \citenamefont {Niehues}, \citenamefont {Schneider}, \citenamefont
  {Dr\"uppel}, \citenamefont {Deilmann}, \citenamefont {Rohlfing},
  \citenamefont {de~Vasconcellos}, \citenamefont {Castellanos-Gomez},\ and\
  \citenamefont {Bratschitsch}}]{Schmidt16}%
  \BibitemOpen
  \bibfield  {author} {\bibinfo {author} {\bibfnamefont {R.}~\bibnamefont
  {Schmidt}}, \bibinfo {author} {\bibfnamefont {I.}~\bibnamefont {Niehues}},
  \bibinfo {author} {\bibfnamefont {R.}~\bibnamefont {Schneider}}, \bibinfo
  {author} {\bibfnamefont {M.}~\bibnamefont {Dr\"uppel}}, \bibinfo {author}
  {\bibfnamefont {T.}~\bibnamefont {Deilmann}}, \bibinfo {author}
  {\bibfnamefont {M.}~\bibnamefont {Rohlfing}}, \bibinfo {author}
  {\bibfnamefont {S.~M.}\ \bibnamefont {de~Vasconcellos}}, \bibinfo {author}
  {\bibfnamefont {A.}~\bibnamefont {Castellanos-Gomez}},\ and\ \bibinfo
  {author} {\bibfnamefont {R.}~\bibnamefont {Bratschitsch}},\ }\bibfield
  {title} {\bibinfo {title} {Reversible uniaxial strain tuning in atomically
  thin {WS}e$_2$},\ }\href {https://doi.org/10.1088/2053-1583/3/2/021011}
  {\bibfield  {journal} {\bibinfo  {journal} {2D Mater.}\ }\textbf {\bibinfo
  {volume} {3}},\ \bibinfo {pages} {021011} (\bibinfo {year}
  {2016})}\BibitemShut {NoStop}%
\bibitem [{\citenamefont {Deilmann}\ and\ \citenamefont
  {Thygesen}(2019)}]{Deilmann19}%
  \BibitemOpen
  \bibfield  {author} {\bibinfo {author} {\bibfnamefont {T.}~\bibnamefont
  {Deilmann}}\ and\ \bibinfo {author} {\bibfnamefont {K.~S.}\ \bibnamefont
  {Thygesen}},\ }\bibfield  {title} {\bibinfo {title} {Finite-momentum exciton
  landscape in mono- and bilayer transition metal dichalcogenides},\ }\href
  {https://doi.org/10.1088/2053-1583/ab0e1d} {\bibfield  {journal} {\bibinfo
  {journal} {2D Mater.}\ }\textbf {\bibinfo {volume} {6}},\ \bibinfo {pages}
  {035003} (\bibinfo {year} {2019})}\BibitemShut {NoStop}%
\bibitem [{\citenamefont {Blundo}\ \emph {et~al.}(2020)\citenamefont {Blundo},
  \citenamefont {Felici}, \citenamefont {Yildirim}, \citenamefont {Pettinari},
  \citenamefont {Tedeschi}, \citenamefont {Miriametro}, \citenamefont {Liu},
  \citenamefont {Ma}, \citenamefont {Lu},\ and\ \citenamefont
  {Polimeni}}]{Blundo20}%
  \BibitemOpen
  \bibfield  {author} {\bibinfo {author} {\bibfnamefont {E.}~\bibnamefont
  {Blundo}}, \bibinfo {author} {\bibfnamefont {M.}~\bibnamefont {Felici}},
  \bibinfo {author} {\bibfnamefont {T.}~\bibnamefont {Yildirim}}, \bibinfo
  {author} {\bibfnamefont {G.}~\bibnamefont {Pettinari}}, \bibinfo {author}
  {\bibfnamefont {D.}~\bibnamefont {Tedeschi}}, \bibinfo {author}
  {\bibfnamefont {A.}~\bibnamefont {Miriametro}}, \bibinfo {author}
  {\bibfnamefont {B.}~\bibnamefont {Liu}}, \bibinfo {author} {\bibfnamefont
  {W.}~\bibnamefont {Ma}}, \bibinfo {author} {\bibfnamefont {Y.}~\bibnamefont
  {Lu}},\ and\ \bibinfo {author} {\bibfnamefont {A.}~\bibnamefont {Polimeni}},\
  }\bibfield  {title} {\bibinfo {title} {Evidence of the direct-to-indirect
  band gap transition in strained two-dimensional {WS}$_{2}$, {MoS}$_{2}$, and
  {WSe}$_{2}$},\ }\href {https://doi.org/10.1103/PhysRevResearch.2.012024}
  {\bibfield  {journal} {\bibinfo  {journal} {Phys. Rev. Res.}\ }\textbf
  {\bibinfo {volume} {2}},\ \bibinfo {pages} {012024(R)} (\bibinfo {year}
  {2020})}\BibitemShut {NoStop}%
\bibitem [{\citenamefont {Li}\ \emph {et~al.}(2023)\citenamefont {Li},
  \citenamefont {Gillen}, \citenamefont {Palummo}, \citenamefont
  {Milošević},\ and\ \citenamefont {Peeters}}]{Li23}%
  \BibitemOpen
  \bibfield  {author} {\bibinfo {author} {\bibfnamefont {L.~L.}\ \bibnamefont
  {Li}}, \bibinfo {author} {\bibfnamefont {R.}~\bibnamefont {Gillen}}, \bibinfo
  {author} {\bibfnamefont {M.}~\bibnamefont {Palummo}}, \bibinfo {author}
  {\bibfnamefont {M.~V.}\ \bibnamefont {Milošević}},\ and\ \bibinfo {author}
  {\bibfnamefont {F.~M.}\ \bibnamefont {Peeters}},\ }\bibfield  {title}
  {\bibinfo {title} {Strain tunable interlayer and intralayer excitons in
  vertically stacked {MoSe}$_2$/{WSe}$_2$ heterobilayers},\ }\href
  {https://doi.org/10.1063/5.0147761} {\bibfield  {journal} {\bibinfo
  {journal} {Appl. Phys. Lett.}\ }\textbf {\bibinfo {volume} {123}},\ \bibinfo
  {pages} {033102} (\bibinfo {year} {2023})}\BibitemShut {NoStop}%
\bibitem [{\citenamefont {Faria~Junior}\ and\ \citenamefont
  {Fabian}(2023)}]{Faria23}%
  \BibitemOpen
  \bibfield  {author} {\bibinfo {author} {\bibfnamefont {P.~E.}\ \bibnamefont
  {Faria~Junior}}\ and\ \bibinfo {author} {\bibfnamefont {J.}~\bibnamefont
  {Fabian}},\ }\bibfield  {title} {\bibinfo {title} {Signatures of electric
  field and layer separation effects on the spin-valley physics of
  {MoSe}$_2$/{WSe}$_2$ heterobilayers: From energy bands to dipolar excitons},\
  }\bibfield  {journal} {\bibinfo  {journal} {Nanomateri.}\ }\textbf {\bibinfo
  {volume} {13}},\ \href {https://doi.org/10.3390/nano13071187}
  {10.3390/nano13071187} (\bibinfo {year} {2023})\BibitemShut {NoStop}%
\bibitem [{\citenamefont {Cordovilla~Leon}\ \emph {et~al.}(2018)\citenamefont
  {Cordovilla~Leon}, \citenamefont {Li}, \citenamefont {Jang}, \citenamefont
  {Cheng},\ and\ \citenamefont {Deotare}}]{Cordovilla18}%
  \BibitemOpen
  \bibfield  {author} {\bibinfo {author} {\bibfnamefont {D.~F.}\ \bibnamefont
  {Cordovilla~Leon}}, \bibinfo {author} {\bibfnamefont {Z.}~\bibnamefont {Li}},
  \bibinfo {author} {\bibfnamefont {S.~W.}\ \bibnamefont {Jang}}, \bibinfo
  {author} {\bibfnamefont {C.-H.}\ \bibnamefont {Cheng}},\ and\ \bibinfo
  {author} {\bibfnamefont {P.~B.}\ \bibnamefont {Deotare}},\ }\bibfield
  {title} {\bibinfo {title} {Exciton transport in strained monolayer
  {WS}e$_2$},\ }\href {https://doi.org/10.1063/1.5063263} {\bibfield  {journal}
  {\bibinfo  {journal} {Appl. Phys. Lett.}\ }\textbf {\bibinfo {volume}
  {113}},\ \bibinfo {pages} {252101} (\bibinfo {year} {2018})}\BibitemShut
  {NoStop}%
\bibitem [{\citenamefont {Moon}\ \emph {et~al.}(2020)\citenamefont {Moon},
  \citenamefont {Grosso}, \citenamefont {Chakraborty}, \citenamefont {Peng},
  \citenamefont {Taniguchi}, \citenamefont {Watanabe},\ and\ \citenamefont
  {Englund}}]{Moon20}%
  \BibitemOpen
  \bibfield  {author} {\bibinfo {author} {\bibfnamefont {H.}~\bibnamefont
  {Moon}}, \bibinfo {author} {\bibfnamefont {G.}~\bibnamefont {Grosso}},
  \bibinfo {author} {\bibfnamefont {C.}~\bibnamefont {Chakraborty}}, \bibinfo
  {author} {\bibfnamefont {C.}~\bibnamefont {Peng}}, \bibinfo {author}
  {\bibfnamefont {T.}~\bibnamefont {Taniguchi}}, \bibinfo {author}
  {\bibfnamefont {K.}~\bibnamefont {Watanabe}},\ and\ \bibinfo {author}
  {\bibfnamefont {D.}~\bibnamefont {Englund}},\ }\bibfield  {title} {\bibinfo
  {title} {Dynamic exciton funneling by local strain control in a monolayer
  semiconductor},\ }\href {https://doi.org/10.1021/acs.nanolett.0c02757}
  {\bibfield  {journal} {\bibinfo  {journal} {Nano Lett.}\ }\textbf {\bibinfo
  {volume} {20}},\ \bibinfo {pages} {6791} (\bibinfo {year}
  {2020})}\BibitemShut {NoStop}%
\bibitem [{\citenamefont {Harats}\ \emph {et~al.}(2020)\citenamefont {Harats},
  \citenamefont {Kirchhof}, \citenamefont {Qiao}, \citenamefont {Greben},\ and\
  \citenamefont {Bolotin}}]{Harats20}%
  \BibitemOpen
  \bibfield  {author} {\bibinfo {author} {\bibfnamefont {M.~G.}\ \bibnamefont
  {Harats}}, \bibinfo {author} {\bibfnamefont {J.~N.}\ \bibnamefont
  {Kirchhof}}, \bibinfo {author} {\bibfnamefont {M.}~\bibnamefont {Qiao}},
  \bibinfo {author} {\bibfnamefont {K.}~\bibnamefont {Greben}},\ and\ \bibinfo
  {author} {\bibfnamefont {K.~I.}\ \bibnamefont {Bolotin}},\ }\bibfield
  {title} {\bibinfo {title} {Dynamics and efficient conversion of excitons to
  trions in non-uniformly strained monolayer {WS}$_2$},\ }\href
  {https://doi.org/10.1038/s41566-019-0581-5} {\bibfield  {journal} {\bibinfo
  {journal} {Nat. Photonics}\ }\textbf {\bibinfo {volume} {14}},\ \bibinfo
  {pages} {324} (\bibinfo {year} {2020})}\BibitemShut {NoStop}%
\bibitem [{\citenamefont {Su}\ \emph {et~al.}(2022)\citenamefont {Su},
  \citenamefont {Xu}, \citenamefont {Cheng}, \citenamefont {Li}, \citenamefont
  {Liu}, \citenamefont {Watanabe}, \citenamefont {Taniguchi}, \citenamefont
  {Berkelbach}, \citenamefont {Hone},\ and\ \citenamefont {Delor}}]{Su22}%
  \BibitemOpen
  \bibfield  {author} {\bibinfo {author} {\bibfnamefont {H.}~\bibnamefont
  {Su}}, \bibinfo {author} {\bibfnamefont {D.}~\bibnamefont {Xu}}, \bibinfo
  {author} {\bibfnamefont {S.-W.}\ \bibnamefont {Cheng}}, \bibinfo {author}
  {\bibfnamefont {B.}~\bibnamefont {Li}}, \bibinfo {author} {\bibfnamefont
  {S.}~\bibnamefont {Liu}}, \bibinfo {author} {\bibfnamefont {K.}~\bibnamefont
  {Watanabe}}, \bibinfo {author} {\bibfnamefont {T.}~\bibnamefont {Taniguchi}},
  \bibinfo {author} {\bibfnamefont {T.~C.}\ \bibnamefont {Berkelbach}},
  \bibinfo {author} {\bibfnamefont {J.~C.}\ \bibnamefont {Hone}},\ and\
  \bibinfo {author} {\bibfnamefont {M.}~\bibnamefont {Delor}},\ }\bibfield
  {title} {\bibinfo {title} {Dark-exciton driven energy funneling into
  dielectric inhomogeneities in two-dimensional semiconductors},\ }\href
  {https://doi.org/10.1021/acs.nanolett.1c04997} {\bibfield  {journal}
  {\bibinfo  {journal} {Nano Lett.}\ }\textbf {\bibinfo {volume} {22}},\
  \bibinfo {pages} {2843} (\bibinfo {year} {2022})}\BibitemShut {NoStop}%
\bibitem [{\citenamefont {Gelly}\ \emph {et~al.}(2022)\citenamefont {Gelly},
  \citenamefont {Renaud}, \citenamefont {Liao}, \citenamefont {Pingault},
  \citenamefont {Bogdanovic}, \citenamefont {Scuri}, \citenamefont {Watanabe},
  \citenamefont {Taniguchi}, \citenamefont {Urbaszek}, \citenamefont {Park},\
  and\ \citenamefont {Lon{\v{c}}ar}}]{Gelly22}%
  \BibitemOpen
  \bibfield  {author} {\bibinfo {author} {\bibfnamefont {R.~J.}\ \bibnamefont
  {Gelly}}, \bibinfo {author} {\bibfnamefont {D.}~\bibnamefont {Renaud}},
  \bibinfo {author} {\bibfnamefont {X.}~\bibnamefont {Liao}}, \bibinfo {author}
  {\bibfnamefont {B.}~\bibnamefont {Pingault}}, \bibinfo {author}
  {\bibfnamefont {S.}~\bibnamefont {Bogdanovic}}, \bibinfo {author}
  {\bibfnamefont {G.}~\bibnamefont {Scuri}}, \bibinfo {author} {\bibfnamefont
  {K.}~\bibnamefont {Watanabe}}, \bibinfo {author} {\bibfnamefont
  {T.}~\bibnamefont {Taniguchi}}, \bibinfo {author} {\bibfnamefont
  {B.}~\bibnamefont {Urbaszek}}, \bibinfo {author} {\bibfnamefont
  {H.}~\bibnamefont {Park}},\ and\ \bibinfo {author} {\bibfnamefont
  {M.}~\bibnamefont {Lon{\v{c}}ar}},\ }\bibfield  {title} {\bibinfo {title}
  {Probing dark exciton navigation through a local strain landscape in a
  {WS}e$_2$ monolayer},\ }\href {https://doi.org/10.1038/s41467-021-27877-2}
  {\bibfield  {journal} {\bibinfo  {journal} {Nat. Commun.}\ }\textbf {\bibinfo
  {volume} {13}},\ \bibinfo {pages} {232} (\bibinfo {year} {2022})}\BibitemShut
  {NoStop}%
\bibitem [{\citenamefont {Lee}\ \emph {et~al.}(2022)\citenamefont {Lee},
  \citenamefont {Koo}, \citenamefont {Choi}, \citenamefont {Kumar},
  \citenamefont {Lee}, \citenamefont {Ji}, \citenamefont {Choi}, \citenamefont
  {Kang}, \citenamefont {Kim}, \citenamefont {Park}, \citenamefont {Choo},\
  and\ \citenamefont {Park}}]{Lee22}%
  \BibitemOpen
  \bibfield  {author} {\bibinfo {author} {\bibfnamefont {H.}~\bibnamefont
  {Lee}}, \bibinfo {author} {\bibfnamefont {Y.}~\bibnamefont {Koo}}, \bibinfo
  {author} {\bibfnamefont {J.}~\bibnamefont {Choi}}, \bibinfo {author}
  {\bibfnamefont {S.}~\bibnamefont {Kumar}}, \bibinfo {author} {\bibfnamefont
  {H.-T.}\ \bibnamefont {Lee}}, \bibinfo {author} {\bibfnamefont
  {G.}~\bibnamefont {Ji}}, \bibinfo {author} {\bibfnamefont {S.~H.}\
  \bibnamefont {Choi}}, \bibinfo {author} {\bibfnamefont {M.}~\bibnamefont
  {Kang}}, \bibinfo {author} {\bibfnamefont {K.~K.}\ \bibnamefont {Kim}},
  \bibinfo {author} {\bibfnamefont {H.-R.}\ \bibnamefont {Park}}, \bibinfo
  {author} {\bibfnamefont {H.}~\bibnamefont {Choo}},\ and\ \bibinfo {author}
  {\bibfnamefont {K.-D.}\ \bibnamefont {Park}},\ }\bibfield  {title} {\bibinfo
  {title} {Drift-dominant exciton funneling and trion conversion in {2D}
  semiconductors on the nanogap},\ }\href
  {https://doi.org/10.1126/sciadv.abm5236} {\bibfield  {journal} {\bibinfo
  {journal} {Sci. Adv.}\ }\textbf {\bibinfo {volume} {8}},\ \bibinfo {pages}
  {eabm5236} (\bibinfo {year} {2022})}\BibitemShut {NoStop}%
\bibitem [{\citenamefont {Branny}\ \emph {et~al.}(2017)\citenamefont {Branny},
  \citenamefont {Kumar}, \citenamefont {Proux},\ and\ \citenamefont
  {Gerardot}}]{Branny17}%
  \BibitemOpen
  \bibfield  {author} {\bibinfo {author} {\bibfnamefont {A.}~\bibnamefont
  {Branny}}, \bibinfo {author} {\bibfnamefont {S.}~\bibnamefont {Kumar}},
  \bibinfo {author} {\bibfnamefont {R.}~\bibnamefont {Proux}},\ and\ \bibinfo
  {author} {\bibfnamefont {B.~D.}\ \bibnamefont {Gerardot}},\ }\bibfield
  {title} {\bibinfo {title} {Deterministic strain-induced arrays of quantum
  emitters in a two-dimensional semiconductor},\ }\href
  {https://doi.org/10.1038/ncomms15053} {\bibfield  {journal} {\bibinfo
  {journal} {Nat. Commun.}\ }\textbf {\bibinfo {volume} {8}},\ \bibinfo {pages}
  {15053} (\bibinfo {year} {2017})}\BibitemShut {NoStop}%
\bibitem [{\citenamefont {Palacios-Berraquero}\ \emph
  {et~al.}(2017)\citenamefont {Palacios-Berraquero}, \citenamefont {Kara},
  \citenamefont {Montblanch}, \citenamefont {Barbone}, \citenamefont
  {Latawiec}, \citenamefont {Yoon}, \citenamefont {Ott}, \citenamefont
  {Loncar}, \citenamefont {Ferrari},\ and\ \citenamefont
  {Atat{\"u}re}}]{Palacios17}%
  \BibitemOpen
  \bibfield  {author} {\bibinfo {author} {\bibfnamefont {C.}~\bibnamefont
  {Palacios-Berraquero}}, \bibinfo {author} {\bibfnamefont {D.~M.}\
  \bibnamefont {Kara}}, \bibinfo {author} {\bibfnamefont {A.~R.-P.}\
  \bibnamefont {Montblanch}}, \bibinfo {author} {\bibfnamefont
  {M.}~\bibnamefont {Barbone}}, \bibinfo {author} {\bibfnamefont
  {P.}~\bibnamefont {Latawiec}}, \bibinfo {author} {\bibfnamefont
  {D.}~\bibnamefont {Yoon}}, \bibinfo {author} {\bibfnamefont {A.~K.}\
  \bibnamefont {Ott}}, \bibinfo {author} {\bibfnamefont {M.}~\bibnamefont
  {Loncar}}, \bibinfo {author} {\bibfnamefont {A.~C.}\ \bibnamefont
  {Ferrari}},\ and\ \bibinfo {author} {\bibfnamefont {M.}~\bibnamefont
  {Atat{\"u}re}},\ }\bibfield  {title} {\bibinfo {title} {Large-scale
  quantum-emitter arrays in atomically thin semiconductors},\ }\href
  {https://doi.org/10.1038/ncomms15093} {\bibfield  {journal} {\bibinfo
  {journal} {Nat. Commun.}\ }\textbf {\bibinfo {volume} {8}},\ \bibinfo {pages}
  {15093} (\bibinfo {year} {2017})}\BibitemShut {NoStop}%
\bibitem [{\citenamefont {Kern}\ \emph {et~al.}(2016)\citenamefont {Kern},
  \citenamefont {Niehues}, \citenamefont {Tonndorf}, \citenamefont {Schmidt},
  \citenamefont {Wigger}, \citenamefont {Schneider}, \citenamefont {Stiehm},
  \citenamefont {Michaelis~de Vasconcellos}, \citenamefont {Reiter},
  \citenamefont {Kuhn},\ and\ \citenamefont {Bratschitsch}}]{Kern16}%
  \BibitemOpen
  \bibfield  {author} {\bibinfo {author} {\bibfnamefont {J.}~\bibnamefont
  {Kern}}, \bibinfo {author} {\bibfnamefont {I.}~\bibnamefont {Niehues}},
  \bibinfo {author} {\bibfnamefont {P.}~\bibnamefont {Tonndorf}}, \bibinfo
  {author} {\bibfnamefont {R.}~\bibnamefont {Schmidt}}, \bibinfo {author}
  {\bibfnamefont {D.}~\bibnamefont {Wigger}}, \bibinfo {author} {\bibfnamefont
  {R.}~\bibnamefont {Schneider}}, \bibinfo {author} {\bibfnamefont
  {T.}~\bibnamefont {Stiehm}}, \bibinfo {author} {\bibfnamefont
  {S.}~\bibnamefont {Michaelis~de Vasconcellos}}, \bibinfo {author}
  {\bibfnamefont {D.~E.}\ \bibnamefont {Reiter}}, \bibinfo {author}
  {\bibfnamefont {T.}~\bibnamefont {Kuhn}},\ and\ \bibinfo {author}
  {\bibfnamefont {R.}~\bibnamefont {Bratschitsch}},\ }\bibfield  {title}
  {\bibinfo {title} {Nanoscale positioning of single-photon emitters in
  atomically thin {WS}e$_2$},\ }\href {https://doi.org/10.1002/adma.201600560}
  {\bibfield  {journal} {\bibinfo  {journal} {Adv. Mater.}\ }\textbf {\bibinfo
  {volume} {28}},\ \bibinfo {pages} {7101} (\bibinfo {year}
  {2016})}\BibitemShut {NoStop}%
\bibitem [{\citenamefont {Rosati}\ \emph
  {et~al.}(2021{\natexlab{a}})\citenamefont {Rosati}, \citenamefont {Schmidt},
  \citenamefont {Brem}, \citenamefont {Perea-Caus{\'i}n}, \citenamefont
  {Niehues}, \citenamefont {Kern}, \citenamefont {Preu{\ss}}, \citenamefont
  {Schneider}, \citenamefont {Michaelis~de Vasconcellos}, \citenamefont
  {Bratschitsch},\ and\ \citenamefont {Malic}}]{Rosati21e}%
  \BibitemOpen
  \bibfield  {author} {\bibinfo {author} {\bibfnamefont {R.}~\bibnamefont
  {Rosati}}, \bibinfo {author} {\bibfnamefont {R.}~\bibnamefont {Schmidt}},
  \bibinfo {author} {\bibfnamefont {S.}~\bibnamefont {Brem}}, \bibinfo {author}
  {\bibfnamefont {R.}~\bibnamefont {Perea-Caus{\'i}n}}, \bibinfo {author}
  {\bibfnamefont {I.}~\bibnamefont {Niehues}}, \bibinfo {author} {\bibfnamefont
  {J.}~\bibnamefont {Kern}}, \bibinfo {author} {\bibfnamefont {J.~A.}\
  \bibnamefont {Preu{\ss}}}, \bibinfo {author} {\bibfnamefont {R.}~\bibnamefont
  {Schneider}}, \bibinfo {author} {\bibfnamefont {S.}~\bibnamefont
  {Michaelis~de Vasconcellos}}, \bibinfo {author} {\bibfnamefont
  {R.}~\bibnamefont {Bratschitsch}},\ and\ \bibinfo {author} {\bibfnamefont
  {E.}~\bibnamefont {Malic}},\ }\bibfield  {title} {\bibinfo {title} {Dark
  exciton anti-funneling in atomically thin semiconductors},\ }\href
  {https://doi.org/10.1038/s41467-021-27425-y} {\bibfield  {journal} {\bibinfo
  {journal} {Nat. Commun.}\ }\textbf {\bibinfo {volume} {12}},\ \bibinfo
  {pages} {7221} (\bibinfo {year} {2021}{\natexlab{a}})}\BibitemShut {NoStop}%
\bibitem [{\citenamefont {Rosati}\ \emph
  {et~al.}(2021{\natexlab{b}})\citenamefont {Rosati}, \citenamefont {Brem},
  \citenamefont {Perea-Caus{\'{\i}}n}, \citenamefont {Schmidt}, \citenamefont
  {Niehues}, \citenamefont {de~Vasconcellos}, \citenamefont {Bratschitsch},\
  and\ \citenamefont {Malic}}]{Rosati21a}%
  \BibitemOpen
  \bibfield  {author} {\bibinfo {author} {\bibfnamefont {R.}~\bibnamefont
  {Rosati}}, \bibinfo {author} {\bibfnamefont {S.}~\bibnamefont {Brem}},
  \bibinfo {author} {\bibfnamefont {R.}~\bibnamefont {Perea-Caus{\'{\i}}n}},
  \bibinfo {author} {\bibfnamefont {R.}~\bibnamefont {Schmidt}}, \bibinfo
  {author} {\bibfnamefont {I.}~\bibnamefont {Niehues}}, \bibinfo {author}
  {\bibfnamefont {S.~M.}\ \bibnamefont {de~Vasconcellos}}, \bibinfo {author}
  {\bibfnamefont {R.}~\bibnamefont {Bratschitsch}},\ and\ \bibinfo {author}
  {\bibfnamefont {E.}~\bibnamefont {Malic}},\ }\bibfield  {title} {\bibinfo
  {title} {Strain-dependent exciton diffusion in transition metal
  dichalcogenides},\ }\href {https://doi.org/10.1088/2053-1583/abbd51}
  {\bibfield  {journal} {\bibinfo  {journal} {2D Mater.}\ }\textbf {\bibinfo
  {volume} {8}},\ \bibinfo {pages} {015030} (\bibinfo {year}
  {2021}{\natexlab{b}})}\BibitemShut {NoStop}%
\bibitem [{\citenamefont {Uddin}\ \emph {et~al.}(2022)\citenamefont {Uddin},
  \citenamefont {Higashitarumizu}, \citenamefont {Kim}, \citenamefont {Yi},
  \citenamefont {Zhang}, \citenamefont {Chrzan},\ and\ \citenamefont
  {Javey}}]{Uddin22}%
  \BibitemOpen
  \bibfield  {author} {\bibinfo {author} {\bibfnamefont {S.~Z.}\ \bibnamefont
  {Uddin}}, \bibinfo {author} {\bibfnamefont {N.}~\bibnamefont
  {Higashitarumizu}}, \bibinfo {author} {\bibfnamefont {H.}~\bibnamefont
  {Kim}}, \bibinfo {author} {\bibfnamefont {J.}~\bibnamefont {Yi}}, \bibinfo
  {author} {\bibfnamefont {X.}~\bibnamefont {Zhang}}, \bibinfo {author}
  {\bibfnamefont {D.}~\bibnamefont {Chrzan}},\ and\ \bibinfo {author}
  {\bibfnamefont {A.}~\bibnamefont {Javey}},\ }\bibfield  {title} {\bibinfo
  {title} {Enhanced neutral exciton diffusion in monolayer {WS}$_2$ by
  exciton–exciton annihilation},\ }\href
  {https://doi.org/10.1021/acsnano.2c00956} {\bibfield  {journal} {\bibinfo
  {journal} {ACS Nano}\ }\textbf {\bibinfo {volume} {16}},\ \bibinfo {pages}
  {8005} (\bibinfo {year} {2022})}\BibitemShut {NoStop}%
\bibitem [{\citenamefont {Kumar}\ \emph {et~al.}(2024)\citenamefont {Kumar},
  \citenamefont {Yagodkin}, \citenamefont {Rosati}, \citenamefont {Bock},
  \citenamefont {Schattauer}, \citenamefont {Tobisch}, \citenamefont {Hagel},
  \citenamefont {H{\"o}fer}, \citenamefont {Kirchhof}, \citenamefont
  {Hern{\'a}ndez~L{\'o}pez}, \citenamefont {Burfeindt}, \citenamefont {Heeg},
  \citenamefont {Gahl}, \citenamefont {Libisch}, \citenamefont {Malic},\ and\
  \citenamefont {Bolotin}}]{Kumar24}%
  \BibitemOpen
  \bibfield  {author} {\bibinfo {author} {\bibfnamefont {A.~M.}\ \bibnamefont
  {Kumar}}, \bibinfo {author} {\bibfnamefont {D.}~\bibnamefont {Yagodkin}},
  \bibinfo {author} {\bibfnamefont {R.}~\bibnamefont {Rosati}}, \bibinfo
  {author} {\bibfnamefont {D.~J.}\ \bibnamefont {Bock}}, \bibinfo {author}
  {\bibfnamefont {C.}~\bibnamefont {Schattauer}}, \bibinfo {author}
  {\bibfnamefont {S.}~\bibnamefont {Tobisch}}, \bibinfo {author} {\bibfnamefont
  {J.}~\bibnamefont {Hagel}}, \bibinfo {author} {\bibfnamefont
  {B.}~\bibnamefont {H{\"o}fer}}, \bibinfo {author} {\bibfnamefont {J.~N.}\
  \bibnamefont {Kirchhof}}, \bibinfo {author} {\bibfnamefont {P.}~\bibnamefont
  {Hern{\'a}ndez~L{\'o}pez}}, \bibinfo {author} {\bibfnamefont
  {K.}~\bibnamefont {Burfeindt}}, \bibinfo {author} {\bibfnamefont
  {S.}~\bibnamefont {Heeg}}, \bibinfo {author} {\bibfnamefont {C.}~\bibnamefont
  {Gahl}}, \bibinfo {author} {\bibfnamefont {F.}~\bibnamefont {Libisch}},
  \bibinfo {author} {\bibfnamefont {E.}~\bibnamefont {Malic}},\ and\ \bibinfo
  {author} {\bibfnamefont {K.~I.}\ \bibnamefont {Bolotin}},\ }\bibfield
  {title} {\bibinfo {title} {Strain fingerprinting of exciton valley character
  in {2D} semiconductors},\ }\href {https://doi.org/10.1038/s41467-024-51195-y}
  {\bibfield  {journal} {\bibinfo  {journal} {Nat. Commun.}\ }\textbf {\bibinfo
  {volume} {15}},\ \bibinfo {pages} {7546} (\bibinfo {year}
  {2024})}\BibitemShut {NoStop}%
\bibitem [{\citenamefont {Kumar}\ \emph {et~al.}(2025)\citenamefont {Kumar},
  \citenamefont {Bock}, \citenamefont {Yagodkin}, \citenamefont {Wietek},
  \citenamefont {H{\"o}fer}, \citenamefont {Sinner}, \citenamefont
  {Dewambrechies}, \citenamefont {L{\'o}pez}, \citenamefont {Kovalchuk},
  \citenamefont {Dhingra}, \citenamefont {Heeg}, \citenamefont {Gahl},
  \citenamefont {Libisch}, \citenamefont {Chernikov}, \citenamefont {Malic},
  \citenamefont {Rosati},\ and\ \citenamefont {Bolotin}}]{Kumar25}%
  \BibitemOpen
  \bibfield  {author} {\bibinfo {author} {\bibfnamefont {A.~M.}\ \bibnamefont
  {Kumar}}, \bibinfo {author} {\bibfnamefont {D.~J.}\ \bibnamefont {Bock}},
  \bibinfo {author} {\bibfnamefont {D.}~\bibnamefont {Yagodkin}}, \bibinfo
  {author} {\bibfnamefont {E.}~\bibnamefont {Wietek}}, \bibinfo {author}
  {\bibfnamefont {B.}~\bibnamefont {H{\"o}fer}}, \bibinfo {author}
  {\bibfnamefont {M.}~\bibnamefont {Sinner}}, \bibinfo {author} {\bibfnamefont
  {A.}~\bibnamefont {Dewambrechies}}, \bibinfo {author} {\bibfnamefont {P.~H.}\
  \bibnamefont {L{\'o}pez}}, \bibinfo {author} {\bibfnamefont {S.}~\bibnamefont
  {Kovalchuk}}, \bibinfo {author} {\bibfnamefont {R.}~\bibnamefont {Dhingra}},
  \bibinfo {author} {\bibfnamefont {S.}~\bibnamefont {Heeg}}, \bibinfo {author}
  {\bibfnamefont {C.}~\bibnamefont {Gahl}}, \bibinfo {author} {\bibfnamefont
  {F.}~\bibnamefont {Libisch}}, \bibinfo {author} {\bibfnamefont
  {A.}~\bibnamefont {Chernikov}}, \bibinfo {author} {\bibfnamefont
  {E.}~\bibnamefont {Malic}}, \bibinfo {author} {\bibfnamefont
  {R.}~\bibnamefont {Rosati}},\ and\ \bibinfo {author} {\bibfnamefont {K.~I.}\
  \bibnamefont {Bolotin}},\ }\bibfield  {title} {\bibinfo {title} {Strain
  control of valley polarization dynamics in a {2D} semiconductor via exciton
  hybridization},\ }\href {https://doi.org/10.1021/acs.nanolett.5c02636}
  {\bibfield  {journal} {\bibinfo  {journal} {Nano Lett.}\ }\textbf {\bibinfo
  {volume} {25}},\ \bibinfo {pages} {15164} (\bibinfo {year}
  {2025})}\BibitemShut {NoStop}%
\bibitem [{\citenamefont {Blundo}\ \emph {et~al.}(2022)\citenamefont {Blundo},
  \citenamefont {Junior}, \citenamefont {Surrente}, \citenamefont {Pettinari},
  \citenamefont {Prosnikov}, \citenamefont {Olkowska-Pucko}, \citenamefont
  {Zollner}, \citenamefont {Wo\ifmmode~\acute{z}\else \'{z}\fi{}niak},
  \citenamefont {Chaves}, \citenamefont {Kazimierczuk}, \citenamefont {Felici},
  \citenamefont {Babi\ifmmode~\acute{n}\else \'{n}\fi{}ski}, \citenamefont
  {Molas}, \citenamefont {Christianen}, \citenamefont {Fabian},\ and\
  \citenamefont {Polimeni}}]{Blundo22}%
  \BibitemOpen
  \bibfield  {author} {\bibinfo {author} {\bibfnamefont {E.}~\bibnamefont
  {Blundo}}, \bibinfo {author} {\bibfnamefont {P.~E.~F.}\ \bibnamefont
  {Junior}}, \bibinfo {author} {\bibfnamefont {A.}~\bibnamefont {Surrente}},
  \bibinfo {author} {\bibfnamefont {G.}~\bibnamefont {Pettinari}}, \bibinfo
  {author} {\bibfnamefont {M.~A.}\ \bibnamefont {Prosnikov}}, \bibinfo {author}
  {\bibfnamefont {K.}~\bibnamefont {Olkowska-Pucko}}, \bibinfo {author}
  {\bibfnamefont {K.}~\bibnamefont {Zollner}}, \bibinfo {author} {\bibfnamefont
  {T.}~\bibnamefont {Wo\ifmmode~\acute{z}\else \'{z}\fi{}niak}}, \bibinfo
  {author} {\bibfnamefont {A.}~\bibnamefont {Chaves}}, \bibinfo {author}
  {\bibfnamefont {T.}~\bibnamefont {Kazimierczuk}}, \bibinfo {author}
  {\bibfnamefont {M.}~\bibnamefont {Felici}}, \bibinfo {author} {\bibfnamefont
  {A.}~\bibnamefont {Babi\ifmmode~\acute{n}\else \'{n}\fi{}ski}}, \bibinfo
  {author} {\bibfnamefont {M.~R.}\ \bibnamefont {Molas}}, \bibinfo {author}
  {\bibfnamefont {P.~C.~M.}\ \bibnamefont {Christianen}}, \bibinfo {author}
  {\bibfnamefont {J.}~\bibnamefont {Fabian}},\ and\ \bibinfo {author}
  {\bibfnamefont {A.}~\bibnamefont {Polimeni}},\ }\bibfield  {title} {\bibinfo
  {title} {Strain-induced exciton hybridization in {WS}$_{2}$ monolayers
  unveiled by {Z}eeman-splitting measurements},\ }\href
  {https://doi.org/10.1103/PhysRevLett.129.067402} {\bibfield  {journal}
  {\bibinfo  {journal} {Phys. Rev. Lett.}\ }\textbf {\bibinfo {volume} {129}},\
  \bibinfo {pages} {067402} (\bibinfo {year} {2022})}\BibitemShut {NoStop}%
\bibitem [{\citenamefont {Rosati}\ \emph
  {et~al.}(2020{\natexlab{a}})\citenamefont {Rosati}, \citenamefont
  {Perea-Caus\'in}, \citenamefont {Brem},\ and\ \citenamefont
  {Malic}}]{Rosati20}%
  \BibitemOpen
  \bibfield  {author} {\bibinfo {author} {\bibfnamefont {R.}~\bibnamefont
  {Rosati}}, \bibinfo {author} {\bibfnamefont {R.}~\bibnamefont
  {Perea-Caus\'in}}, \bibinfo {author} {\bibfnamefont {S.}~\bibnamefont
  {Brem}},\ and\ \bibinfo {author} {\bibfnamefont {E.}~\bibnamefont {Malic}},\
  }\bibfield  {title} {\bibinfo {title} {Negative effective excitonic diffusion
  in monolayer transition metal dichalcogenides},\ }\href
  {https://doi.org/10.1039/C9NR07056G} {\bibfield  {journal} {\bibinfo
  {journal} {Nanoscale}\ }\textbf {\bibinfo {volume} {12}},\ \bibinfo {pages}
  {356} (\bibinfo {year} {2020}{\natexlab{a}})}\BibitemShut {NoStop}%
\bibitem [{\citenamefont {Rosati}\ \emph
  {et~al.}(2020{\natexlab{b}})\citenamefont {Rosati}, \citenamefont {Wagner},
  \citenamefont {Brem}, \citenamefont {Perea-Caus{\'i}n}, \citenamefont
  {Wietek}, \citenamefont {Zipfel}, \citenamefont {Ziegler}, \citenamefont
  {Selig}, \citenamefont {Taniguchi}, \citenamefont {Watanabe}, \citenamefont
  {Knorr}, \citenamefont {Chernikov},\ and\ \citenamefont {Malic}}]{Rosati20b}%
  \BibitemOpen
  \bibfield  {author} {\bibinfo {author} {\bibfnamefont {R.}~\bibnamefont
  {Rosati}}, \bibinfo {author} {\bibfnamefont {K.}~\bibnamefont {Wagner}},
  \bibinfo {author} {\bibfnamefont {S.}~\bibnamefont {Brem}}, \bibinfo {author}
  {\bibfnamefont {R.}~\bibnamefont {Perea-Caus{\'i}n}}, \bibinfo {author}
  {\bibfnamefont {E.}~\bibnamefont {Wietek}}, \bibinfo {author} {\bibfnamefont
  {J.}~\bibnamefont {Zipfel}}, \bibinfo {author} {\bibfnamefont {J.~D.}\
  \bibnamefont {Ziegler}}, \bibinfo {author} {\bibfnamefont {M.}~\bibnamefont
  {Selig}}, \bibinfo {author} {\bibfnamefont {T.}~\bibnamefont {Taniguchi}},
  \bibinfo {author} {\bibfnamefont {K.}~\bibnamefont {Watanabe}}, \bibinfo
  {author} {\bibfnamefont {A.}~\bibnamefont {Knorr}}, \bibinfo {author}
  {\bibfnamefont {A.}~\bibnamefont {Chernikov}},\ and\ \bibinfo {author}
  {\bibfnamefont {E.}~\bibnamefont {Malic}},\ }\bibfield  {title} {\bibinfo
  {title} {Temporal evolution of low-temperature phonon sidebands in transition
  metal dichalcogenides},\ }\href
  {https://doi.org/10.1021/acsphotonics.0c00866} {\bibfield  {journal}
  {\bibinfo  {journal} {ACS Photonics}\ }\textbf {\bibinfo {volume} {7}},\
  \bibinfo {pages} {2756} (\bibinfo {year} {2020}{\natexlab{b}})}\BibitemShut
  {NoStop}%
\bibitem [{\citenamefont {Rosati}\ \emph
  {et~al.}(2021{\natexlab{c}})\citenamefont {Rosati}, \citenamefont {Wagner},
  \citenamefont {Brem}, \citenamefont {Perea-Caus\'in}, \citenamefont {Ziegler},
  \citenamefont {Zipfel}, \citenamefont {Taniguchi}, \citenamefont {Watanabe},
  \citenamefont {Chernikov},\ and\ \citenamefont {Malic}}]{Rosati21c}%
  \BibitemOpen
  \bibfield  {author} {\bibinfo {author} {\bibfnamefont {R.}~\bibnamefont
  {Rosati}}, \bibinfo {author} {\bibfnamefont {K.}~\bibnamefont {Wagner}},
  \bibinfo {author} {\bibfnamefont {S.}~\bibnamefont {Brem}}, \bibinfo {author}
  {\bibfnamefont {R.}~\bibnamefont {Perea-Caus\'in}}, \bibinfo {author}
  {\bibfnamefont {J.~D.}\ \bibnamefont {Ziegler}}, \bibinfo {author}
  {\bibfnamefont {J.}~\bibnamefont {Zipfel}}, \bibinfo {author} {\bibfnamefont
  {T.}~\bibnamefont {Taniguchi}}, \bibinfo {author} {\bibfnamefont
  {K.}~\bibnamefont {Watanabe}}, \bibinfo {author} {\bibfnamefont
  {A.}~\bibnamefont {Chernikov}},\ and\ \bibinfo {author} {\bibfnamefont
  {E.}~\bibnamefont {Malic}},\ }\bibfield  {title} {\bibinfo {title}
  {Non-equilibrium diffusion of dark excitons in atomically thin
  semiconductors},\ }\href {https://doi.org/10.1039/D1NR06230A} {\bibfield
  {journal} {\bibinfo  {journal} {Nanoscale}\ }\textbf {\bibinfo {volume}
  {13}},\ \bibinfo {pages} {19966} (\bibinfo {year}
  {2021}{\natexlab{c}})}\BibitemShut {NoStop}%
\bibitem [{\citenamefont {Jin}\ \emph {et~al.}(2014)\citenamefont {Jin},
  \citenamefont {Li}, \citenamefont {Mullen},\ and\ \citenamefont
  {Kim}}]{Jin14}%
  \BibitemOpen
  \bibfield  {author} {\bibinfo {author} {\bibfnamefont {Z.}~\bibnamefont
  {Jin}}, \bibinfo {author} {\bibfnamefont {X.}~\bibnamefont {Li}}, \bibinfo
  {author} {\bibfnamefont {J.~T.}\ \bibnamefont {Mullen}},\ and\ \bibinfo
  {author} {\bibfnamefont {K.~W.}\ \bibnamefont {Kim}},\ }\bibfield  {title}
  {\bibinfo {title} {Intrinsic transport properties of electrons and holes in
  monolayer transition-metal dichalcogenides},\ }\href
  {https://doi.org/10.1103/PhysRevB.90.045422} {\bibfield  {journal} {\bibinfo
  {journal} {Phys. Rev. B}\ }\textbf {\bibinfo {volume} {90}},\ \bibinfo
  {pages} {045422} (\bibinfo {year} {2014})}\BibitemShut {NoStop}%
\bibitem [{\citenamefont {Ziegler}\ \emph {et~al.}(2020)\citenamefont
  {Ziegler}, \citenamefont {Zipfel}, \citenamefont {Meisinger}, \citenamefont
  {Menahem}, \citenamefont {Zhu}, \citenamefont {Taniguchi}, \citenamefont
  {Watanabe}, \citenamefont {Yaffe}, \citenamefont {Egger},\ and\ \citenamefont
  {Chernikov}}]{Ziegler20}%
  \BibitemOpen
  \bibfield  {author} {\bibinfo {author} {\bibfnamefont {J.~D.}\ \bibnamefont
  {Ziegler}}, \bibinfo {author} {\bibfnamefont {J.}~\bibnamefont {Zipfel}},
  \bibinfo {author} {\bibfnamefont {B.}~\bibnamefont {Meisinger}}, \bibinfo
  {author} {\bibfnamefont {M.}~\bibnamefont {Menahem}}, \bibinfo {author}
  {\bibfnamefont {X.}~\bibnamefont {Zhu}}, \bibinfo {author} {\bibfnamefont
  {T.}~\bibnamefont {Taniguchi}}, \bibinfo {author} {\bibfnamefont
  {K.}~\bibnamefont {Watanabe}}, \bibinfo {author} {\bibfnamefont
  {O.}~\bibnamefont {Yaffe}}, \bibinfo {author} {\bibfnamefont {D.~A.}\
  \bibnamefont {Egger}},\ and\ \bibinfo {author} {\bibfnamefont
  {A.}~\bibnamefont {Chernikov}},\ }\bibfield  {title} {\bibinfo {title} {Fast
  and anomalous exciton diffusion in two-dimensional hybrid perovskites},\
  }\href {https://doi.org/10.1021/acs.nanolett.0c02472} {\bibfield  {journal}
  {\bibinfo  {journal} {Nano Lett.}\ }\textbf {\bibinfo {volume} {20}},\
  \bibinfo {pages} {6674} (\bibinfo {year} {2020})}\BibitemShut {NoStop}%
\bibitem [{\citenamefont {Rosati}\ \emph {et~al.}(2026)\citenamefont {Rosati},
  \citenamefont {K{\"o}nig}, \citenamefont {Terres}, \citenamefont {Thompson},
  \citenamefont {Baranowski}, \citenamefont {P{\l}ochocka}, \citenamefont
  {Chernikov},\ and\ \citenamefont {Malic}}]{Rosati26}%
  \BibitemOpen
  \bibfield  {author} {\bibinfo {author} {\bibfnamefont {R.}~\bibnamefont
  {Rosati}}, \bibinfo {author} {\bibfnamefont {J.~K.}\ \bibnamefont
  {K{\"o}nig}}, \bibinfo {author} {\bibfnamefont {S.}~\bibnamefont {Terres}},
  \bibinfo {author} {\bibfnamefont {J.~J.~P.}\ \bibnamefont {Thompson}},
  \bibinfo {author} {\bibfnamefont {M.}~\bibnamefont {Baranowski}}, \bibinfo
  {author} {\bibfnamefont {P.}~\bibnamefont {P{\l}ochocka}}, \bibinfo {author}
  {\bibfnamefont {A.}~\bibnamefont {Chernikov}},\ and\ \bibinfo {author}
  {\bibfnamefont {E.}~\bibnamefont {Malic}},\ }\bibfield  {title} {\bibinfo
  {title} {Microscopic insights into magneto-optics and magneto-transport in
  2{D} perovskites},\ }\href {https://doi.org/10.1021/acs.nanolett.6c00182}
  {\bibfield  {journal} {\bibinfo  {journal} {Nano Lett.}\ }\textbf {\bibinfo
  {volume} {26}},\ \bibinfo {pages} {5468} (\bibinfo {year}
  {2026})}\BibitemShut {NoStop}%
\bibitem [{\citenamefont {Saris}\ \emph {et~al.}(2026)\citenamefont {Saris},
  \citenamefont {Rosati}, \citenamefont {Bruevich}, \citenamefont {Sheehan},
  \citenamefont {Roman}, \citenamefont {Podzorov}, \citenamefont {Malic},\ and\
  \citenamefont {Tisdale}}]{Saris26}%
  \BibitemOpen
  \bibfield  {author} {\bibinfo {author} {\bibfnamefont {S.}~\bibnamefont
  {Saris}}, \bibinfo {author} {\bibfnamefont {R.}~\bibnamefont {Rosati}},
  \bibinfo {author} {\bibfnamefont {V.}~\bibnamefont {Bruevich}}, \bibinfo
  {author} {\bibfnamefont {T.~J.}\ \bibnamefont {Sheehan}}, \bibinfo {author}
  {\bibfnamefont {M.}~\bibnamefont {Roman}}, \bibinfo {author} {\bibfnamefont
  {V.}~\bibnamefont {Podzorov}}, \bibinfo {author} {\bibfnamefont
  {E.}~\bibnamefont {Malic}},\ and\ \bibinfo {author} {\bibfnamefont {W.~A.}\
  \bibnamefont {Tisdale}},\ }\href {https://arxiv.org/abs/2606.02460} {\bibinfo
  {title} {Nonequilibrium transport in epitaxial {CsPbBr}$_3$ single crystals}}
  (\bibinfo {year} {2026}),\ \Eprint {https://arxiv.org/abs/2606.02460}
  {arXiv:2606.02460 [cond-mat.mes-hall]} \BibitemShut {NoStop}%
\bibitem [{\citenamefont {Korm{\'{a}}nyos}\ \emph {et~al.}(2015)\citenamefont
  {Korm{\'{a}}nyos}, \citenamefont {Burkard}, \citenamefont {Gmitra},
  \citenamefont {Fabian}, \citenamefont {Z{\'{o}}lyomi}, \citenamefont
  {Drummond},\ and\ \citenamefont {Fal'ko}}]{Kormanyos15}%
  \BibitemOpen
  \bibfield  {author} {\bibinfo {author} {\bibfnamefont {A.}~\bibnamefont
  {Korm{\'{a}}nyos}}, \bibinfo {author} {\bibfnamefont {G.}~\bibnamefont
  {Burkard}}, \bibinfo {author} {\bibfnamefont {M.}~\bibnamefont {Gmitra}},
  \bibinfo {author} {\bibfnamefont {J.}~\bibnamefont {Fabian}}, \bibinfo
  {author} {\bibfnamefont {V.}~\bibnamefont {Z{\'{o}}lyomi}}, \bibinfo {author}
  {\bibfnamefont {N.~D.}\ \bibnamefont {Drummond}},\ and\ \bibinfo {author}
  {\bibfnamefont {V.}~\bibnamefont {Fal'ko}},\ }\bibfield  {title} {\bibinfo
  {title} {k$\cdot$p theory for two-dimensional transition metal dichalcogenide
  semiconductors},\ }\href {https://doi.org/10.1088/2053-1583/2/2/022001}
  {\bibfield  {journal} {\bibinfo  {journal} {2D Mater.}\ }\textbf {\bibinfo
  {volume} {2}},\ \bibinfo {pages} {022001} (\bibinfo {year}
  {2015})}\BibitemShut {NoStop}%
\bibitem [{\citenamefont {Haug}\ and\ \citenamefont {Koch}(2009)}]{Haug09}%
  \BibitemOpen
  \bibfield  {author} {\bibinfo {author} {\bibfnamefont {H.}~\bibnamefont
  {Haug}}\ and\ \bibinfo {author} {\bibfnamefont {S.~W.}\ \bibnamefont
  {Koch}},\ }\href@noop {} {\emph {\bibinfo {title} {Quantum Theory of the
  Optical and Electronic Properties of Semiconductors: Fifth Edition}}}\
  (\bibinfo  {publisher} {World Scientific Publishing Company},\ \bibinfo
  {year} {2009})\BibitemShut {NoStop}%
\bibitem [{\citenamefont {Selig}\ \emph {et~al.}(2016)\citenamefont {Selig},
  \citenamefont {Bergh{\"a}user}, \citenamefont {Raja}, \citenamefont {Nagler},
  \citenamefont {Sch{\"u}ller}, \citenamefont {Heinz}, \citenamefont {Korn},
  \citenamefont {Chernikov}, \citenamefont {Malic},\ and\ \citenamefont
  {Knorr}}]{Selig16}%
  \BibitemOpen
  \bibfield  {author} {\bibinfo {author} {\bibfnamefont {M.}~\bibnamefont
  {Selig}}, \bibinfo {author} {\bibfnamefont {G.}~\bibnamefont
  {Bergh{\"a}user}}, \bibinfo {author} {\bibfnamefont {A.}~\bibnamefont
  {Raja}}, \bibinfo {author} {\bibfnamefont {P.}~\bibnamefont {Nagler}},
  \bibinfo {author} {\bibfnamefont {C.}~\bibnamefont {Sch{\"u}ller}}, \bibinfo
  {author} {\bibfnamefont {T.~F.}\ \bibnamefont {Heinz}}, \bibinfo {author}
  {\bibfnamefont {T.}~\bibnamefont {Korn}}, \bibinfo {author} {\bibfnamefont
  {A.}~\bibnamefont {Chernikov}}, \bibinfo {author} {\bibfnamefont
  {E.}~\bibnamefont {Malic}},\ and\ \bibinfo {author} {\bibfnamefont
  {A.}~\bibnamefont {Knorr}},\ }\bibfield  {title} {\bibinfo {title} {Excitonic
  linewidth and coherence lifetime in monolayer transition metal
  dichalcogenides},\ }\href {https://doi.org/10.1038/ncomms13279} {\bibfield
  {journal} {\bibinfo  {journal} {Nat. Commun.}\ }\textbf {\bibinfo {volume}
  {7}},\ \bibinfo {pages} {13279} (\bibinfo {year} {2016})}\BibitemShut
  {NoStop}%
\bibitem [{\citenamefont {Selig}\ \emph {et~al.}(2018)\citenamefont {Selig},
  \citenamefont {Bergh\"auser}, \citenamefont {Richter}, \citenamefont
  {Bratschitsch}, \citenamefont {Knorr},\ and\ \citenamefont
  {Malic}}]{Selig18}%
  \BibitemOpen
  \bibfield  {author} {\bibinfo {author} {\bibfnamefont {M.}~\bibnamefont
  {Selig}}, \bibinfo {author} {\bibfnamefont {G.}~\bibnamefont {Bergh\"auser}},
  \bibinfo {author} {\bibfnamefont {M.}~\bibnamefont {Richter}}, \bibinfo
  {author} {\bibfnamefont {R.}~\bibnamefont {Bratschitsch}}, \bibinfo {author}
  {\bibfnamefont {A.}~\bibnamefont {Knorr}},\ and\ \bibinfo {author}
  {\bibfnamefont {E.}~\bibnamefont {Malic}},\ }\bibfield  {title} {\bibinfo
  {title} {Dark and bright exciton formation, thermalization, and
  photoluminescence in monolayer transition metal dichalcogenides},\ }\href
  {https://doi.org/10.1088/2053-1583/aabea3} {\bibfield  {journal} {\bibinfo
  {journal} {2D Mater.}\ }\textbf {\bibinfo {volume} {5}},\ \bibinfo {pages}
  {035017} (\bibinfo {year} {2018})}\BibitemShut {NoStop}%
\bibitem [{\citenamefont {Brem}\ \emph {et~al.}(2018)\citenamefont {Brem},
  \citenamefont {Selig}, \citenamefont {Berghaeuser},\ and\ \citenamefont
  {Malic}}]{Brem18}%
  \BibitemOpen
  \bibfield  {author} {\bibinfo {author} {\bibfnamefont {S.}~\bibnamefont
  {Brem}}, \bibinfo {author} {\bibfnamefont {M.}~\bibnamefont {Selig}},
  \bibinfo {author} {\bibfnamefont {G.}~\bibnamefont {Berghaeuser}},\ and\
  \bibinfo {author} {\bibfnamefont {E.}~\bibnamefont {Malic}},\ }\bibfield
  {title} {\bibinfo {title} {Exciton relaxation cascade in two-dimensional
  transition metal dichalcogenides},\ }\href
  {https://doi.org/10.1038/s41598-018-25906-7} {\bibfield  {journal} {\bibinfo
  {journal} {Sci. Rep.}\ }\textbf {\bibinfo {volume} {8}},\ \bibinfo {pages}
  {8238} (\bibinfo {year} {2018})}\BibitemShut {NoStop}%
\bibitem [{\citenamefont {Brem}\ \emph {et~al.}(2019)\citenamefont {Brem},
  \citenamefont {Zipfel}, \citenamefont {Selig}, \citenamefont {Raja},
  \citenamefont {Waldecker}, \citenamefont {Ziegler}, \citenamefont
  {Taniguchi}, \citenamefont {Watanabe}, \citenamefont {Chernikov},\ and\
  \citenamefont {Malic}}]{Brem19b}%
  \BibitemOpen
  \bibfield  {author} {\bibinfo {author} {\bibfnamefont {S.}~\bibnamefont
  {Brem}}, \bibinfo {author} {\bibfnamefont {J.}~\bibnamefont {Zipfel}},
  \bibinfo {author} {\bibfnamefont {M.}~\bibnamefont {Selig}}, \bibinfo
  {author} {\bibfnamefont {A.}~\bibnamefont {Raja}}, \bibinfo {author}
  {\bibfnamefont {L.}~\bibnamefont {Waldecker}}, \bibinfo {author}
  {\bibfnamefont {J.~D.}\ \bibnamefont {Ziegler}}, \bibinfo {author}
  {\bibfnamefont {T.}~\bibnamefont {Taniguchi}}, \bibinfo {author}
  {\bibfnamefont {K.}~\bibnamefont {Watanabe}}, \bibinfo {author}
  {\bibfnamefont {A.}~\bibnamefont {Chernikov}},\ and\ \bibinfo {author}
  {\bibfnamefont {E.}~\bibnamefont {Malic}},\ }\bibfield  {title} {\bibinfo
  {title} {Intrinsic lifetime of higher excitonic states in tungsten diselenide
  monolayers},\ }\href {https://doi.org/10.1039/C9NR04211C} {\bibfield
  {journal} {\bibinfo  {journal} {Nanoscale}\ }\textbf {\bibinfo {volume}
  {11}},\ \bibinfo {pages} {12381} (\bibinfo {year} {2019})}\BibitemShut
  {NoStop}%
\bibitem [{\citenamefont {Malic}\ \emph {et~al.}(2018)\citenamefont {Malic},
  \citenamefont {Selig}, \citenamefont {Feierabend}, \citenamefont {Brem},
  \citenamefont {Christiansen}, \citenamefont {Wendler}, \citenamefont
  {Knorr},\ and\ \citenamefont {Bergh\"auser}}]{Malic18}%
  \BibitemOpen
  \bibfield  {author} {\bibinfo {author} {\bibfnamefont {E.}~\bibnamefont
  {Malic}}, \bibinfo {author} {\bibfnamefont {M.}~\bibnamefont {Selig}},
  \bibinfo {author} {\bibfnamefont {M.}~\bibnamefont {Feierabend}}, \bibinfo
  {author} {\bibfnamefont {S.}~\bibnamefont {Brem}}, \bibinfo {author}
  {\bibfnamefont {D.}~\bibnamefont {Christiansen}}, \bibinfo {author}
  {\bibfnamefont {F.}~\bibnamefont {Wendler}}, \bibinfo {author} {\bibfnamefont
  {A.}~\bibnamefont {Knorr}},\ and\ \bibinfo {author} {\bibfnamefont
  {G.}~\bibnamefont {Bergh\"auser}},\ }\bibfield  {title} {\bibinfo {title}
  {Dark excitons in transition metal dichalcogenides},\ }\href
  {https://doi.org/10.1103/PhysRevMaterials.2.014002} {\bibfield  {journal}
  {\bibinfo  {journal} {Phys. Rev. Mater.}\ }\textbf {\bibinfo {volume} {2}},\
  \bibinfo {pages} {014002} (\bibinfo {year} {2018})}\BibitemShut {NoStop}%
\bibitem [{\citenamefont {Rosati}\ \emph {et~al.}(2014)\citenamefont {Rosati},
  \citenamefont {Iotti}, \citenamefont {Dolcini},\ and\ \citenamefont
  {Rossi}}]{Rosati14}%
  \BibitemOpen
  \bibfield  {author} {\bibinfo {author} {\bibfnamefont {R.}~\bibnamefont
  {Rosati}}, \bibinfo {author} {\bibfnamefont {R.~C.}\ \bibnamefont {Iotti}},
  \bibinfo {author} {\bibfnamefont {F.}~\bibnamefont {Dolcini}},\ and\ \bibinfo
  {author} {\bibfnamefont {F.}~\bibnamefont {Rossi}},\ }\bibfield  {title}
  {\bibinfo {title} {Derivation of nonlinear single-particle equations via
  many-body lindblad superoperators: A density-matrix approach},\ }\href
  {https://doi.org/10.1103/PhysRevB.90.125140} {\bibfield  {journal} {\bibinfo
  {journal} {Phys. Rev. B}\ }\textbf {\bibinfo {volume} {90}},\ \bibinfo
  {pages} {125140} (\bibinfo {year} {2014})}\BibitemShut {NoStop}%
\bibitem [{\citenamefont {Kulig}\ \emph {et~al.}(2018)\citenamefont {Kulig},
  \citenamefont {Zipfel}, \citenamefont {Nagler}, \citenamefont {Blanter},
  \citenamefont {Sch\"uller}, \citenamefont {Korn}, \citenamefont {Paradiso},
  \citenamefont {Glazov},\ and\ \citenamefont {Chernikov}}]{Kulig18}%
  \BibitemOpen
  \bibfield  {author} {\bibinfo {author} {\bibfnamefont {M.}~\bibnamefont
  {Kulig}}, \bibinfo {author} {\bibfnamefont {J.}~\bibnamefont {Zipfel}},
  \bibinfo {author} {\bibfnamefont {P.}~\bibnamefont {Nagler}}, \bibinfo
  {author} {\bibfnamefont {S.}~\bibnamefont {Blanter}}, \bibinfo {author}
  {\bibfnamefont {C.}~\bibnamefont {Sch\"uller}}, \bibinfo {author}
  {\bibfnamefont {T.}~\bibnamefont {Korn}}, \bibinfo {author} {\bibfnamefont
  {N.}~\bibnamefont {Paradiso}}, \bibinfo {author} {\bibfnamefont {M.~M.}\
  \bibnamefont {Glazov}},\ and\ \bibinfo {author} {\bibfnamefont
  {A.}~\bibnamefont {Chernikov}},\ }\bibfield  {title} {\bibinfo {title}
  {Exciton diffusion and halo effects in monolayer semiconductors},\ }\href
  {https://doi.org/10.1103/PhysRevLett.120.207401} {\bibfield  {journal}
  {\bibinfo  {journal} {Phys. Rev. Lett.}\ }\textbf {\bibinfo {volume} {120}},\
  \bibinfo {pages} {207401} (\bibinfo {year} {2018})}\BibitemShut {NoStop}%
\bibitem [{\citenamefont {Perea-Caus{\'i}n}\ \emph {et~al.}(2019)\citenamefont
  {Perea-Caus{\'i}n}, \citenamefont {Brem}, \citenamefont {Rosati},
  \citenamefont {Jago}, \citenamefont {Kulig}, \citenamefont {Ziegler},
  \citenamefont {Zipfel}, \citenamefont {Chernikov},\ and\ \citenamefont
  {Malic}}]{Perea19}%
  \BibitemOpen
  \bibfield  {author} {\bibinfo {author} {\bibfnamefont {R.}~\bibnamefont
  {Perea-Caus{\'i}n}}, \bibinfo {author} {\bibfnamefont {S.}~\bibnamefont
  {Brem}}, \bibinfo {author} {\bibfnamefont {R.}~\bibnamefont {Rosati}},
  \bibinfo {author} {\bibfnamefont {R.}~\bibnamefont {Jago}}, \bibinfo {author}
  {\bibfnamefont {M.}~\bibnamefont {Kulig}}, \bibinfo {author} {\bibfnamefont
  {J.~D.}\ \bibnamefont {Ziegler}}, \bibinfo {author} {\bibfnamefont
  {J.}~\bibnamefont {Zipfel}}, \bibinfo {author} {\bibfnamefont
  {A.}~\bibnamefont {Chernikov}},\ and\ \bibinfo {author} {\bibfnamefont
  {E.}~\bibnamefont {Malic}},\ }\bibfield  {title} {\bibinfo {title} {Exciton
  propagation and halo formation in two-dimensional materials},\ }\href
  {https://doi.org/10.1021/acs.nanolett.9b02948} {\bibfield  {journal}
  {\bibinfo  {journal} {Nano Lett.}\ }\textbf {\bibinfo {volume} {19}},\
  \bibinfo {pages} {7317} (\bibinfo {year} {2019})}\BibitemShut {NoStop}%
\bibitem [{\citenamefont {Gao}\ \emph {et~al.}(2016)\citenamefont {Gao},
  \citenamefont {Gong}, \citenamefont {Titze}, \citenamefont {Almeida},
  \citenamefont {Ajayan},\ and\ \citenamefont {Li}}]{Gao16}%
  \BibitemOpen
  \bibfield  {author} {\bibinfo {author} {\bibfnamefont {F.}~\bibnamefont
  {Gao}}, \bibinfo {author} {\bibfnamefont {Y.}~\bibnamefont {Gong}}, \bibinfo
  {author} {\bibfnamefont {M.}~\bibnamefont {Titze}}, \bibinfo {author}
  {\bibfnamefont {R.}~\bibnamefont {Almeida}}, \bibinfo {author} {\bibfnamefont
  {P.~M.}\ \bibnamefont {Ajayan}},\ and\ \bibinfo {author} {\bibfnamefont
  {H.}~\bibnamefont {Li}},\ }\bibfield  {title} {\bibinfo {title} {Valley trion
  dynamics in monolayer {MoSe}$_{2}$},\ }\href
  {https://doi.org/10.1103/PhysRevB.94.245413} {\bibfield  {journal} {\bibinfo
  {journal} {Phys. Rev. B}\ }\textbf {\bibinfo {volume} {94}},\ \bibinfo
  {pages} {245413} (\bibinfo {year} {2016})}\BibitemShut {NoStop}%
\bibitem [{\citenamefont {Kato}\ and\ \citenamefont {Kaneko}(2016)}]{Kato16}%
  \BibitemOpen
  \bibfield  {author} {\bibinfo {author} {\bibfnamefont {T.}~\bibnamefont
  {Kato}}\ and\ \bibinfo {author} {\bibfnamefont {T.}~\bibnamefont {Kaneko}},\
  }\bibfield  {title} {\bibinfo {title} {Transport dynamics of neutral excitons
  and trions in monolayer {WS}$_2$},\ }\href
  {https://doi.org/10.1021/acsnano.6b05580} {\bibfield  {journal} {\bibinfo
  {journal} {ACS Nano}\ }\textbf {\bibinfo {volume} {10}},\ \bibinfo {pages}
  {9687} (\bibinfo {year} {2016})}\BibitemShut {NoStop}%
\bibitem [{\citenamefont {Cadiz}\ \emph {et~al.}(2018)\citenamefont {Cadiz},
  \citenamefont {Robert}, \citenamefont {Courtade}, \citenamefont {Manca},
  \citenamefont {Martinelli}, \citenamefont {Taniguchi}, \citenamefont
  {Watanabe}, \citenamefont {Amand}, \citenamefont {Rowe}, \citenamefont
  {Paget}, \citenamefont {Urbaszek},\ and\ \citenamefont {Marie}}]{Cadiz17}%
  \BibitemOpen
  \bibfield  {author} {\bibinfo {author} {\bibfnamefont {F.}~\bibnamefont
  {Cadiz}}, \bibinfo {author} {\bibfnamefont {C.}~\bibnamefont {Robert}},
  \bibinfo {author} {\bibfnamefont {E.}~\bibnamefont {Courtade}}, \bibinfo
  {author} {\bibfnamefont {M.}~\bibnamefont {Manca}}, \bibinfo {author}
  {\bibfnamefont {L.}~\bibnamefont {Martinelli}}, \bibinfo {author}
  {\bibfnamefont {T.}~\bibnamefont {Taniguchi}}, \bibinfo {author}
  {\bibfnamefont {K.}~\bibnamefont {Watanabe}}, \bibinfo {author}
  {\bibfnamefont {T.}~\bibnamefont {Amand}}, \bibinfo {author} {\bibfnamefont
  {A.~C.~H.}\ \bibnamefont {Rowe}}, \bibinfo {author} {\bibfnamefont
  {D.}~\bibnamefont {Paget}}, \bibinfo {author} {\bibfnamefont
  {B.}~\bibnamefont {Urbaszek}},\ and\ \bibinfo {author} {\bibfnamefont
  {X.}~\bibnamefont {Marie}},\ }\bibfield  {title} {\bibinfo {title} {Exciton
  diffusion in {WS}e$_2$ monolayers embedded in a van der {W}aals
  heterostructure},\ }\href {https://doi.org/10.1063/1.5026478} {\bibfield
  {journal} {\bibinfo  {journal} {Appl. Phys. Lett.}\ }\textbf {\bibinfo
  {volume} {112}},\ \bibinfo {pages} {152106} (\bibinfo {year} {2018})},\
  \Eprint {https://arxiv.org/abs/https://doi.org/10.1063/1.5026478}
  {https://doi.org/10.1063/1.5026478} \BibitemShut {NoStop}%
\bibitem [{\citenamefont {Titze}\ \emph {et~al.}(2018)\citenamefont {Titze},
  \citenamefont {Li}, \citenamefont {Zhang}, \citenamefont {Ajayan},\ and\
  \citenamefont {Li}}]{Titze18}%
  \BibitemOpen
  \bibfield  {author} {\bibinfo {author} {\bibfnamefont {M.}~\bibnamefont
  {Titze}}, \bibinfo {author} {\bibfnamefont {B.}~\bibnamefont {Li}}, \bibinfo
  {author} {\bibfnamefont {X.}~\bibnamefont {Zhang}}, \bibinfo {author}
  {\bibfnamefont {P.~M.}\ \bibnamefont {Ajayan}},\ and\ \bibinfo {author}
  {\bibfnamefont {H.}~\bibnamefont {Li}},\ }\bibfield  {title} {\bibinfo
  {title} {Intrinsic coherence time of trions in monolayer {MoSe}$_{2}$
  measured via two-dimensional coherent spectroscopy},\ }\href
  {https://doi.org/10.1103/PhysRevMaterials.2.054001} {\bibfield  {journal}
  {\bibinfo  {journal} {Phys. Rev. Mater.}\ }\textbf {\bibinfo {volume} {2}},\
  \bibinfo {pages} {054001} (\bibinfo {year} {2018})}\BibitemShut {NoStop}%
\bibitem [{\citenamefont {Hess}\ and\ \citenamefont {Kuhn}(1996)}]{Hess96}%
  \BibitemOpen
  \bibfield  {author} {\bibinfo {author} {\bibfnamefont {O.}~\bibnamefont
  {Hess}}\ and\ \bibinfo {author} {\bibfnamefont {T.}~\bibnamefont {Kuhn}},\
  }\bibfield  {title} {\bibinfo {title} {Maxwell-bloch equations for spatially
  inhomogeneous semiconductor lasers. i. theoretical formulation},\ }\href
  {https://doi.org/10.1103/PhysRevA.54.3347} {\bibfield  {journal} {\bibinfo
  {journal} {Phys. Rev. A}\ }\textbf {\bibinfo {volume} {54}},\ \bibinfo
  {pages} {3347} (\bibinfo {year} {1996})}\BibitemShut {NoStop}%
\bibitem [{\citenamefont {Dadgar}\ \emph {et~al.}(2018)\citenamefont {Dadgar},
  \citenamefont {Scullion}, \citenamefont {Kang}, \citenamefont {Esposito},
  \citenamefont {Yang}, \citenamefont {Herman}, \citenamefont {Pimenta},
  \citenamefont {Santos},\ and\ \citenamefont {Pasupathy}}]{Dadgar18}%
  \BibitemOpen
  \bibfield  {author} {\bibinfo {author} {\bibfnamefont {A.~M.}\ \bibnamefont
  {Dadgar}}, \bibinfo {author} {\bibfnamefont {D.}~\bibnamefont {Scullion}},
  \bibinfo {author} {\bibfnamefont {K.}~\bibnamefont {Kang}}, \bibinfo {author}
  {\bibfnamefont {D.}~\bibnamefont {Esposito}}, \bibinfo {author}
  {\bibfnamefont {E.~H.}\ \bibnamefont {Yang}}, \bibinfo {author}
  {\bibfnamefont {I.~P.}\ \bibnamefont {Herman}}, \bibinfo {author}
  {\bibfnamefont {M.~A.}\ \bibnamefont {Pimenta}}, \bibinfo {author}
  {\bibfnamefont {E.-J.~G.}\ \bibnamefont {Santos}},\ and\ \bibinfo {author}
  {\bibfnamefont {A.~N.}\ \bibnamefont {Pasupathy}},\ }\bibfield  {title}
  {\bibinfo {title} {Strain engineering and {R}aman spectroscopy of monolayer
  transition metal dichalcogenides},\ }\href
  {https://doi.org/10.1021/acs.chemmater.8b01672} {\bibfield  {journal}
  {\bibinfo  {journal} {Chem. Mater.}\ }\textbf {\bibinfo {volume} {30}},\
  \bibinfo {pages} {5148} (\bibinfo {year} {2018})}\BibitemShut {NoStop}%
\bibitem [{\citenamefont {Aslan}\ \emph {et~al.}(2018)\citenamefont {Aslan},
  \citenamefont {Deng},\ and\ \citenamefont {Heinz}}]{Aslan18}%
  \BibitemOpen
  \bibfield  {author} {\bibinfo {author} {\bibfnamefont {O.~B.}\ \bibnamefont
  {Aslan}}, \bibinfo {author} {\bibfnamefont {M.}~\bibnamefont {Deng}},\ and\
  \bibinfo {author} {\bibfnamefont {T.~F.}\ \bibnamefont {Heinz}},\ }\bibfield
  {title} {\bibinfo {title} {Strain tuning of excitons in monolayer
  {WS}e$_{2}$},\ }\href {https://doi.org/10.1103/PhysRevB.98.115308} {\bibfield
   {journal} {\bibinfo  {journal} {Phys. Rev. B}\ }\textbf {\bibinfo {volume}
  {98}},\ \bibinfo {pages} {115308} (\bibinfo {year} {2018})}\BibitemShut
  {NoStop}%
\bibitem [{\citenamefont {Brem}\ \emph {et~al.}(2020)\citenamefont {Brem},
  \citenamefont {Ekman}, \citenamefont {Christiansen}, \citenamefont {Katsch},
  \citenamefont {Selig}, \citenamefont {Robert}, \citenamefont {Marie},
  \citenamefont {Urbaszek}, \citenamefont {Knorr},\ and\ \citenamefont
  {Malic}}]{Brem20}%
  \BibitemOpen
  \bibfield  {author} {\bibinfo {author} {\bibfnamefont {S.}~\bibnamefont
  {Brem}}, \bibinfo {author} {\bibfnamefont {A.}~\bibnamefont {Ekman}},
  \bibinfo {author} {\bibfnamefont {D.}~\bibnamefont {Christiansen}}, \bibinfo
  {author} {\bibfnamefont {F.}~\bibnamefont {Katsch}}, \bibinfo {author}
  {\bibfnamefont {M.}~\bibnamefont {Selig}}, \bibinfo {author} {\bibfnamefont
  {C.}~\bibnamefont {Robert}}, \bibinfo {author} {\bibfnamefont
  {X.}~\bibnamefont {Marie}}, \bibinfo {author} {\bibfnamefont
  {B.}~\bibnamefont {Urbaszek}}, \bibinfo {author} {\bibfnamefont
  {A.}~\bibnamefont {Knorr}},\ and\ \bibinfo {author} {\bibfnamefont
  {E.}~\bibnamefont {Malic}},\ }\bibfield  {title} {\bibinfo {title}
  {Phonon-assisted photoluminescence from indirect excitons in monolayers of
  transition-metal dichalcogenides},\ }\href
  {https://doi.org/10.1021/acs.nanolett.0c00633} {\bibfield  {journal}
  {\bibinfo  {journal} {Nano Lett.}\ }\textbf {\bibinfo {volume} {20}},\
  \bibinfo {pages} {2849} (\bibinfo {year} {2020})}\BibitemShut {NoStop}%
\bibitem [{Note1()}]{Note1}%
  \BibitemOpen
  \bibinfo {note} {\label {SI}See Supplemental Material at [URL will be
  inserted by publisher] for additional details on the microscopic evaluation
  of photoluminescence and diffusion as well as for additional time-resolved
  spectra and diffusion at temperature $T=50\protect \,$K or with artificially
  stronger bright-to-KK$^\prime $ scattering.}\BibitemShut {Stop}%
\bibitem [{\citenamefont {Cordovilla~Leon}\ \emph {et~al.}(2019)\citenamefont
  {Cordovilla~Leon}, \citenamefont {Li}, \citenamefont {Jang},\ and\
  \citenamefont {Deotare}}]{Cordovilla19}%
  \BibitemOpen
  \bibfield  {author} {\bibinfo {author} {\bibfnamefont {D.~F.}\ \bibnamefont
  {Cordovilla~Leon}}, \bibinfo {author} {\bibfnamefont {Z.}~\bibnamefont {Li}},
  \bibinfo {author} {\bibfnamefont {S.~W.}\ \bibnamefont {Jang}},\ and\
  \bibinfo {author} {\bibfnamefont {P.~B.}\ \bibnamefont {Deotare}},\
  }\bibfield  {title} {\bibinfo {title} {Hot exciton transport in
  $\mathrm{WS}{\mathrm{e}}_{2}$ monolayers},\ }\href
  {https://doi.org/10.1103/PhysRevB.100.241401} {\bibfield  {journal} {\bibinfo
   {journal} {Phys. Rev. B}\ }\textbf {\bibinfo {volume} {100}},\ \bibinfo
  {pages} {241401} (\bibinfo {year} {2019})}\BibitemShut {NoStop}%
\bibitem [{\citenamefont {Ginsberg}\ and\ \citenamefont
  {Tisdale}(2020)}]{Ginsberg20}%
  \BibitemOpen
  \bibfield  {author} {\bibinfo {author} {\bibfnamefont {N.~S.}\ \bibnamefont
  {Ginsberg}}\ and\ \bibinfo {author} {\bibfnamefont {W.~A.}\ \bibnamefont
  {Tisdale}},\ }\bibfield  {title} {\bibinfo {title} {Spatially resolved
  photogenerated exciton and charge transport in emerging semiconductors},\
  }\href
  {https://doi.org/https://doi.org/10.1146/annurev-physchem-052516-050703}
  {\bibfield  {journal} {\bibinfo  {journal} {Annu. Rev. Phys. Chem.}\ }\textbf
  {\bibinfo {volume} {71}},\ \bibinfo {pages} {1} (\bibinfo {year}
  {2020})}\BibitemShut {NoStop}%
\bibitem [{\citenamefont {Yuan}\ \emph {et~al.}(2020)\citenamefont {Yuan},
  \citenamefont {Zheng}, \citenamefont {Kunstmann}, \citenamefont {Brumme},
  \citenamefont {Kuc}, \citenamefont {Ma}, \citenamefont {Deng}, \citenamefont
  {Blach}, \citenamefont {Pan},\ and\ \citenamefont {Huang}}]{Yuan20}%
  \BibitemOpen
  \bibfield  {author} {\bibinfo {author} {\bibfnamefont {L.}~\bibnamefont
  {Yuan}}, \bibinfo {author} {\bibfnamefont {B.}~\bibnamefont {Zheng}},
  \bibinfo {author} {\bibfnamefont {J.}~\bibnamefont {Kunstmann}}, \bibinfo
  {author} {\bibfnamefont {T.}~\bibnamefont {Brumme}}, \bibinfo {author}
  {\bibfnamefont {A.~B.}\ \bibnamefont {Kuc}}, \bibinfo {author} {\bibfnamefont
  {C.}~\bibnamefont {Ma}}, \bibinfo {author} {\bibfnamefont {S.}~\bibnamefont
  {Deng}}, \bibinfo {author} {\bibfnamefont {D.}~\bibnamefont {Blach}},
  \bibinfo {author} {\bibfnamefont {A.}~\bibnamefont {Pan}},\ and\ \bibinfo
  {author} {\bibfnamefont {L.}~\bibnamefont {Huang}},\ }\bibfield  {title}
  {\bibinfo {title} {Twist-angle-dependent interlayer exciton diffusion in
  {WS}$_2$--{WSe}$_2$ heterobilayers},\ }\href
  {https://doi.org/10.1038/s41563-020-0670-3} {\bibfield  {journal} {\bibinfo
  {journal} {Nat. Mater.}\ }\textbf {\bibinfo {volume} {19}},\ \bibinfo {pages}
  {617} (\bibinfo {year} {2020})}\BibitemShut {NoStop}%
\bibitem [{\citenamefont {Lo~Gerfo~Morganti}\ \emph {et~al.}(2025)\citenamefont
  {Lo~Gerfo~Morganti}, \citenamefont {Rosati}, \citenamefont
  {Brinatti~Vazquez}, \citenamefont {Varghese}, \citenamefont {Saleta~Reig},
  \citenamefont {Malic}, \citenamefont {van Hulst},\ and\ \citenamefont
  {Tielrooij}}]{LoGerfoMorganti25}%
  \BibitemOpen
  \bibfield  {author} {\bibinfo {author} {\bibfnamefont {G.}~\bibnamefont
  {Lo~Gerfo~Morganti}}, \bibinfo {author} {\bibfnamefont {R.}~\bibnamefont
  {Rosati}}, \bibinfo {author} {\bibfnamefont {G.~D.}\ \bibnamefont
  {Brinatti~Vazquez}}, \bibinfo {author} {\bibfnamefont {S.}~\bibnamefont
  {Varghese}}, \bibinfo {author} {\bibfnamefont {D.}~\bibnamefont
  {Saleta~Reig}}, \bibinfo {author} {\bibfnamefont {E.}~\bibnamefont {Malic}},
  \bibinfo {author} {\bibfnamefont {N.~F.}\ \bibnamefont {van Hulst}},\ and\
  \bibinfo {author} {\bibfnamefont {K.-J.}\ \bibnamefont {Tielrooij}},\
  }\bibfield  {title} {\bibinfo {title} {Transient ultrafast and negative
  diffusion of charge carriers in suspended {MoSe}$_2$ from multilayer to
  monolayer},\ }\href {https://doi.org/10.1038/s41467-025-60197-3} {\bibfield
  {journal} {\bibinfo  {journal} {Nat. Commun.}\ }\textbf {\bibinfo {volume}
  {16}},\ \bibinfo {pages} {5184} (\bibinfo {year} {2025})}\BibitemShut
  {NoStop}%
\bibitem [{\citenamefont {Beret}\ \emph {et~al.}(2022)\citenamefont {Beret},
  \citenamefont {Paradisanos}, \citenamefont {Lamsaadi}, \citenamefont {Gan},
  \citenamefont {Najafidehaghani}, \citenamefont {George}, \citenamefont
  {Lehnert}, \citenamefont {Biskupek}, \citenamefont {Kaiser}, \citenamefont
  {Shree}, \citenamefont {Estrada-Real}, \citenamefont {Lagarde}, \citenamefont
  {Marie}, \citenamefont {Renucci}, \citenamefont {Watanabe}, \citenamefont
  {Taniguchi}, \citenamefont {Weber}, \citenamefont {Paillard}, \citenamefont
  {Lombez}, \citenamefont {Poumirol}, \citenamefont {Turchanin},\ and\
  \citenamefont {Urbaszek}}]{Beret22}%
  \BibitemOpen
  \bibfield  {author} {\bibinfo {author} {\bibfnamefont {D.}~\bibnamefont
  {Beret}}, \bibinfo {author} {\bibfnamefont {I.}~\bibnamefont {Paradisanos}},
  \bibinfo {author} {\bibfnamefont {H.}~\bibnamefont {Lamsaadi}}, \bibinfo
  {author} {\bibfnamefont {Z.}~\bibnamefont {Gan}}, \bibinfo {author}
  {\bibfnamefont {E.}~\bibnamefont {Najafidehaghani}}, \bibinfo {author}
  {\bibfnamefont {A.}~\bibnamefont {George}}, \bibinfo {author} {\bibfnamefont
  {T.}~\bibnamefont {Lehnert}}, \bibinfo {author} {\bibfnamefont
  {J.}~\bibnamefont {Biskupek}}, \bibinfo {author} {\bibfnamefont
  {U.}~\bibnamefont {Kaiser}}, \bibinfo {author} {\bibfnamefont
  {S.}~\bibnamefont {Shree}}, \bibinfo {author} {\bibfnamefont
  {A.}~\bibnamefont {Estrada-Real}}, \bibinfo {author} {\bibfnamefont
  {D.}~\bibnamefont {Lagarde}}, \bibinfo {author} {\bibfnamefont
  {X.}~\bibnamefont {Marie}}, \bibinfo {author} {\bibfnamefont
  {P.}~\bibnamefont {Renucci}}, \bibinfo {author} {\bibfnamefont
  {K.}~\bibnamefont {Watanabe}}, \bibinfo {author} {\bibfnamefont
  {T.}~\bibnamefont {Taniguchi}}, \bibinfo {author} {\bibfnamefont
  {S.}~\bibnamefont {Weber}}, \bibinfo {author} {\bibfnamefont
  {V.}~\bibnamefont {Paillard}}, \bibinfo {author} {\bibfnamefont
  {L.}~\bibnamefont {Lombez}}, \bibinfo {author} {\bibfnamefont {J.-M.}\
  \bibnamefont {Poumirol}}, \bibinfo {author} {\bibfnamefont {A.}~\bibnamefont
  {Turchanin}},\ and\ \bibinfo {author} {\bibfnamefont {B.}~\bibnamefont
  {Urbaszek}},\ }\bibfield  {title} {\bibinfo {title} {Exciton spectroscopy and
  unidirectional transport in {MoSe}$_2$-{WSe}$_2$ lateral heterostructures
  encapsulated in hexagonal boron nitride},\ }\href
  {https://doi.org/10.1038/s41699-022-00354-0} {\bibfield  {journal} {\bibinfo
  {journal} {npj 2D Mater. Appl.}\ }\textbf {\bibinfo {volume} {6}},\ \bibinfo
  {pages} {84} (\bibinfo {year} {2022})}\BibitemShut {NoStop}%
\bibitem [{\citenamefont {Lamsaadi}\ \emph {et~al.}(2023)\citenamefont
  {Lamsaadi}, \citenamefont {Beret}, \citenamefont {Paradisanos}, \citenamefont
  {Renucci}, \citenamefont {Lagarde}, \citenamefont {Marie}, \citenamefont
  {Urbaszek}, \citenamefont {Gan}, \citenamefont {George}, \citenamefont
  {Watanabe}, \citenamefont {Taniguchi}, \citenamefont {Turchanin},
  \citenamefont {Lombez}, \citenamefont {Combe}, \citenamefont {Paillard},\
  and\ \citenamefont {Poumirol}}]{Lamsaadi2023}%
  \BibitemOpen
  \bibfield  {author} {\bibinfo {author} {\bibfnamefont {H.}~\bibnamefont
  {Lamsaadi}}, \bibinfo {author} {\bibfnamefont {D.}~\bibnamefont {Beret}},
  \bibinfo {author} {\bibfnamefont {I.}~\bibnamefont {Paradisanos}}, \bibinfo
  {author} {\bibfnamefont {P.}~\bibnamefont {Renucci}}, \bibinfo {author}
  {\bibfnamefont {D.}~\bibnamefont {Lagarde}}, \bibinfo {author} {\bibfnamefont
  {X.}~\bibnamefont {Marie}}, \bibinfo {author} {\bibfnamefont
  {B.}~\bibnamefont {Urbaszek}}, \bibinfo {author} {\bibfnamefont
  {Z.}~\bibnamefont {Gan}}, \bibinfo {author} {\bibfnamefont {A.}~\bibnamefont
  {George}}, \bibinfo {author} {\bibfnamefont {K.}~\bibnamefont {Watanabe}},
  \bibinfo {author} {\bibfnamefont {T.}~\bibnamefont {Taniguchi}}, \bibinfo
  {author} {\bibfnamefont {A.}~\bibnamefont {Turchanin}}, \bibinfo {author}
  {\bibfnamefont {L.}~\bibnamefont {Lombez}}, \bibinfo {author} {\bibfnamefont
  {N.}~\bibnamefont {Combe}}, \bibinfo {author} {\bibfnamefont
  {V.}~\bibnamefont {Paillard}},\ and\ \bibinfo {author} {\bibfnamefont
  {J.-M.}\ \bibnamefont {Poumirol}},\ }\bibfield  {title} {\bibinfo {title}
  {Kapitza-resistance-like exciton dynamics in atomically flat
  {MoSe}$_2${-WSe}$_2$ lateral heterojunction},\ }\href
  {https://doi.org/10.1038/s41467-023-41538-6} {\bibfield  {journal} {\bibinfo
  {journal} {Nat. Commun.}\ }\textbf {\bibinfo {volume} {14}},\ \bibinfo
  {pages} {5881} (\bibinfo {year} {2023})}\BibitemShut {NoStop}%
\bibitem [{\citenamefont {Rosati}\ \emph {et~al.}(2025)\citenamefont {Rosati},
  \citenamefont {Shradha}, \citenamefont {Picker}, \citenamefont {Turchanin},
  \citenamefont {Urbaszek},\ and\ \citenamefont {Malic}}]{Rosati25}%
  \BibitemOpen
  \bibfield  {author} {\bibinfo {author} {\bibfnamefont {R.}~\bibnamefont
  {Rosati}}, \bibinfo {author} {\bibfnamefont {S.}~\bibnamefont {Shradha}},
  \bibinfo {author} {\bibfnamefont {J.}~\bibnamefont {Picker}}, \bibinfo
  {author} {\bibfnamefont {A.}~\bibnamefont {Turchanin}}, \bibinfo {author}
  {\bibfnamefont {B.}~\bibnamefont {Urbaszek}},\ and\ \bibinfo {author}
  {\bibfnamefont {E.}~\bibnamefont {Malic}},\ }\bibfield  {title} {\bibinfo
  {title} {Impact of charge-transfer excitons on unidirectional exciton
  transport in lateral {TMD} heterostructures},\ }\href
  {https://doi.org/10.1021/acs.nanolett.5c02447} {\bibfield  {journal}
  {\bibinfo  {journal} {Nano Lett.}\ }\textbf {\bibinfo {volume} {25}},\
  \bibinfo {pages} {11319} (\bibinfo {year} {2025})}\BibitemShut {NoStop}%
\bibitem [{\citenamefont {Rieland}\ \emph {et~al.}(2024)\citenamefont
  {Rieland}, \citenamefont {Wagner}, \citenamefont {Bernhardt}, \citenamefont
  {Wang}, \citenamefont {Abdul-Aziz}, \citenamefont {Stein}, \citenamefont
  {Pogna}, \citenamefont {Dal~Conte}, \citenamefont {Cerullo}, \citenamefont
  {Hedayat},\ and\ \citenamefont {van Loosdrecht}}]{Rieland24}%
  \BibitemOpen
  \bibfield  {author} {\bibinfo {author} {\bibfnamefont {L.}~\bibnamefont
  {Rieland}}, \bibinfo {author} {\bibfnamefont {J.}~\bibnamefont {Wagner}},
  \bibinfo {author} {\bibfnamefont {R.}~\bibnamefont {Bernhardt}}, \bibinfo
  {author} {\bibfnamefont {T.}~\bibnamefont {Wang}}, \bibinfo {author}
  {\bibfnamefont {O.}~\bibnamefont {Abdul-Aziz}}, \bibinfo {author}
  {\bibfnamefont {P.}~\bibnamefont {Stein}}, \bibinfo {author} {\bibfnamefont
  {E.~A.~A.}\ \bibnamefont {Pogna}}, \bibinfo {author} {\bibfnamefont
  {S.}~\bibnamefont {Dal~Conte}}, \bibinfo {author} {\bibfnamefont
  {G.}~\bibnamefont {Cerullo}}, \bibinfo {author} {\bibfnamefont
  {H.}~\bibnamefont {Hedayat}},\ and\ \bibinfo {author} {\bibfnamefont
  {P.~H.~M.}\ \bibnamefont {van Loosdrecht}},\ }\bibfield  {title} {\bibinfo
  {title} {Ultrafast optical control of exciton diffusion in {WSe}$_2$/graphene
  heterostructures revealed by heterodyne transient grating spectroscopy},\
  }\href {https://doi.org/10.1021/acs.nanolett.4c01516} {\bibfield  {journal}
  {\bibinfo  {journal} {Nano Lett.}\ }\textbf {\bibinfo {volume} {24}},\
  \bibinfo {pages} {9824} (\bibinfo {year} {2024})}\BibitemShut {NoStop}%
\bibitem [{\citenamefont {Bi}\ \emph {et~al.}(2025)\citenamefont {Bi},
  \citenamefont {Bai}, \citenamefont {Shi}, \citenamefont {Sun}, \citenamefont
  {Wu}, \citenamefont {Zhang}, \citenamefont {Cui}, \citenamefont {Zhu},
  \citenamefont {Wang}, \citenamefont {Lin}, \citenamefont {Wang},
  \citenamefont {Ma}, \citenamefont {Tan}, \citenamefont {Meng}, \citenamefont
  {Xiong},\ and\ \citenamefont {Yang}}]{Bi25}%
  \BibitemOpen
  \bibfield  {author} {\bibinfo {author} {\bibfnamefont {Z.}~\bibnamefont
  {Bi}}, \bibinfo {author} {\bibfnamefont {Y.}~\bibnamefont {Bai}}, \bibinfo
  {author} {\bibfnamefont {Y.}~\bibnamefont {Shi}}, \bibinfo {author}
  {\bibfnamefont {T.}~\bibnamefont {Sun}}, \bibinfo {author} {\bibfnamefont
  {H.}~\bibnamefont {Wu}}, \bibinfo {author} {\bibfnamefont {H.}~\bibnamefont
  {Zhang}}, \bibinfo {author} {\bibfnamefont {Y.}~\bibnamefont {Cui}}, \bibinfo
  {author} {\bibfnamefont {D.}~\bibnamefont {Zhu}}, \bibinfo {author}
  {\bibfnamefont {Y.}~\bibnamefont {Wang}}, \bibinfo {author} {\bibfnamefont
  {M.-L.}\ \bibnamefont {Lin}}, \bibinfo {author} {\bibfnamefont
  {Y.}~\bibnamefont {Wang}}, \bibinfo {author} {\bibfnamefont {D.}~\bibnamefont
  {Ma}}, \bibinfo {author} {\bibfnamefont {P.-H.}\ \bibnamefont {Tan}},
  \bibinfo {author} {\bibfnamefont {S.}~\bibnamefont {Meng}}, \bibinfo {author}
  {\bibfnamefont {Q.}~\bibnamefont {Xiong}},\ and\ \bibinfo {author}
  {\bibfnamefont {L.}~\bibnamefont {Yang}},\ }\bibfield  {title} {\bibinfo
  {title} {Intrinsic exciton transport and recombination in single-crystal lead
  bromide perovskite},\ }\href {https://doi.org/10.1021/acsnano.5c03274}
  {\bibfield  {journal} {\bibinfo  {journal} {ACS Nano}\ }\textbf {\bibinfo
  {volume} {19}},\ \bibinfo {pages} {19989} (\bibinfo {year}
  {2025})}\BibitemShut {NoStop}%
\bibitem [{\citenamefont {Schmitt}\ \emph {et~al.}(2025)\citenamefont
  {Schmitt}, \citenamefont {Bange}, \citenamefont {Bennecke}, \citenamefont
  {Meneghini}, \citenamefont {AlMutairi}, \citenamefont {Merboldt},
  \citenamefont {P{\"o}hls}, \citenamefont {Watanabe}, \citenamefont
  {Taniguchi}, \citenamefont {Steil}, \citenamefont {Steil}, \citenamefont
  {Weitz}, \citenamefont {Hofmann}, \citenamefont {Brem}, \citenamefont
  {Jansen}, \citenamefont {Malic}, \citenamefont {Mathias},\ and\ \citenamefont
  {Reutzel}}]{Schmitt25}%
  \BibitemOpen
  \bibfield  {author} {\bibinfo {author} {\bibfnamefont {D.}~\bibnamefont
  {Schmitt}}, \bibinfo {author} {\bibfnamefont {J.~P.}\ \bibnamefont {Bange}},
  \bibinfo {author} {\bibfnamefont {W.}~\bibnamefont {Bennecke}}, \bibinfo
  {author} {\bibfnamefont {G.}~\bibnamefont {Meneghini}}, \bibinfo {author}
  {\bibfnamefont {A.}~\bibnamefont {AlMutairi}}, \bibinfo {author}
  {\bibfnamefont {M.}~\bibnamefont {Merboldt}}, \bibinfo {author}
  {\bibfnamefont {J.}~\bibnamefont {P{\"o}hls}}, \bibinfo {author}
  {\bibfnamefont {K.}~\bibnamefont {Watanabe}}, \bibinfo {author}
  {\bibfnamefont {T.}~\bibnamefont {Taniguchi}}, \bibinfo {author}
  {\bibfnamefont {S.}~\bibnamefont {Steil}}, \bibinfo {author} {\bibfnamefont
  {D.}~\bibnamefont {Steil}}, \bibinfo {author} {\bibfnamefont {R.~T.}\
  \bibnamefont {Weitz}}, \bibinfo {author} {\bibfnamefont {S.}~\bibnamefont
  {Hofmann}}, \bibinfo {author} {\bibfnamefont {S.}~\bibnamefont {Brem}},
  \bibinfo {author} {\bibfnamefont {G.~S.~M.}\ \bibnamefont {Jansen}}, \bibinfo
  {author} {\bibfnamefont {E.}~\bibnamefont {Malic}}, \bibinfo {author}
  {\bibfnamefont {S.}~\bibnamefont {Mathias}},\ and\ \bibinfo {author}
  {\bibfnamefont {M.}~\bibnamefont {Reutzel}},\ }\bibfield  {title} {\bibinfo
  {title} {Ultrafast nano-imaging of dark excitons},\ }\href
  {https://doi.org/10.1038/s41566-024-01568-y} {\bibfield  {journal} {\bibinfo
  {journal} {Nat. Photonics}\ }\textbf {\bibinfo {volume} {19}},\ \bibinfo
  {pages} {187} (\bibinfo {year} {2025})}\BibitemShut {NoStop}%
\bibitem [{\citenamefont {Mad{\'e}o}\ \emph {et~al.}(2020)\citenamefont
  {Mad{\'e}o}, \citenamefont {Man}, \citenamefont {Sahoo}, \citenamefont
  {Campbell}, \citenamefont {Pareek}, \citenamefont {Wong}, \citenamefont
  {Al-Mahboob}, \citenamefont {Chan}, \citenamefont {Karmakar}, \citenamefont
  {Mariserla}, \citenamefont {Li}, \citenamefont {Heinz}, \citenamefont {Cao},\
  and\ \citenamefont {Dani}}]{Madeo20}%
  \BibitemOpen
  \bibfield  {author} {\bibinfo {author} {\bibfnamefont {J.}~\bibnamefont
  {Mad{\'e}o}}, \bibinfo {author} {\bibfnamefont {M.~K.~L.}\ \bibnamefont
  {Man}}, \bibinfo {author} {\bibfnamefont {C.}~\bibnamefont {Sahoo}}, \bibinfo
  {author} {\bibfnamefont {M.}~\bibnamefont {Campbell}}, \bibinfo {author}
  {\bibfnamefont {V.}~\bibnamefont {Pareek}}, \bibinfo {author} {\bibfnamefont
  {E.~L.}\ \bibnamefont {Wong}}, \bibinfo {author} {\bibfnamefont
  {A.}~\bibnamefont {Al-Mahboob}}, \bibinfo {author} {\bibfnamefont {N.~S.}\
  \bibnamefont {Chan}}, \bibinfo {author} {\bibfnamefont {A.}~\bibnamefont
  {Karmakar}}, \bibinfo {author} {\bibfnamefont {B.~M.~K.}\ \bibnamefont
  {Mariserla}}, \bibinfo {author} {\bibfnamefont {X.}~\bibnamefont {Li}},
  \bibinfo {author} {\bibfnamefont {T.~F.}\ \bibnamefont {Heinz}}, \bibinfo
  {author} {\bibfnamefont {T.}~\bibnamefont {Cao}},\ and\ \bibinfo {author}
  {\bibfnamefont {K.~M.}\ \bibnamefont {Dani}},\ }\bibfield  {title} {\bibinfo
  {title} {Directly visualizing the momentum-forbidden dark excitons and their
  dynamics in atomically thin semiconductors},\ }\href
  {https://doi.org/10.1126/science.aba1029} {\bibfield  {journal} {\bibinfo
  {journal} {Science}\ }\textbf {\bibinfo {volume} {370}},\ \bibinfo {pages}
  {1199} (\bibinfo {year} {2020})}\BibitemShut {NoStop}%
\bibitem [{\citenamefont {Schmitt}\ \emph {et~al.}(2022)\citenamefont
  {Schmitt}, \citenamefont {Bange}, \citenamefont {Bennecke}, \citenamefont
  {AlMutairi}, \citenamefont {Meneghini}, \citenamefont {Watanabe},
  \citenamefont {Taniguchi}, \citenamefont {Steil}, \citenamefont {Luke},
  \citenamefont {Weitz}, \citenamefont {Steil}, \citenamefont {Jansen},
  \citenamefont {Brem}, \citenamefont {Malic}, \citenamefont {Hofmann},
  \citenamefont {Reutzel},\ and\ \citenamefont {Mathias}}]{Schmitt22}%
  \BibitemOpen
  \bibfield  {author} {\bibinfo {author} {\bibfnamefont {D.}~\bibnamefont
  {Schmitt}}, \bibinfo {author} {\bibfnamefont {J.~P.}\ \bibnamefont {Bange}},
  \bibinfo {author} {\bibfnamefont {W.}~\bibnamefont {Bennecke}}, \bibinfo
  {author} {\bibfnamefont {A.}~\bibnamefont {AlMutairi}}, \bibinfo {author}
  {\bibfnamefont {G.}~\bibnamefont {Meneghini}}, \bibinfo {author}
  {\bibfnamefont {K.}~\bibnamefont {Watanabe}}, \bibinfo {author}
  {\bibfnamefont {T.}~\bibnamefont {Taniguchi}}, \bibinfo {author}
  {\bibfnamefont {D.}~\bibnamefont {Steil}}, \bibinfo {author} {\bibfnamefont
  {D.~R.}\ \bibnamefont {Luke}}, \bibinfo {author} {\bibfnamefont {R.~T.}\
  \bibnamefont {Weitz}}, \bibinfo {author} {\bibfnamefont {S.}~\bibnamefont
  {Steil}}, \bibinfo {author} {\bibfnamefont {G.~S.~M.}\ \bibnamefont
  {Jansen}}, \bibinfo {author} {\bibfnamefont {S.}~\bibnamefont {Brem}},
  \bibinfo {author} {\bibfnamefont {E.}~\bibnamefont {Malic}}, \bibinfo
  {author} {\bibfnamefont {S.}~\bibnamefont {Hofmann}}, \bibinfo {author}
  {\bibfnamefont {M.}~\bibnamefont {Reutzel}},\ and\ \bibinfo {author}
  {\bibfnamefont {S.}~\bibnamefont {Mathias}},\ }\bibfield  {title} {\bibinfo
  {title} {Formation of moir{\'e} interlayer excitons in space and time},\
  }\href {https://doi.org/10.1038/s41586-022-04977-7} {\bibfield  {journal}
  {\bibinfo  {journal} {Nature}\ }\textbf {\bibinfo {volume} {608}},\ \bibinfo
  {pages} {499} (\bibinfo {year} {2022})}\BibitemShut {NoStop}%
\bibitem [{\citenamefont {Aeschlimann}\ \emph {et~al.}(2020)\citenamefont
  {Aeschlimann}, \citenamefont {Rossi}, \citenamefont {Ch\'avez-Cervantes},
  \citenamefont {Krause}, \citenamefont {Arnoldi}, \citenamefont
  {Stadtm\"uller}, \citenamefont {Aeschlimann}, \citenamefont {Forti},
  \citenamefont {Fabbri}, \citenamefont {Coletti},\ and\ \citenamefont
  {Gierz}}]{Aeschlimann20}%
  \BibitemOpen
  \bibfield  {author} {\bibinfo {author} {\bibfnamefont {S.}~\bibnamefont
  {Aeschlimann}}, \bibinfo {author} {\bibfnamefont {A.}~\bibnamefont {Rossi}},
  \bibinfo {author} {\bibfnamefont {M.}~\bibnamefont {Ch\'avez-Cervantes}},
  \bibinfo {author} {\bibfnamefont {R.}~\bibnamefont {Krause}}, \bibinfo
  {author} {\bibfnamefont {B.}~\bibnamefont {Arnoldi}}, \bibinfo {author}
  {\bibfnamefont {B.}~\bibnamefont {Stadtm\"uller}}, \bibinfo {author}
  {\bibfnamefont {M.}~\bibnamefont {Aeschlimann}}, \bibinfo {author}
  {\bibfnamefont {S.}~\bibnamefont {Forti}}, \bibinfo {author} {\bibfnamefont
  {F.}~\bibnamefont {Fabbri}}, \bibinfo {author} {\bibfnamefont
  {C.}~\bibnamefont {Coletti}},\ and\ \bibinfo {author} {\bibfnamefont
  {I.}~\bibnamefont {Gierz}},\ }\bibfield  {title} {\bibinfo {title} {Direct
  evidence for efficient ultrafast charge separation in epitaxial
  {WS}$_2$/graphene heterostructures},\ }\href
  {https://doi.org/10.1126/sciadv.aay0761} {\bibfield  {journal} {\bibinfo
  {journal} {Sci. Adv.}\ }\textbf {\bibinfo {volume} {6}},\ \bibinfo {pages}
  {eaay0761} (\bibinfo {year} {2020})}\BibitemShut {NoStop}%
\bibitem [{\citenamefont {Dong}\ \emph {et~al.}(2021)\citenamefont {Dong},
  \citenamefont {Puppin}, \citenamefont {Pincelli}, \citenamefont {Beaulieu},
  \citenamefont {Christiansen}, \citenamefont {H\"ubener}, \citenamefont
  {Nicholson}, \citenamefont {Xian}, \citenamefont {Dendzik}, \citenamefont
  {Deng}, \citenamefont {Windsor}, \citenamefont {Selig}, \citenamefont
  {Malic}, \citenamefont {Rubio}, \citenamefont {Knorr}, \citenamefont {Wolf},
  \citenamefont {Rettig},\ and\ \citenamefont {Ernstorfer}}]{Dong21}%
  \BibitemOpen
  \bibfield  {author} {\bibinfo {author} {\bibfnamefont {S.}~\bibnamefont
  {Dong}}, \bibinfo {author} {\bibfnamefont {M.}~\bibnamefont {Puppin}},
  \bibinfo {author} {\bibfnamefont {T.}~\bibnamefont {Pincelli}}, \bibinfo
  {author} {\bibfnamefont {S.}~\bibnamefont {Beaulieu}}, \bibinfo {author}
  {\bibfnamefont {D.}~\bibnamefont {Christiansen}}, \bibinfo {author}
  {\bibfnamefont {H.}~\bibnamefont {H\"ubener}}, \bibinfo {author}
  {\bibfnamefont {C.~W.}\ \bibnamefont {Nicholson}}, \bibinfo {author}
  {\bibfnamefont {R.~P.}\ \bibnamefont {Xian}}, \bibinfo {author}
  {\bibfnamefont {M.}~\bibnamefont {Dendzik}}, \bibinfo {author} {\bibfnamefont
  {Y.}~\bibnamefont {Deng}}, \bibinfo {author} {\bibfnamefont {Y.~W.}\
  \bibnamefont {Windsor}}, \bibinfo {author} {\bibfnamefont {M.}~\bibnamefont
  {Selig}}, \bibinfo {author} {\bibfnamefont {E.}~\bibnamefont {Malic}},
  \bibinfo {author} {\bibfnamefont {A.}~\bibnamefont {Rubio}}, \bibinfo
  {author} {\bibfnamefont {A.}~\bibnamefont {Knorr}}, \bibinfo {author}
  {\bibfnamefont {M.}~\bibnamefont {Wolf}}, \bibinfo {author} {\bibfnamefont
  {L.}~\bibnamefont {Rettig}},\ and\ \bibinfo {author} {\bibfnamefont
  {R.}~\bibnamefont {Ernstorfer}},\ }\bibfield  {title} {\bibinfo {title}
  {Direct measurement of key exciton properties: Energy, dynamics, and spatial
  distribution of the wave function},\ }\href
  {https://doi.org/https://doi.org/10.1002/ntls.10010} {\bibfield  {journal}
  {\bibinfo  {journal} {Nat. Sci.}\ }\textbf {\bibinfo {volume} {1}},\ \bibinfo
  {pages} {e10010} (\bibinfo {year} {2021})}\BibitemShut {NoStop}%
\bibitem [{\citenamefont {Bange}\ \emph {et~al.}(2024)\citenamefont {Bange},
  \citenamefont {Schmitt}, \citenamefont {Bennecke}, \citenamefont {Meneghini},
  \citenamefont {AlMutairi}, \citenamefont {Watanabe}, \citenamefont
  {Taniguchi}, \citenamefont {Steil}, \citenamefont {Steil}, \citenamefont
  {Weitz}, \citenamefont {Jansen}, \citenamefont {Hofmann}, \citenamefont
  {Brem}, \citenamefont {Malic}, \citenamefont {Reutzel},\ and\ \citenamefont
  {Mathias}}]{Bange24}%
  \BibitemOpen
  \bibfield  {author} {\bibinfo {author} {\bibfnamefont {J.~P.}\ \bibnamefont
  {Bange}}, \bibinfo {author} {\bibfnamefont {D.}~\bibnamefont {Schmitt}},
  \bibinfo {author} {\bibfnamefont {W.}~\bibnamefont {Bennecke}}, \bibinfo
  {author} {\bibfnamefont {G.}~\bibnamefont {Meneghini}}, \bibinfo {author}
  {\bibfnamefont {A.}~\bibnamefont {AlMutairi}}, \bibinfo {author}
  {\bibfnamefont {K.}~\bibnamefont {Watanabe}}, \bibinfo {author}
  {\bibfnamefont {T.}~\bibnamefont {Taniguchi}}, \bibinfo {author}
  {\bibfnamefont {D.}~\bibnamefont {Steil}}, \bibinfo {author} {\bibfnamefont
  {S.}~\bibnamefont {Steil}}, \bibinfo {author} {\bibfnamefont {R.~T.}\
  \bibnamefont {Weitz}}, \bibinfo {author} {\bibfnamefont {G.~S.~M.}\
  \bibnamefont {Jansen}}, \bibinfo {author} {\bibfnamefont {S.}~\bibnamefont
  {Hofmann}}, \bibinfo {author} {\bibfnamefont {S.}~\bibnamefont {Brem}},
  \bibinfo {author} {\bibfnamefont {E.}~\bibnamefont {Malic}}, \bibinfo
  {author} {\bibfnamefont {M.}~\bibnamefont {Reutzel}},\ and\ \bibinfo {author}
  {\bibfnamefont {S.}~\bibnamefont {Mathias}},\ }\bibfield  {title} {\bibinfo
  {title} {Probing electron-hole {C}oulomb correlations in the exciton
  landscape of a twisted semiconductor heterostructure},\ }\href
  {https://doi.org/10.1126/sciadv.adi1323} {\bibfield  {journal} {\bibinfo
  {journal} {Sci. Adv.}\ }\textbf {\bibinfo {volume} {10}},\ \bibinfo {pages}
  {eadi1323} (\bibinfo {year} {2024})}\BibitemShut {NoStop}%
\bibitem [{\citenamefont {Boschini}\ \emph {et~al.}(2024)\citenamefont
  {Boschini}, \citenamefont {Zonno},\ and\ \citenamefont
  {Damascelli}}]{Boschini24}%
  \BibitemOpen
  \bibfield  {author} {\bibinfo {author} {\bibfnamefont {F.}~\bibnamefont
  {Boschini}}, \bibinfo {author} {\bibfnamefont {M.}~\bibnamefont {Zonno}},\
  and\ \bibinfo {author} {\bibfnamefont {A.}~\bibnamefont {Damascelli}},\
  }\bibfield  {title} {\bibinfo {title} {Time-resolved {ARPES} studies of
  quantum materials},\ }\href {https://doi.org/10.1103/RevModPhys.96.015003}
  {\bibfield  {journal} {\bibinfo  {journal} {Rev. Mod. Phys.}\ }\textbf
  {\bibinfo {volume} {96}},\ \bibinfo {pages} {015003} (\bibinfo {year}
  {2024})}\BibitemShut {NoStop}%
\bibitem [{\citenamefont {Meneghini}\ \emph {et~al.}(2026)\citenamefont
  {Meneghini}, \citenamefont {L\"owe}, \citenamefont {Perea-Causin},
  \citenamefont {Bange}, \citenamefont {Bennecke}, \citenamefont {Reutzel},
  \citenamefont {Mathias},\ and\ \citenamefont {Malic}}]{Meneghini26}%
  \BibitemOpen
  \bibfield  {author} {\bibinfo {author} {\bibfnamefont {G.}~\bibnamefont
  {Meneghini}}, \bibinfo {author} {\bibfnamefont {M.}~\bibnamefont {L\"owe}},
  \bibinfo {author} {\bibfnamefont {R.}~\bibnamefont {Perea-Causin}}, \bibinfo
  {author} {\bibfnamefont {J.~P.}\ \bibnamefont {Bange}}, \bibinfo {author}
  {\bibfnamefont {W.}~\bibnamefont {Bennecke}}, \bibinfo {author}
  {\bibfnamefont {M.}~\bibnamefont {Reutzel}}, \bibinfo {author} {\bibfnamefont
  {S.}~\bibnamefont {Mathias}},\ and\ \bibinfo {author} {\bibfnamefont
  {E.}~\bibnamefont {Malic}},\ }\bibfield  {title} {\bibinfo {title} {A{RPES}
  signatures of trions in van der {W}aals materials},\ }\href
  {https://doi.org/10.1021/acs.nanolett.6c00729} {\bibfield  {journal}
  {\bibinfo  {journal} {Nano Lett.}\ }\textbf {\bibinfo {volume} {26}},\
  \bibinfo {pages} {5866} (\bibinfo {year} {2026})}\BibitemShut {NoStop}%
\bibitem [{\citenamefont {Hern{\'a}ndez~L{\'o}pez}\ \emph
  {et~al.}(2022)\citenamefont {Hern{\'a}ndez~L{\'o}pez}, \citenamefont {Heeg},
  \citenamefont {Schattauer}, \citenamefont {Kovalchuk}, \citenamefont {Kumar},
  \citenamefont {Bock}, \citenamefont {Kirchhof}, \citenamefont {H{\"o}fer},
  \citenamefont {Greben}, \citenamefont {Yagodkin}, \citenamefont {Linhart},
  \citenamefont {Libisch},\ and\ \citenamefont {Bolotin}}]{Lopez22}%
  \BibitemOpen
  \bibfield  {author} {\bibinfo {author} {\bibfnamefont {P.}~\bibnamefont
  {Hern{\'a}ndez~L{\'o}pez}}, \bibinfo {author} {\bibfnamefont
  {S.}~\bibnamefont {Heeg}}, \bibinfo {author} {\bibfnamefont {C.}~\bibnamefont
  {Schattauer}}, \bibinfo {author} {\bibfnamefont {S.}~\bibnamefont
  {Kovalchuk}}, \bibinfo {author} {\bibfnamefont {A.}~\bibnamefont {Kumar}},
  \bibinfo {author} {\bibfnamefont {D.~J.}\ \bibnamefont {Bock}}, \bibinfo
  {author} {\bibfnamefont {J.~N.}\ \bibnamefont {Kirchhof}}, \bibinfo {author}
  {\bibfnamefont {B.}~\bibnamefont {H{\"o}fer}}, \bibinfo {author}
  {\bibfnamefont {K.}~\bibnamefont {Greben}}, \bibinfo {author} {\bibfnamefont
  {D.}~\bibnamefont {Yagodkin}}, \bibinfo {author} {\bibfnamefont
  {L.}~\bibnamefont {Linhart}}, \bibinfo {author} {\bibfnamefont
  {F.}~\bibnamefont {Libisch}},\ and\ \bibinfo {author} {\bibfnamefont {K.~I.}\
  \bibnamefont {Bolotin}},\ }\bibfield  {title} {\bibinfo {title} {Strain
  control of hybridization between dark and localized excitons in a 2{D}
  semiconductor},\ }\href {https://doi.org/10.1038/s41467-022-35352-9}
  {\bibfield  {journal} {\bibinfo  {journal} {Nat. Commun.}\ }\textbf {\bibinfo
  {volume} {13}},\ \bibinfo {pages} {7691} (\bibinfo {year}
  {2022})}\BibitemShut {NoStop}%
\bibitem [{\citenamefont {Rosati}\ \emph
  {et~al.}(2021{\natexlab{d}})\citenamefont {Rosati}, \citenamefont {Lengers},
  \citenamefont {Carmesin}, \citenamefont {Florian}, \citenamefont {Kuhn},
  \citenamefont {Jahnke}, \citenamefont {Lorke},\ and\ \citenamefont
  {Reiter}}]{Rosati21d}%
  \BibitemOpen
  \bibfield  {author} {\bibinfo {author} {\bibfnamefont {R.}~\bibnamefont
  {Rosati}}, \bibinfo {author} {\bibfnamefont {F.}~\bibnamefont {Lengers}},
  \bibinfo {author} {\bibfnamefont {C.}~\bibnamefont {Carmesin}}, \bibinfo
  {author} {\bibfnamefont {M.}~\bibnamefont {Florian}}, \bibinfo {author}
  {\bibfnamefont {T.}~\bibnamefont {Kuhn}}, \bibinfo {author} {\bibfnamefont
  {F.}~\bibnamefont {Jahnke}}, \bibinfo {author} {\bibfnamefont
  {M.}~\bibnamefont {Lorke}},\ and\ \bibinfo {author} {\bibfnamefont {D.~E.}\
  \bibnamefont {Reiter}},\ }\bibfield  {title} {\bibinfo {title} {Electron
  dynamics in a two-dimensional nanobubble: A two-level system based on spatial
  density},\ }\href {https://doi.org/10.1021/acs.nanolett.1c02864} {\bibfield
  {journal} {\bibinfo  {journal} {Nano Lett.}\ }\textbf {\bibinfo {volume}
  {21}},\ \bibinfo {pages} {9896} (\bibinfo {year}
  {2021}{\natexlab{d}})}\BibitemShut {NoStop}%
\bibitem [{\citenamefont {Fitzgerald}\ \emph {et~al.}(2025)\citenamefont
  {Fitzgerald}, \citenamefont {Rosati},\ and\ \citenamefont
  {Malic}}]{Fitzgerald25}%
  \BibitemOpen
  \bibfield  {author} {\bibinfo {author} {\bibfnamefont {J.~M.}\ \bibnamefont
  {Fitzgerald}}, \bibinfo {author} {\bibfnamefont {R.}~\bibnamefont {Rosati}},\
  and\ \bibinfo {author} {\bibfnamefont {E.}~\bibnamefont {Malic}},\ }\bibfield
   {title} {\bibinfo {title} {Polariton transport in 2d semiconductors:
  Phonon-mediated transitions between ballistic, superdiffusive, and
  exciton-limited regimes},\ }\href {https://doi.org/10.1126/sciadv.aea3495}
  {\bibfield  {journal} {\bibinfo  {journal} {Sci. Adv.}\ }\textbf {\bibinfo
  {volume} {11}},\ \bibinfo {pages} {eaea3495} (\bibinfo {year}
  {2025})}\BibitemShut {NoStop}%
\bibitem [{\citenamefont {Kuhn}\ and\ \citenamefont {Rossi}(1992)}]{Kuhn92}%
  \BibitemOpen
  \bibfield  {author} {\bibinfo {author} {\bibfnamefont {T.}~\bibnamefont
  {Kuhn}}\ and\ \bibinfo {author} {\bibfnamefont {F.}~\bibnamefont {Rossi}},\
  }\bibfield  {title} {\bibinfo {title} {Monte carlo simulation of ultrafast
  processes in photoexcited semiconductors: Coherent and incoherent dynamics},\
  }\href {https://doi.org/10.1103/PhysRevB.46.7496} {\bibfield  {journal}
  {\bibinfo  {journal} {Phys. Rev. B}\ }\textbf {\bibinfo {volume} {46}},\
  \bibinfo {pages} {7496} (\bibinfo {year} {1992})}\BibitemShut {NoStop}%
\bibitem [{\citenamefont {Rosati}\ \emph {et~al.}(2015)\citenamefont {Rosati},
  \citenamefont {Dolcini},\ and\ \citenamefont {Rossi}}]{Rosati15b}%
  \BibitemOpen
  \bibfield  {author} {\bibinfo {author} {\bibfnamefont {R.}~\bibnamefont
  {Rosati}}, \bibinfo {author} {\bibfnamefont {F.}~\bibnamefont {Dolcini}},\
  and\ \bibinfo {author} {\bibfnamefont {F.}~\bibnamefont {Rossi}},\ }\bibfield
   {title} {\bibinfo {title} {Electron-phonon coupling in metallic carbon
  nanotubes: Dispersionless electron propagation despite dissipation},\ }\href
  {https://doi.org/10.1103/PhysRevB.92.235423} {\bibfield  {journal} {\bibinfo
  {journal} {Phys. Rev. B}\ }\textbf {\bibinfo {volume} {92}},\ \bibinfo
  {pages} {235423} (\bibinfo {year} {2015})}\BibitemShut {NoStop}%
\bibitem [{\citenamefont {Iff}\ \emph {et~al.}(2019)\citenamefont {Iff},
  \citenamefont {Tedeschi}, \citenamefont {Mart{\'i}n-S{\'a}nchez},
  \citenamefont {Mocza{\l}a-Dusanowska}, \citenamefont {Tongay}, \citenamefont
  {Yumigeta}, \citenamefont {Taboada-Guti{\'e}rrez}, \citenamefont {Savaresi},
  \citenamefont {Rastelli}, \citenamefont {Alonso-Gonz{\'a}lez}, \citenamefont
  {H{\"o}fling}, \citenamefont {Trotta},\ and\ \citenamefont
  {Schneider}}]{Iff19}%
  \BibitemOpen
  \bibfield  {author} {\bibinfo {author} {\bibfnamefont {O.}~\bibnamefont
  {Iff}}, \bibinfo {author} {\bibfnamefont {D.}~\bibnamefont {Tedeschi}},
  \bibinfo {author} {\bibfnamefont {J.}~\bibnamefont {Mart{\'i}n-S{\'a}nchez}},
  \bibinfo {author} {\bibfnamefont {M.}~\bibnamefont {Mocza{\l}a-Dusanowska}},
  \bibinfo {author} {\bibfnamefont {S.}~\bibnamefont {Tongay}}, \bibinfo
  {author} {\bibfnamefont {K.}~\bibnamefont {Yumigeta}}, \bibinfo {author}
  {\bibfnamefont {J.}~\bibnamefont {Taboada-Guti{\'e}rrez}}, \bibinfo {author}
  {\bibfnamefont {M.}~\bibnamefont {Savaresi}}, \bibinfo {author}
  {\bibfnamefont {A.}~\bibnamefont {Rastelli}}, \bibinfo {author}
  {\bibfnamefont {P.}~\bibnamefont {Alonso-Gonz{\'a}lez}}, \bibinfo {author}
  {\bibfnamefont {S.}~\bibnamefont {H{\"o}fling}}, \bibinfo {author}
  {\bibfnamefont {R.}~\bibnamefont {Trotta}},\ and\ \bibinfo {author}
  {\bibfnamefont {C.}~\bibnamefont {Schneider}},\ }\bibfield  {title} {\bibinfo
  {title} {Strain-tunable single photon sources in {WSe}$_2$ monolayers},\
  }\href {https://doi.org/10.1021/acs.nanolett.9b02221} {\bibfield  {journal}
  {\bibinfo  {journal} {Nano Lett.}\ }\textbf {\bibinfo {volume} {19}},\
  \bibinfo {pages} {6931} (\bibinfo {year} {2019})}\BibitemShut {NoStop}%
\bibitem [{\citenamefont {An}\ \emph {et~al.}(2023)\citenamefont {An},
  \citenamefont {Soubelet}, \citenamefont {Zhumagulov}, \citenamefont {Zopf},
  \citenamefont {Delhomme}, \citenamefont {Qian}, \citenamefont {Faria~Junior},
  \citenamefont {Fabian}, \citenamefont {Cao}, \citenamefont {Yang},
  \citenamefont {Stier}, \citenamefont {Ding},\ and\ \citenamefont
  {Finley}}]{An23}%
  \BibitemOpen
  \bibfield  {author} {\bibinfo {author} {\bibfnamefont {Z.}~\bibnamefont
  {An}}, \bibinfo {author} {\bibfnamefont {P.}~\bibnamefont {Soubelet}},
  \bibinfo {author} {\bibfnamefont {Y.}~\bibnamefont {Zhumagulov}}, \bibinfo
  {author} {\bibfnamefont {M.}~\bibnamefont {Zopf}}, \bibinfo {author}
  {\bibfnamefont {A.}~\bibnamefont {Delhomme}}, \bibinfo {author}
  {\bibfnamefont {C.}~\bibnamefont {Qian}}, \bibinfo {author} {\bibfnamefont
  {P.~E.}\ \bibnamefont {Faria~Junior}}, \bibinfo {author} {\bibfnamefont
  {J.}~\bibnamefont {Fabian}}, \bibinfo {author} {\bibfnamefont
  {X.}~\bibnamefont {Cao}}, \bibinfo {author} {\bibfnamefont {J.}~\bibnamefont
  {Yang}}, \bibinfo {author} {\bibfnamefont {A.~V.}\ \bibnamefont {Stier}},
  \bibinfo {author} {\bibfnamefont {F.}~\bibnamefont {Ding}},\ and\ \bibinfo
  {author} {\bibfnamefont {J.~J.}\ \bibnamefont {Finley}},\ }\bibfield  {title}
  {\bibinfo {title} {Strain control of exciton and trion spin-valley dynamics
  in monolayer transition metal dichalcogenides},\ }\href
  {https://doi.org/10.1103/PhysRevB.108.L041404} {\bibfield  {journal}
  {\bibinfo  {journal} {Phys. Rev. B}\ }\textbf {\bibinfo {volume} {108}},\
  \bibinfo {pages} {L041404} (\bibinfo {year} {2023})}\BibitemShut {NoStop}%
\bibitem [{\citenamefont {Gant}\ \emph {et~al.}(2019)\citenamefont {Gant},
  \citenamefont {Huang}, \citenamefont {{Perez de Lara}}, \citenamefont {Guo},
  \citenamefont {Frisenda},\ and\ \citenamefont {Castellanos-Gomez}}]{Gant19}%
  \BibitemOpen
  \bibfield  {author} {\bibinfo {author} {\bibfnamefont {P.}~\bibnamefont
  {Gant}}, \bibinfo {author} {\bibfnamefont {P.}~\bibnamefont {Huang}},
  \bibinfo {author} {\bibfnamefont {D.}~\bibnamefont {{Perez de Lara}}},
  \bibinfo {author} {\bibfnamefont {D.}~\bibnamefont {Guo}}, \bibinfo {author}
  {\bibfnamefont {R.}~\bibnamefont {Frisenda}},\ and\ \bibinfo {author}
  {\bibfnamefont {A.}~\bibnamefont {Castellanos-Gomez}},\ }\bibfield  {title}
  {\bibinfo {title} {A strain tunable single-layer {MoS}$_2$ photodetector},\
  }\href {https://doi.org/https://doi.org/10.1016/j.mattod.2019.04.019}
  {\bibfield  {journal} {\bibinfo  {journal} {Mater. Today}\ }\textbf {\bibinfo
  {volume} {27}},\ \bibinfo {pages} {8} (\bibinfo {year} {2019})}\BibitemShut
  {NoStop}%
\bibitem [{\citenamefont {Henriquez-Guerra}\ \emph {et~al.}(2023)\citenamefont
  {Henriquez-Guerra}, \citenamefont {Li}, \citenamefont {Pasques-Gramage},
  \citenamefont {Gos\'albez-Mart\'inez}, \citenamefont {D’Agosta},
  \citenamefont {Castellanos-Gomez},\ and\ \citenamefont
  {Calvo}}]{Henriquez23}%
  \BibitemOpen
  \bibfield  {author} {\bibinfo {author} {\bibfnamefont {E.}~\bibnamefont
  {Henriquez-Guerra}}, \bibinfo {author} {\bibfnamefont {H.}~\bibnamefont
  {Li}}, \bibinfo {author} {\bibfnamefont {P.}~\bibnamefont
  {Pasques-Gramage}}, \bibinfo {author} {\bibfnamefont {D.}~\bibnamefont
  {Gos\'albez-Mart\'inez}}, \bibinfo {author} {\bibfnamefont {R.}~\bibnamefont
  {D’Agosta}}, \bibinfo {author} {\bibfnamefont {A.}~\bibnamefont
  {Castellanos-Gomez}},\ and\ \bibinfo {author} {\bibfnamefont {M.~R.}\
  \bibnamefont {Calvo}},\ }\bibfield  {title} {\bibinfo {title} {Large biaxial
  compressive strain tuning of neutral and charged excitons in single-layer
  transition metal dichalcogenides},\ }\href
  {https://doi.org/10.1021/acsami.3c13281} {\bibfield  {journal} {\bibinfo
  {journal} {ACS Appl. Mater. Interfaces}\ }\textbf {\bibinfo {volume} {15}},\
  \bibinfo {pages} {57369} (\bibinfo {year} {2023})}\BibitemShut {NoStop}%
\bibitem [{\citenamefont {Dirnberger}\ \emph {et~al.}(2021)\citenamefont
  {Dirnberger}, \citenamefont {Ziegler}, \citenamefont {Junior}, \citenamefont
  {Bushati}, \citenamefont {Taniguchi}, \citenamefont {Watanabe}, \citenamefont
  {Fabian}, \citenamefont {Bougeard}, \citenamefont {Chernikov},\ and\
  \citenamefont {Menon}}]{Dirnberger21}%
  \BibitemOpen
  \bibfield  {author} {\bibinfo {author} {\bibfnamefont {F.}~\bibnamefont
  {Dirnberger}}, \bibinfo {author} {\bibfnamefont {J.~D.}\ \bibnamefont
  {Ziegler}}, \bibinfo {author} {\bibfnamefont {P.~E.~F.}\ \bibnamefont
  {Junior}}, \bibinfo {author} {\bibfnamefont {R.}~\bibnamefont {Bushati}},
  \bibinfo {author} {\bibfnamefont {T.}~\bibnamefont {Taniguchi}}, \bibinfo
  {author} {\bibfnamefont {K.}~\bibnamefont {Watanabe}}, \bibinfo {author}
  {\bibfnamefont {J.}~\bibnamefont {Fabian}}, \bibinfo {author} {\bibfnamefont
  {D.}~\bibnamefont {Bougeard}}, \bibinfo {author} {\bibfnamefont
  {A.}~\bibnamefont {Chernikov}},\ and\ \bibinfo {author} {\bibfnamefont
  {V.~M.}\ \bibnamefont {Menon}},\ }\bibfield  {title} {\bibinfo {title}
  {Quasi-{1D} exciton channels in strain-engineered {2D} materials},\ }\href
  {https://doi.org/10.1126/sciadv.abj3066} {\bibfield  {journal} {\bibinfo
  {journal} {Sci. Adv.}\ }\textbf {\bibinfo {volume} {7}},\ \bibinfo {pages}
  {eabj3066} (\bibinfo {year} {2021})}\BibitemShut {NoStop}%
\end{thebibliography}
\end{document}


\title{Supplemental Material for \\ Strain engineering of transient exciton diffusion in WSe$_2$ monolayers\\ at cryogenic temperatures}

\author{Roberto Rosati}
\email{rosatir@staff.uni-marburg.de}
\affiliation{Department of Physics, Philipps-Universit\"at Marburg, Renthof 7, D-35032 Marburg, Germany}

\author{Mohammed Adel Aly}
\affiliation{Department of Physics, University of M\"unster, Wilhelm-Klemm-Strasse 10, D-48149 M\"unster, Germany}
\affiliation{Department of Physics, Faculty of Science, Ain Shams University, 11566 Cairo, Egypt}

\author{Robert Schmidt}
\affiliation{Department of Physics, University of M\"unster, Wilhelm-Klemm-Strasse 10, D-48149 M\"unster, Germany}

\author{Rudolf Bratschitsch}
\affiliation{Department of Physics, University of M\"unster, Wilhelm-Klemm-Strasse 10, D-48149 M\"unster, Germany}

\author{Ermin Malic}
\affiliation{Department of Physics, Philipps-Universit\"at Marburg, Renthof 7, D-35032 Marburg, Germany}

\maketitle



\section{Abrupt strain-induced change in PL}\label{Sec:theory-landscape}

The time-resolved photoluminescence  is abruptly modified by strain, as discussed in Fig. 2 of the main paper. Bright KK and dark K$\Lambda$ states have opposite strain gauge factors. As a consequence, tensile strain deactivates K$\Lambda$ states by increasing their separation to optically excited KK states. This significantly modifies the formation of  energetically lowest dark KK$^\prime$ excitons. For $s\gtrsim s_0\approx 0.3\,\%$, hot dark KK$^\prime$ excitons with a larger excess energy are formed, as KK excitons do not scatter anymore via K$\Lambda$ states. This value originates from the energy exciton landscape, which we microscopically calculate by solving the Wannier equations including first-principle input on effective electron and hole masses \cite{Kormanyos15,Khatibi18}. Note that we phenomenologically include a 10\,meV redshift of KK excitons to obtain a better agreement with previous experiments \cite{Kumar24}. 

\begin{figure}[b!]
 \centering
 \includegraphics[width=0.8\textwidth]{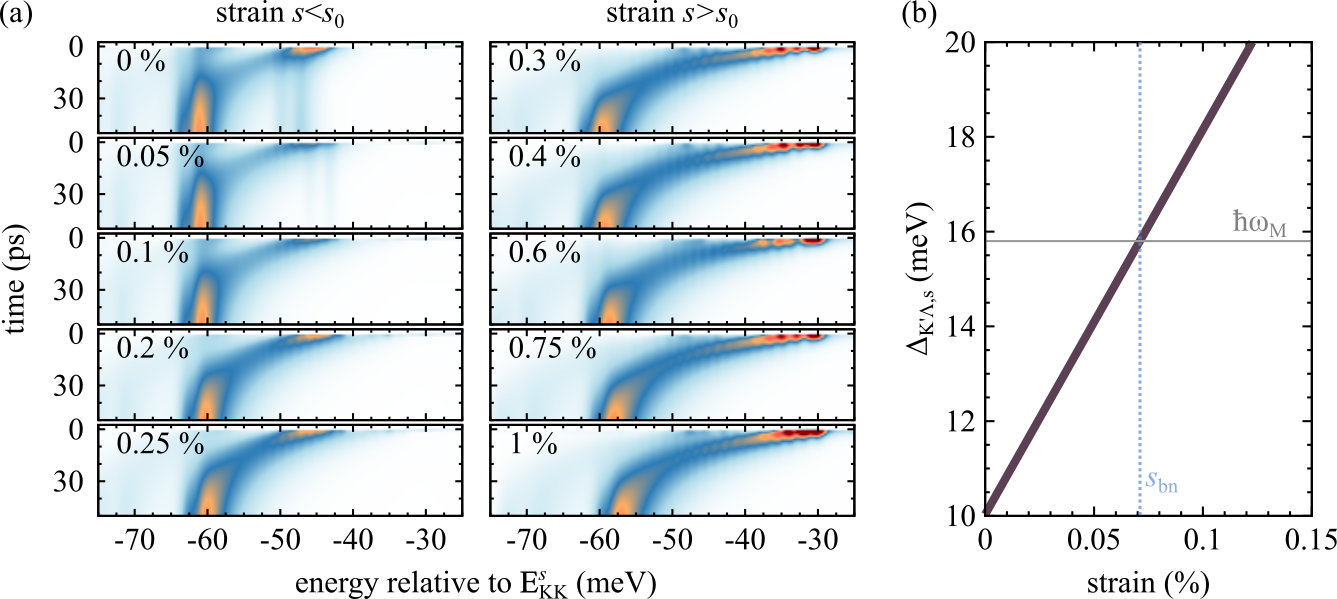}
 \caption{\textbf{Abrupt strain-induced change of phonon sidebands.}
(a) Time- and energy-resolved photoluminescence spectrum at different values of strain smaller (left column) and larger (right column) than the critical strain $s_0\approx0.3\,\%$.  The initial signal abruptly increases from 15\,meV to almost 30\,meV for strain values above $s_0$, while the PL is very similar for all values smaller (left) and larger than $s_0$ (right).  (b) Energy difference $\Delta_{\text{K}^\prime\Lambda,s}=$E$_{\text{K}\Lambda,s}-$E$_{\text{KK}^\prime,s}$ between  K$\Lambda$ and KK$^\prime$ excitons: only tensile strain larger than $s_{\text{bn}}\approx 0.05\,\%$ allows to open intervalley scattering  via emission of acoustic M phonons with the average energy $\hbar \omega_\text{M}$.}
 \label{figS1}
\end{figure}

A higher excess energy of excitons results in  an initially higher emission energy in PL spectra, as shown in Fig. 2 of the main paper and in Fig. S\ref{figS1}(a) for more strain values. For all considered strain values  smaller than $s_0$, the initial PL signature is located 15 \, meV above the equilibrium PL peak. This changes abruptly for  $s>s_0$, where the initial signal is shifted up to 30\, meV above the equilibrium resonance.
This abrupt change reflects the closing of the scattering channel from KK to K$\Lambda$ excitons for strain above $s_0$.  In contrast, strain values away from $s_0$ do not qualitatively change the exciton energy landscape nor the phonon-mediated scattering, and thus the PL is largely independent of strain. 

Furthermore, the energy separation between phonon sidebands at equilibrium and the bright exciton peak at  E$_{\text{KK}, s}$ decreases with strain from about 60\,meV at $s=0\,\%$ to approximately 57\,meV at $s=1\,\%$. This occurs due to the slightly different strain gauge factors of KK and KK$^\prime$ excitons, similarly to what was observed for the energy separation between spin-bright and spin-dark KK excitons  (the latter behaving as KK$^\prime$ states) \cite{Dirnberger21}. In the unstrained case, we find a long-lasting signal approximately 12\,meV above the equilibrium emission, which is attributed to  the phonon-sideband of K$\Lambda$ excitons \cite{Rosati20b}. This reflects a transient overpopulation of these states due to a phonon bottleneck, which inhibits  intervalley scattering from K$\Lambda$ to KK$^\prime$ excions, when the energy separation $\Delta_{\text{K}^\prime\Lambda,s}=$E$_{\text{K}\Lambda,s}-$E$_{\text{KK}^\prime,s}$ between K$\Lambda$ and KK$^\prime$ is smaller than the average energy $\hbar \omega_{\text{M}}$ of acoustic M phonons. Such a bottleneck is circumvented by tensile strain larger than approximately $s_{\text{bn}}\approx0.05\,\%$, see \fig{figS1}(b).

\begin{figure}[t!]
 \centering
 \includegraphics[width=\textwidth]{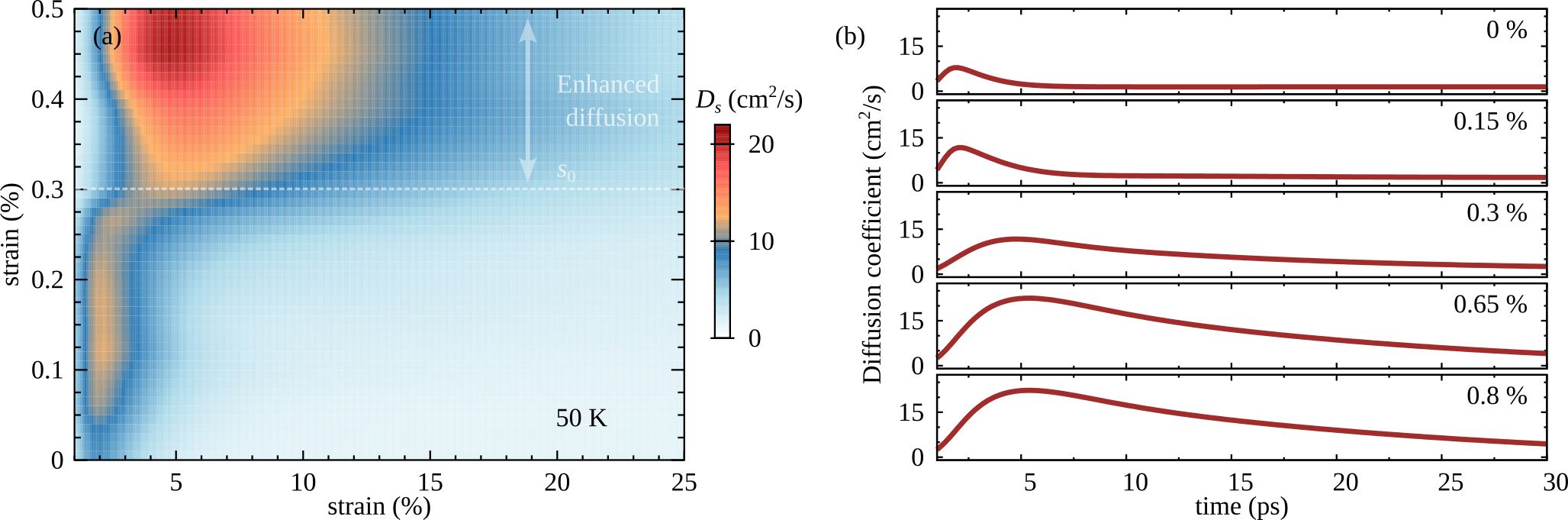}
 \caption{\textbf{Strain-induced change in exciton diffusion at 50\,K.}
(a) Time- and strain-resolved diffusion coefficient at $T=$50\,K and (b) cuts at given strain values. Compared to the case at $T=20\,$K in Fig. 3(a) in the main manuscript, the transient diffusion  is suppressed at small strain values and shows a smoother, two-fold speed-up for strain increasing from $s_0$ to 0.4-0.5\,\%.}
 \label{figS2}
\end{figure}

As discussed in the main paper, the strain-induced deactivation of K$\Lambda$ excitons has also an important impact on exciton propagation. At $T=$20\,K, the maximum exciton diffusion becomes abruptly larger for strain values above $s_0$, see Fig. 2(a) of the main text. Furthermore, for $T=50\,$K a suppression of diffusion for  $s=0.15\,\%$ was found, but not for $s=0.5\,\%$. In Fig. S\ref{figS2}, we show the  effective diffusion coefficient for different values of strain. This is evaluated from the full spatiotemporal PL, from which the squared width  reads
 $\sigma_s^2(t)=\int_0^{E_{\text{c},s}} dE \int \mathbf{r}^2I_s(E,\mathbf{r},t) d\mathbf{r} /2\int_0^{E_{\text{c},s}} dE \int I_s(E,\mathbf{r},t) d\mathbf{r}$ for $E\leq E_{\text{c},s}\equiv\text{E}_{\text{KK},s}-\Delta_E$ with the spectral cut-off $E_{\text{c},s}$ located $\Delta_E=10$\,meV below the bright-exciton energy to mimic experimental broadening \cite{Rosati20b}. The diffusion coefficients is numerically evaluated as $D_s(t)=(\sigma_s^2(t)-\sigma_s^2(t-\Delta_t))/(2\Delta_t)$ with  $\Delta_t=1\,$ps reflecting  realistic experimental sensitivity \cite{Rosati21c} (and about three orders or magnitudes larger than the time-step adopted to solve the spatiotemporal exciton dynamics).
The predicted suppression in diffusion at  $T=50\,$K occurs due to the thermal activation of the scattering from KK$^\prime$ excitons to the energetically higher K$\Lambda$ states. 
At $T=$20\,K, the mechanisms is inefficient because it requires absorption of intervalley acoustic phonons with an energy of 15-16\,meV. At a slightly larger temperature of $T=$50\,K these scattering channels become thermally activated, resulting in a smaller diffusion at small strain values, see the top two panels of \fig{figS2}(b) as well as the full surface plot in \fig{figS2}(a). In contrast, at much larger strain values, the transient diffusion remains unaffected also at 50\,K, see the low two panels in \fig{figS2}(b). Here, strain closes the scattering channel from KK$^\prime$ to K$\Lambda$ excitons  through an increased energy separation between K$\Lambda$ and KK$^\prime$ states. As a result, we have a qualitative difference in diffusion for strain values smaller and larger than $s_0$.
In the intermediate range of strain values $s_0\lesssim s\lesssim 0.4-0.5\,\%$, K$\Lambda$ is energetically too close to KK to allow KK-to-K$\Lambda$ scattering, but still close enough to KK$^\prime$ to allow  intervalley scattering of hot KK$^\prime$ excitons. Together with the larger thermal broadening of the transient distributions, this leads to
a variation of the transient diffusion with strain that is less abrupt at $T=50\,$K than at 20\,K. In particular, the maximum diffusion smoothly increases from 10 to 20 cm$^2$/s for strain values going from $s_0$ to 0.4-0.5\%, cf. Fig. S\ref{figS2}.

\section{Rise time of fast exciton diffusion}

\begin{figure}[t!]
 \centering
 \includegraphics[width=0.6\textwidth]{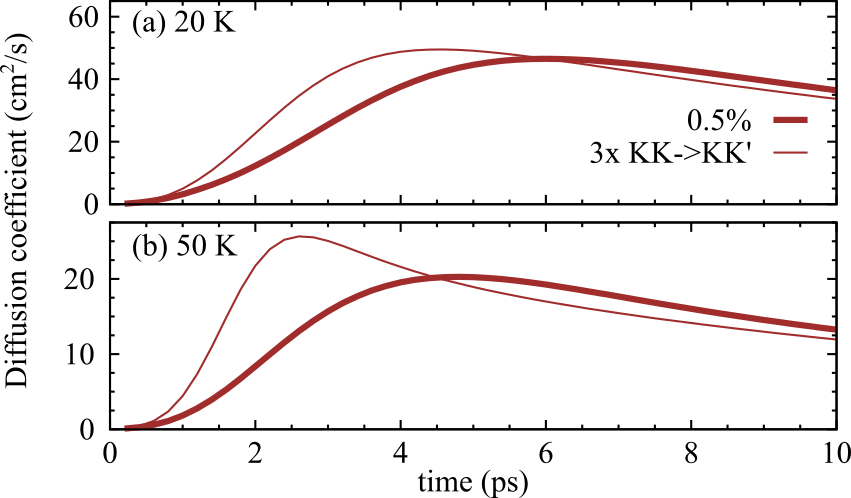}
 \caption{\textbf{Rise-time of fast exciton diffusion:}
Time-resolved exciton diffusion coefficient at (a) $T=\,$20\,K and (b)  $T=\,$50\,K for an uniaxial strain $s=0.5\,\%$. The realistic case (thick lines) is compared to the artificial situation, where the scattering from KK to KK$^\prime$ excitons is increased by a factor of three (thin lines). A more efficient intervalley scattering leads to a shorter rise-time of fast excitons.}
 \label{figS3}
\end{figure}

In the main manuscript, it has been discussed how the transient exciton diffusion  becomes faster for  $s>s_0$ due to the deactivation of K$\Lambda$ excitons. Interestingly, the opposite takes place in the very first 2-3 ps, where a higher diffusion is found for $s<s_0$, see Fig.  3(b) in the main manuscript. This reflects a shorter  formation time of hot KK$^\prime$ excitons  for $s<s_0$ because of the more efficient scattering from KK to K$\Lambda$ excitons (due to the three-fold degeneracy of K$\Lambda$ states). In \fig{figS3} we explore the  situation, where the KK-to-KK$^\prime$ scattering  is artificially increased by a factor of three (thin lines).
We  focus on the case of $s=0.5\,\%>s_0$, where the impact of K$\Lambda$ excitons is negligible. A more efficient direct KK-to-KK$^\prime$ scattering  leads to a faster formation of hot dark KK$^\prime$ excitons. This results in a shorter rise-time of fast exciton diffusion, see thin and solid lines in \fig{figS3}. Furthermore, the faster formation of hot excitons also results in a slight increase of the maximum diffusion.

\section{References}

\bibliography{rosati_bibShort}